\documentclass{article}

\usepackage{amsmath, amsfonts, amssymb, amsxtra, amsopn}
\usepackage{graphicx}
\usepackage{multirow}
\usepackage{algorithmic}
\usepackage{longtable}
\usepackage{booktabs}
\usepackage{listings}
\usepackage{cmap}
\usepackage{colortbl}
\usepackage{pgfplots}
\usepackage{pgfplotstable}
\usepackage{adjustbox}

\usepackage{xstring}%%% needed for \rowvec macros

\PassOptionsToPackage{hyphens}{url}

\usepackage{hyperref}
\hypersetup{colorlinks=true,linkcolor=black,citecolor=black,urlcolor=blue,filecolor=black}
\hypersetup{pdfpagemode=UseNone,pdfstartview=}

\usepackage[tableposition=top]{caption}
\def\z{\phantom{0}}
    
\usepackage{enumitem}
\setlist[itemize]{noitemsep, topsep=0pt}

\advance\oddsidemargin by -0.25in
\advance\textwidth by 0.5in

\advance\topmargin by -0.25in
\advance\textheight by 0.5in

\def\eref#1{(\ref{#1})}

\DeclareMathOperator{\maj}{maj}

\def\S{{S}}
\def\Sb{\widehat{\S}}
\def\dd{\hspace*{0.05em};\hspace*{-0.05em}}

\def\pr{{\hspace*{-0.12em}}'}

\long\def\symbolfootnotetext[#1]#2{\begingroup%
\def\thefootnote{\fnsymbol{footnote}}\footnotetext[#1]{#2}\endgroup}

\def\srowvecc#1#2{(\!\begin{array}{cc} 
      \noexpandarg\IfBeginWith{#1}{-}{\! #1}{#1}
    & #2\kern-0.5pt\end{array}\!)}
\def\rowvecc#1#2{\left(\!\begin{array}{cc} 
      \noexpandarg\IfBeginWith{#1}{-}{\! #1}{#1}
    & #2\kern-0.5pt\end{array}\!\right)}
\def\rowveccc#1#2#3{\left(\!\begin{array}{ccc} 
      \noexpandarg\IfBeginWith{#1}{-}{\! #1}{#1}
    & #2 
    & #3\kern-0.5pt\end{array}\!\right)}
\def\rowvecccc#1#2#3#4{\left(\!\begin{array}{cccc}
      \noexpandarg\IfBeginWith{#1}{-}{\! #1}{#1}
    & #2 
    & #3 
    & #4\kern-0.5pt\end{array}\!\right)}
\def\srowvecccc#1#2#3#4{\bigl(\!\begin{array}{cccc}
      \noexpandarg\IfBeginWith{#1}{-}{\! #1}{#1}
    & #2 
    & #3 
    & #4\kern-0.5pt\end{array}\!\bigr)}
\def\rowveccccc#1#2#3#4#5{\left(\!\begin{array}{ccccc} 
      \noexpandarg\IfBeginWith{#1}{-}{\! #1}{#1}
    & #2
    & #3
    & #4
    & #5\kern-0.5pt\end{array}\!\right)}
\def\srowvecccccc#1#2#3#4#5#6{(\!\begin{array}{cccccc} 
      \noexpandarg\IfBeginWith{#1}{-}{\! #1}{#1}
    & #2
    & #3
    & #4
    & #5
    & #6\kern-0.5pt\end{array}\!)}
\def\rowvecccccc#1#2#3#4#5#6{\left(\!\begin{array}{cccccc} 
      \noexpandarg\IfBeginWith{#1}{-}{\! #1}{#1}
    & #2
    & #3
    & #4
    & #5
    & #6\kern-0.5pt\end{array}\!\right)}

\def\w{\omega}
\def\p{\lambda}
\def\P{\Lambda}

\def\sscoin{%
  \leavevmode
  \vtop{\offinterlineskip %\bfseries
    \setbox0=\hbox{\scriptsize S}%
    \setbox2=\hbox to\wd0{\hfil\hskip-.03em
    \vrule height .3ex width .08ex\hskip .08em
    \vrule height .3ex width .08ex\hfil}
    \vbox{\copy2\box0}\box2}}
\newcommand\affil[2]{%
  \begingroup
  \renewcommand\thefootnote{}\footnote{\llap{$\hbox{}^{#1}\hbox{}$}#2}%
  \addtocounter{footnote}{-1}%
  \endgroup
}
\newcommand\markonly[1]{%
$\hbox{}^{\mbox{\kern4.5pt,\kern0.75pt #1}}$
}

\advance\oddsidemargin by -0.525in
\advance\textwidth by 1.05in
\advance\topmargin by -0.625in
\advance\textheight by 1.25in

\title{\vspace{-0.5in}On Ensemble Learning}

\author{Mark Stamp\footnote{mark.stamp@sjsu.edu}\markonly{\sscoin}\ \ 
\setcounter{footnote}{2}
Aniket Chandak\footnote{chandakaniket537@gmail.com}\ \ \ \ 
Gavin Wong\footnote{gavinmwong@gmail.com}\ \ \ \ 
Allen Ye\footnote{allenye66@gmail.com}
}
\date{}

\begin{document}

\maketitle

\vglue-0.35in

\affil{\sscoin}{Department 
of Computer Science,
San Jose State University,
San Jose, California}

\abstract
In this paper, 
we consider ensemble classifiers, that is, machine learning based
classifiers that utilize a combination of scoring functions. We provide a
framework for categorizing such classifiers, and we outline several 
ensemble techniques, discussing how each fits into our framework. From this
general introduction, we then pivot to the topic of
ensemble learning within the context of malware analysis.
We present a brief survey of some of the ensemble techniques that have
been used in malware (and related) research. We conclude with an extensive
set of experiments, where we apply ensemble techniques to a large and challenging 
malware dataset. While many of these ensemble techniques have appeared 
in the malware literature, previously there has been no way to directly compare
results such as these, as different datasets and different 
measures of success are typically used. Our common framework and 
empirical results are an effort to bring some sense of order to the chaos
that is evident in the evolving field of ensemble learning---both
within the narrow confines of the malware analysis problem, and in the
larger realm of machine learning in general.

\section{Introduction}\label{chap:intro}

In ensemble learning, multiple learning algorithms are combined, with the goal of improved
accuracy as compared to the individual algorithms. Ensemble techniques are widely used, 
and as a testament to their strength, ensembles have won numerous machine learning contests 
in recent years, including the KDD Cup~\cite{KDD}, the Kaggle competition~\cite{Kaggle}, 
and the Netflix prize~\cite{Netflix}. 

Many such ensembles resemble Frankenstein's monster~\cite{shelley1869frankenstein}, 
in the sense that they are an agglomeration of disparate components, with some 
of the components being of questionable value---an ``everything and the kitchen sink''
approach clearly prevails. This effect can be clearly observed in
the aforementioned machine learning contests, where there
is little (if any) incentive to make systems that are efficient or
practical, as accuracy is typically the only criteria for success. 
In the case of the Netflix prize, the winning team was awarded~\$1,000,000, yet
Netflix never implement the winning scheme,
since the improvements in accuracy ``did not seem to justify the engineering effort 
needed to bring them into a production environment''~\cite{netflixBlog}.
In real-world systems, practicality and efficiency are necessarily crucial factors. 

In this paper, we provide a straightforward framework for 
categorizing ensemble techniques. We then consider specific (and relatively simple)
examples of various categories of such ensembles, and we
show how these fit into our framework. For various examples of 
ensembles, we also provide experimental results, based on a large and diverse malware
dataset. 

While many of the techniques that we consider have previously appeared
in the malware literature, we are not aware of any comparable study
focused on the effectiveness of various ensembles
using a common dataset and common measures of success.
While we believe that these examples are interesting in their own
right, they also provide a basis for discussing various tradeoffs between
measures of accuracy and practical considerations. 

The remainder of this paper is organized as follows. In Section~\ref{sect:EC}
we discuss ensemble classifiers, including our framework for categorizing
such classifiers. 
%Section~\ref{sect:dm} discusses the dataset used
%in our experiments, the features that we extract, 
%and the various metrics we employ to quantify our
%experimental results.
Section~\ref{sect:exp} contains our experimental
results. 
This section also includes a discussion of our dataset, scoring metrics,
software used, and so on.
Finally, Section~\ref{sect:conc} 
concludes the paper and includes suggestions for future work.

\section{Ensemble Classifiers}\label{sect:EC}

In this section, we first give a selective survey of some examples of 
malware (and closely related) research involving ensemble learning.
Then we provide a framework for discussing ensemble classifiers
in general.

\subsection{Examples of Related Work}

The paper~\cite{Kittler1998} discusses various ways to combine classifiers
and provides a theoretical framework for such combinations. The focus is 
on straightforward combinations, such as a maximum, sum, product, majority vote,
and so on. The work in~\cite{Kittler1998} has clearly been influential,
but it seems somewhat dated, given the wide variety of ensemble methods 
that are used today.

The book~\cite{Kuncheva2004} presents the topic of ensemble learning from a similar perspective 
as~\cite{Kittler1998} but in much more detail. Perhaps not surprisingly, the 
more recent book~\cite{Zhou2012} seems to have a somewhat more modern perspective
with respect to ensemble methods, but retains the theoretical flavor
of~\cite{Kuncheva2004} and~\cite{Kittler1998}.
The brief blog at~\cite{Smolyakov2017} provides a highly readable 
(if highly selective) summary of some of the topics covered in the 
books~\cite{Kuncheva2004} and~\cite{Zhou2012}. 
%This paper is recommended
%as a gentle introduction to the topic of ensemble methods.

Here, we take an approach that is, in some sense, 
more concrete than that in~\cite{Kittler1998,Kuncheva2004,Zhou2012}. Our objective
is to provide a relatively straightforward framework for categorizing and discussing ensemble 
techniques. We then use this framework as a frame of reference for experimental
results based on a variety of ensemble methods.

%\subsubsection{Ensemble Classifiers in Information Security}

%In this section, we give selected examples of previous research where ensemble methods
%have been applied to problems in information security. Our primary focus here is on 
%malware detection and classification, but we include a few examples where ensemble techniques
%have been applied to closely related research problems.

Table~\ref{tab:ensemble} provides a summary of several research papers where ensemble
techniques have been applied to security-related problems. The emphasis here is on malware, 
but we have also included a few closely related topics. In any case, this represents a small sample of
the many papers that have been published, and is only intended to provide an indication
as to the types and variety of ensemble strategies that have been considered to date.
On this list, we see examples of ensemble methods based on bagging, boosting, and
stacking, as discussed below in Section~\ref{sect:CEC}.

\begin{table}[!htb]
\centering
\caption{Security research papers using ensemble classifiers}\label{tab:ensemble}
{%\footnotesize
\begin{tabular}{c|ccc} \midrule\midrule
\textbf{Authors} & \textbf{Application} & \textbf{Features} & \textbf{Ensemble} \\ \midrule
Alazab et al.~\cite{alazab2011zero} & Detection & API calls & Neural networks \\ % Self organizing neural networks
Comar et al.~\cite{comar2013combining} & Detection & Network traffic & Random forest \\
Dimja{\v{s}}evic et al.~\cite{dimjavsevic2015android} & Android & System calls & RF and SVM \\
Guo et al.~\cite{guo2010malware} & Detection & API calls & BKS \\ % BKS theory ?????
Idrees et al.~\cite{idrees2017pindroid} & Android & Permissions, intents & RF and others \\
Jain \&\ Meena~\cite{jain2011byte} & Detection & Byte $n$-grams & AdaBoost \\
Khan et al.~\cite{khan2016fractal} & Detection & Network based & Boosting \\ 
Kong \&\ Yan~\cite{kong2013discriminant} & Classification & Function call graph & Boosting \\
Morales et al.~\cite{morales2016native} & Android & Permissions & Several \\ % Ensemble ?????
Narouei et al.~\cite{narouei2015dllminer} & Detection & DLL dependency & Random forest \\ % Ensemble ?????
Shahzad et al.~\cite{shahzad2013comparative} & Detection & Opcodes & Voting \\
Sheen et al.~\cite{sheen2013malware} &  Various & Detection efficiency & Pruning \\
Singh et al.~\cite{SinghTVAS16} & Detection & Opcodes & SVM \\
Smutz \&\ Stavrou~\cite{smutz2012malicious} & Malicious PDF & Metadata & Random forest \\
Toolan \&\ Carthy~\cite{toolan2009phishing} & Phishing & Various & C5.0, boosting \\
Ye et al.~\cite{ye2009sbmds} & Detection & API calls, strings & SVM, bagging \\
Ye et al.~\cite{ye2010automatic} & Categorization & Opcodes & Clustering \\
Yerima et al.~\cite{yerima2015high} & Zero day & 179 features & RF, regression \\ % "bagging" and "boosting"
Zhang et al.~\cite{zhang2007malicious} & Detection & $n$-grams & Dempster-Shafer \\
\midrule\midrule
\end{tabular}
}
\end{table}

\subsection{A Framework for Ensemble Classifiers}

In this section, we consider various means of constructing 
ensemble classifiers, as viewed from a high-level perspective. 
We then provide an equally high level framework that we
find useful in our subsequent discussion of ensemble classifiers in
Sections~\ref{sect:CEC} and, especially, in Section~\ref{sect:examples}.
%We conclude
%this section with a brief overview of some of the many applications of ensemble 
%methods to problems in malware analysis, and other closely related topics.

%\subsection{Overview}

We consider ensemble learners that are based on combinations of scoring functions. 
In the general case, we assume the 
scoring functions are real valued, but the more restricted case of 
zero-one valued ``scoring'' functions (i.e., classifiers) easily fits into our framework. 
%That is, we assume that the individual scores are designed for the regression problem, 
%but the binary classification problem is easily handled as well.
We place no additional restrictions on the
scoring functions and, in particular, they do not necessarily represent
``learning'' algorithms, per se. Hence, we are dealing with ensemble methods
broadly speaking, rather than ensemble learners in a strict sense.
We assume that the ensemble method itself---as opposed to the scoring functions that comprise
the ensemble---is for classification,
and hence ensemble functions are zero-one valued. 

Let~$\w_{1},\w_{2},\ldots,\w_{n}$ be training samples, and 
let~$v_{i}$ be a feature vector of length~$m$, where the 
features that comprise~$v_{i}$ are extracted from sample~$\w_{i}$.
We collect the feature vectors for all~$n$ training samples
into an~$m\times n$ matrix that we denote as
\begin{equation}\label{eq:featureMatrix}
  V=\rowvecccc{v_{1}}{v_{2}}{\cdots}{v_{n}}
\end{equation}
where each~$v_{i}$ is a column of the matrix~$V$. Note that each row of~$V$ 
corresponds to a specific feature type, while column~$i$ of~$V$ 
corresponds to the features extracted from the training sample~$\w_{i}$.

Let~$\S:\mathbb{R}^{m}\to\mathbb{R}$ be a scoring function. Such a scoring function
will be determined based on training data, where this training data is given
by a feature matrix~$V$, as in equation~\eref{eq:featureMatrix}. 
A scoring function~$\S$ will generally
also depend on a set of~$k$ parameters that we denote as
\begin{equation}\label{eq:parmVec}
  \P = \rowvecccc{\p_{1}}{\p_{2}}{\ldots}{\p_{k}}
\end{equation}
The score generated by the scoring function~$\S$ when applied to sample~$x$
is given by
$$
  \S(x\dd V,\P)
$$
where we have explicitly included the dependence on the 
training data~$V$ and the function parameters~$\P$.

For any scoring function~$\S$, there is a corresponding classification 
function that we denote as~$\Sb:\mathbb{R}^{m}\to\{0,1\}$. That is, once we determine
a threshold to apply to the scoring function~$\S$, it provides a binary classification
function that we denote as~$\Sb$. As with~$\S$, we explicitly
indicate the dependence on training data~$V$ and the function parameters~$\P$
by writing
$$
  \Sb(x\dd V,\P) .
$$

For example, each training sample~$\w_{i}$ could be a malware
executable file, where all of the~$\w_{i}$ belong to the same malware family.
Then an example of an extracted feature~$v_{i}$ would be the opcode histogram,
that is, the relative frequencies of the mnemonic opcodes that are 
obtained when~$\w_{i}$ is disassembled. The scoring function~$\S$ could, 
for example, be based on a hidden Markov model that is trained on the feature 
matrix~$V$ as given in equation~\eref{eq:featureMatrix}, 
with the parameters~$\P$ in equation~\eref{eq:parmVec}
being the initial values that are selected when training the HMM.

In its most general form, an ensemble method for a binary classification 
problem can be viewed as a function~$F:\mathbb{R}^{\ell}\to \{0,1\}$ 
of the form
\begin{equation}\label{eq:general}
  F\bigl(\S_{1}(x\dd V_{1},\P_{1}),\S_{2}(x\dd V_{2},\P_{2}),\ldots,\S_{\ell}(x\dd V_{\ell},\P_{\ell})\bigr)
\end{equation}
That is, the ensemble method defined by the function~$F$ produces a classification 
based on the scores~$\S_{1},\S_{2},\ldots,\S_{\ell}$, where scoring function~$\S_{i}$
is trained using the data~$V_{i}$ and parameters~$\P_{i}$.

\subsection{Classifying Ensemble Classifiers}\label{sect:CEC}

From a high level perspective, ensemble classifiers can be 
categorized as bagging, boosting, stacking, or some combination 
thereof~\cite{Kuncheva2004,Smolyakov2017,Zhou2012}. In this section, we
briefly introduce each of these general classes of ensemble 
methods and give their generic formulation in
terms of equation~\eref{eq:general}.

\subsubsection{Bagging}

In bootstrap aggregation (i.e., bagging), different subsets of the data or features (or both) are
used to generate different scores. The results are then
combined in some way, such as a sum of the scores, or a majority vote of the
corresponding classifications. For bagging we assume
that the same scoring method is used for all scores in the ensemble.
For example, bagging is used when generating 
a random forest, where each individual scoring function is
based on a decision tree structure. One benefit of bagging
is that it reduces overfitting, which is a particular problem for decision trees.

For bagging, the general equation~\eref{eq:general} is restricted to
%%%%% Seems too strict of a definition. For example, if use bagging with HMMs.
%%%%% by this definition, we would have to initialize all models with exactly the
%%%%% same values, as opposed to generating random initial values...
\begin{equation}\label{eq:bagging}
  F\bigl(\S(x\dd V_{1},\P),\S(x\dd V_{2},\P),\ldots,\S(x\dd V_{\ell},\P)\bigr)
\end{equation}
That is, in bagging, each scoring function is essentially the same, 
but each is trained on a different feature set. For example, 
suppose that we collect all available feature vectors into a matrix~$V$ as 
in equation~\eref{eq:featureMatrix}. Then bagging based on subsets
of samples would correspond to generating~$V_i$ by deleting a
subset of the columns of~$V$. On the other hand, bagging based
on features would correspond to generating~$V_i$ by deleting
a subset of the rows of~$V$. Of course, we can easily extend this
to bagging based on both the data and features simultaneously,
as in a random forest.
In Section~\ref{sect:examples}, we discuss specific examples of bagging.

\subsubsection{Boosting}

Boosting is a process whereby distinct classifiers are combined to 
produce a stronger classifier. Generally, boosting deals with weak classifiers
that are combined in an adaptive or iterative manner so as to improve
the overall classifier. We restrict our definition of boosting to
cases where the classifiers are closely related, in the sense
that they differ only in terms of parameters.
From this perspective, boosting can be viewed as ``bagging'' based on
classifiers, rather than data or features. That is, all of the scoring
functions are reparameterized versions of the same scoring technique. 
%For example, with an HMM,
%the parameters include the initial values of the matrices~$A$, $B$, and~$\pi$.
Under this definition of boosting, the general equation~\eref{eq:general} becomes
\begin{equation}\label{eq:boosting}
  F\bigl(\S(x\dd V,\P_{1}),\S(x\dd V,\P_{2}),\ldots,\S(x\dd V,\P_{\ell})\bigr)
\end{equation}
That is, the scoring functions differ only by re-parameterization, 
while the scoring data and features do not change.

Below, in Section~\ref{sect:examples},
we discuss specific examples of boosting;
in particular, we discuss the most popular method of boosting, AdaBoost.
In addition, we show that some other popular techniques 
fit our definition of boosting.

\subsubsection{Stacking}

Stacking is an ensemble method that combines disparate
scores using a meta-classifier~\cite{Smolyakov2017}. 
In this generic form, stacking is defined by the general case in 
equation~\eref{eq:general}, where the scoring functions can
be (and typically are) significantly different. Note that from this perspective, 
stacking is easily seen to be a generalization of both bagging and boosting.

Because stacking generalizes both bagging and boosting, it is not
surprising that stacking based ensemble methods
can outperform bagging and boosting methods,
as evidenced by recent
machine learning competitions, including the
KDD Cup~\cite{KDD}, the Kaggle competition~\cite{Kaggle}, 
as well as the infamous Netflix prize~\cite{Netflix}.
However, this is not the end of the story, as efficiency and
practicality are often ignored in such competitions, whereas in practice,
it is virtually always necessary to consider such issues. 
Of course, the appropriate tradeoffs will depend
on the specifics of the problem at hand. Our empirical results
in Section~\ref{sect:exp} 
provide some insights into these tradeoff
issues within the malware analysis domain.

In the next section, we discuss concrete examples of 
bagging, boosting, and stacking techniques.
Then in Section~\ref{sect:exp} we present our experimental 
results, which include selected bagging, boosting, and
stacking architectures.

\subsection{Ensemble Classifier Examples}\label{sect:examples}

Here, we consider a variety of ensemble methods and
discuss how each fits into the general framework presented above.
We begin with a few fairly generic examples, and then 
discuss several more specific examples.

\subsubsection{Maximum}\label{sect:max} %%%%% generic

In this case, we have
\begin{equation}\label{eq:max}
  F\bigl(\S_{1}(x\dd V_{1},\P_{1}),\S_{2}(x\dd V_{2},\P_{2}),\ldots,\S_{\ell}(x\dd V_{\ell},\P_{\ell})\bigr)
       = \max\{\S_{i}(x\dd V_{i},\P_{i})\}
\end{equation}

\subsubsection{Averaging} %%%%% generic

Averaging is defined by
\begin{equation}\label{eq:avg}
  F\bigl(\S_{1}(x\dd V_{1},\P_{1}),\S_{2}(x\dd V_{2},\P_{2}),\ldots,\S_{\ell}(x\dd V_{\ell},\P_{\ell})\bigr)
       = \frac{1}{\ell} \sum_{i=1}^{\ell}\S_{i}(x\dd V_{i},\P_{i})
\end{equation}

\subsubsection{Voting} %%%%% generic

Voting could be used as a form of boosting, provided that no bagging is involved 
(i.e., the same data and features are used in each case). Voting 
is also applicable to stacking, and is generally applied in such a mode,
or at least with significant diversity in the scoring functions,
since we want limited correlation when voting. 

In the case of stacking,
a simple majority vote is of the form
\begin{equation}\nonumber
\begin{split}
  F\bigl(\Sb_{1}(x\dd V_{1},\P_{1})&,\Sb_{2}(x\dd V_{2},\P_{2}),\ldots,\Sb_{\ell}(x\dd V_{\ell},\P_{\ell})\bigr) \\
     &= \maj \bigl(\Sb_{1}(x\dd V_{1},\P_{1}),\Sb_{2}(x\dd V_{2},\P_{2}),\ldots,\Sb_{\ell}(x\dd V_{\ell},\P_{\ell})\bigr)
\end{split}
\end{equation}
where~``$\maj$'' is the majority vote function. Note that the majority vote is well defined
in this case, provided that~$\ell$ is odd---if~$\ell$ is even, we can simply flip a coin in 
case of a tie.

%%%%% Does this make sense?????
%Note that we could also vote based on the scores~$\S_{i}(x\dd V_{i},\P_{i})$,
%as opposed to the classifiers~$\Sb_{i}(x\dd V_{i},\P_{i})$. Such a score-based
%``vote'' based is similar to the~``$\max$'' technique discussed in Section~\ref{sect:max}, above.
%However, it would likely be necessary to scale the scores before such a voting
%technique would be expected to produce reliable results.

As an aside, we note that it is easy
to see why we want to avoid correlation when voting is used
as a combining function. Consider the following
example from~\cite{Kaggle-ensembling-guide}. Suppose that
we have the three highly correlated scores
$$
\left(
\begin{array}{cccccccccc}
\Sb_1\\
\Sb_2\\
\Sb_3
\end{array}
\right)
=
\left(
\begin{array}{cccccccccc}
1&1&1&1&1&1&1&1&0&0\\
1&1&1&1&1&1&1&1&0&0\\
1&0&1&1&1&1&1&1&0&0
\end{array}
\right)
$$
where each~1 indicates correct classification, and each~0 is an incorrect classification. Then,
both~$\Sb_1$ and~$\Sb_2$ are~80\%\ accurate, and~$\Sb_3$ is~70\%\ accurate. If we use a simple
majority vote, then we obtain the classifier
$$
 C = (\begin{array}{cccccccccc}1&1&1&1&1&1&1&1&0&0\end{array})
$$
which is~80\%\ accurate. On the other hand, the less correlated classifiers
$$
\left(
\begin{array}{cccccccccc}
\Sb_1\pr\\
\Sb_2\pr\\
\Sb_3\pr
\end{array}
\right)
=
\left(
\begin{array}{cccccccccc}
1&1&1&1&1&1&1&1&0&0\\
0&1&1&1&0&1&1&1&0&1\\
1&0&0&0&1&0&1&1&1&1
\end{array}
\right)
$$
are only~80\%, 70\% and~60\%\ accurate, respectively, but the majority vote in this case
gives us
$$
 C{\hspace*{0.08em}}' = (\begin{array}{cccccccccc}1&1&1&1&1&1&1&1&0&1\end{array})
$$
which is~90\%\ accurate. 

\subsubsection{ML-Based Combination} %%%%% generic (like SVM meta-classifier)

Recall that the most general formulation of an ensemble classifier is given in equation~\eref{eq:general}.
In this formulation, we can select the function~$F$ based on a machine learning technique, which
is applied to the individual scores~$\S(x\dd V_{i},\P_{i})$. In the remainder of
this section, we consider specific ensemble examples involving machine learning techniques.

\subsubsection{AdaBoost}

Given a collection of (weak) classifiers~$c_{1},c_{2},\ldots,c_{\ell}$,
AdaBoost is an iterative algorithm that generates a series of (generally, stronger)
classifiers, $C_1,C_2,\ldots,C_M$ based on the classifiers~$c_{i}$.
%The final classifier is given by~$C=C_M$. 
Each classifier is determined from the previous classifier by 
the simple linear extension
$$
  C_{m}(x) = C_{m-1}(x) + \alpha_{m} c_{i}(x) 
$$
and the final classifier is given by~$C=C_M$.
Note that at each iteration, we include a previously unused~$c_{i}$ from the 
set of (weak) classifiers and determine a new weight~$\alpha_{i}$. A greedy
approach is used when selecting~$c_{i}$, but it is not a hill climb,
so that results might get worse at any step in the AdaBoost process.

From this description, we see that the AdaBoost algorithm fits the form
in equation~\eref{eq:boosting}, with~$\Sb(x\dd V,\P_{i}) = C_{i}(x)$, and
$$
  F\bigl(\Sb(x\dd V,\P_{1}),\Sb(x\dd V,\P_{2}),\ldots,\Sb(x\dd V,\P_{M})\bigr) = \Sb(x\dd V,\P_{M}) = C_{M}(x)
$$

\subsubsection{SVM as Meta-Classifier}

It is natural to use an SVM as a
meta-classifier to combine scores~\cite{StampML2017}.
For example, in~\cite{SinghTVAS16}, an SVM is used to generate a malware 
classifier based on several machine learning and statistical based malware scores.
In~\cite{SinghTVAS16}, it is shown that the resulting SVM classifier consistently outperforms 
any of the component scores, and the differences are most pronounced
in the most challenging cases.

The use of SVM in this meta-classifier mode can be viewed as a 
general stacking method. Thus, this SVM technique is equivalent
to equation~\eref{eq:general}, where the function~$F$ is 
simply an SVM classifier based on the 
component scores~$\S_{i}(x\dd V_{i},\P_{i})$, for~$i=1,2,\ldots,\ell$.

\subsubsection{HMM with Random Restarts}\label{sect:HMM_rr}

A hidden Markov model can be viewed as a discrete hill climb 
technique~\cite{Stamp04arevealing,StampML2017}.
As with any hill climb, when training an HMM we are only assured of a local
maximum, and we can often significantly improve our results  
by executing the hill climb multiple times with different initial values,
selecting the best of the resulting models.
For example, in~\cite{VobbilisettyTLV17} it is shown that
an HMM can be highly effective for breaking classic substitution 
ciphers and, furthermore, by using a large number of random restarts,
we can significantly increase the success rate in the most difficult cases.
The work in~\cite{VobbilisettyTLV17} is closely related to that
in~\cite{BKK2013}, where such an approach is used to 
analyze the unsolved Zodiac~340 cipher.

From the perspective considered in this paper, an HMM with random restarts 
can be seen as special case of boosting. If we simply select the
best model, then the ``combining'' function is particularly simple, and is given by
\begin{equation}\label{eq:HMM_rr}
  F\bigl(\S(x\dd V,\P_{1}),\S(x\dd V,\P_{2}),\ldots,\S(x\dd V,\P_{\ell})\bigr) = \max\{\S(x\dd V,\P_{i})\}
\end{equation}
Here, each scoring function is an HMM, where the
trained models differ based only on different initial values. 
We see that equation~\eref{eq:HMM_rr} is a special case of
equation~\eref{eq:max}. However, 
the~``$\max$'' in equation~\eref{eq:HMM_rr} is
the maximum over the HMM model scores, 
not the maximum over any particular set of input values.
That is, we select the highest scoring model and use it for scoring.
Of course, we could use other combining functions, such as an
average or majority vote of the corresponding classifiers.
In any case, since there is a score associated with each model generated
by an HMM, any such combining function is well-defined.

\subsubsection{Bagged Perceptron}

Like a linear SVM, a perceptron will separate linearly separable data. However, unlike an
SVM, a perceptron will not necessarily produce the optimal separation, in the sense
of maximizing the margin. If we generate multiple perceptrons, each
with different random initial weights, and then average these models,
the resulting classifier will tend to be nearer to optimal, in the sense of
maximizing the margin~\cite{Michailidis,Kaggle-ensembling-guide}.
That is, we construct a classifier
\begin{equation}\label{eq:bagged_perceptron}
  F\bigl(\S(x\dd V,\P_{1}),\S(x\dd V,\P_{2}),\ldots,\S(x\dd V,\P_{\ell})\bigr) 
  	= \frac{1}{\ell}\sum_{i=1}^{\ell}\S(x\dd V,\P_{i})
\end{equation}
where~$\S$ is a perceptron and each~$P_{i}$ represents a set of
initial values. We see that equation~\eref{eq:bagged_perceptron}
is a special case of the averaging example given in
equation~\eref{eq:avg}. Also, we note that in this sum, we are averaging the perceptron
models, not the classifications generated by the models.

Although this technique is sometimes referred to as 
``bagged'' perceptrons~\cite{Kaggle-ensembling-guide},
by our criteria, it is a boosting scheme. That is, the ``bagging'' here 
is done with respect to parameters of the scoring functions, 
which is our working definition of boosting.

\subsubsection{Bagged Hidden Markov Model}\label{sect:HMM_bag}

Like the HMM with random restarts example given above, in this case,
we generate multiple HMMs. However, here we leave the model parameters
unchanged, and simply train each on a subset of the data. We could then
average the model scores (for example) as a way of combining the HMMs into
a single score, from which we can easily construct a classifier.
%Another alternative would be to combine the results with an SVM. ?????

\subsubsection{Bagged and Boosted Hidden Markov Model}\label{sect:HMM_bb}

Of course, we could combine both the HMM with random restarts
discussed in Section~\ref{sect:HMM_rr} with the bagging approach discussed in
the previous section. This process would yield an HMM-based ensemble
technique that combines both bagging and boosting.

\section{Experiments and Results}\label{sect:exp}

In this section, we consider a variety of experiments that illustrate 
various ensemble techniques. There experiments involve
malware classification, based on a challenging dataset that includes
a large number of samples from a significant number of malware families.

\subsection{Dataset and Features}\label{sect:dm}

Our dataset consists of samples from the~21 malware families listed in Table~\ref{tab:families}.
These families are from various different types of malware, including Trojans, worms, backdoors,
password stealers, so-called VirTools, and so on.

\begin{table}[!htb]
\begin{center}
\caption{Type of each malware family}\label{tab:families}
\resizebox{0.85\textwidth}{!}{%
\begin{tabular}{ccc|ccc}\midrule\midrule
\textbf{Index} & \textbf{Family} & \textbf{Type} & \textbf{Index} & \textbf{Family}  & \textbf{Type} \\
\midrule
\z1 & Adload~\cite{adload}     & Trojan Downloader  & 12 & Renos~\cite{renos} & Trojan Downloader\\
\z2 & Agent~\cite{agent}         & Trojan            & 13 & Rimecud~\cite{rimecud} & Worm \\
\z3 & Allaple~\cite{allaple}       & Worm            & 14 & Small~\cite{small}      & Trojan Downloader\\
\z4 & BHO~\cite{bho}             & Trojan             & 15 & Toga~\cite{toga}       & Trojan \\ 
\z5 & Bifrose~\cite{bifrose}      & Backdoor       & 16 & VB~\cite{vb}          & Backdoor  \\
\z6 & CeeInject~\cite{ceeinject}  & VirTool        & 17 & VBinject~\cite{vbinject} & VirTool  \\  
\z7 & Cycbot~\cite{cycbot}      & Backdoor       & 18 & Vobfus~\cite{vobfus}   & Worm \\             
\z8 & FakeRean~\cite{fakerean} & Rogue        & 19 & Vundo~\cite{vundo}    & Trojan Downloader\\ 
\z9 & Hotbar~\cite{hotbar}          & Adware       & 20 & Winwebsec~\cite{winwebsec}  & Rogue\\ 
10  & Injector~\cite{injector}      & VirTool         & 21 & Zbot~\cite{zbot}           & Password Stealer \\
11 & OnLineGames~\cite{onlinegames} & Password Stealer & --- & --- & ---\\
\midrule\midrule 
\end{tabular}}
\end{center}
\end{table}

Each of the malware families in Table~\ref{tab:families} is summarized below.
\begin{description}
\item[\bf Adload] downloads an executable file, stores it remotely, executes the file, and disables 
proxy settings~\cite{adload}. 
\item[\bf Agent] downloads Trojans or other software from a remote server~\cite{agent}. 
\item[\bf Allaple] is a worm that can be used as part of a denial of service (DoS) attack~\cite{allaple}. 
\item[\bf BHO] can perform a variety of actions, guided by an attacker~\cite{bho}. 
\item[\bf Bifrose] is a backdoor Trojan that enables a variety of attacks~\cite{bifrose}. 
\item[\bf CeeInject] uses advanced obfuscation to avoid being detected by antivirus software~\cite{ceeinject}. 
\item[\bf Cycbot] connects to a remote server, exploits vulnerabilities, and spreads through backdoor 
ports~\cite{cycbot}. 
\item[\bf FakeRean] pretends to scan the system, notifies the user of supposed issues, 
and asks the user to pay to clean the system~\cite{fakerean}. 
\item[\bf Hotbar] is adware that shows ads on webpages and installs additional adware~\cite{hotbar}. 
\item[\bf Injector] loads other processes to perform attacks on its behalf~\cite{injector}. 
\item[\bf OnLineGames] steals login information of online games and tracks user 
keystroke activity~\cite{onlinegames}. 
\item[\bf Renos] downloads software that claims the system has spyware and asks for a payment to 
remove the nonexistent spyware~\cite{renos}. 
\item[\bf Rimecud] is a sophisticated family of worms that perform a variety of activities
and can spread through instant messaging~\cite{rimecud}. 
\item[\bf Small] is a family of Trojans that downloads unwanted software. 
This downloaded software can perform a variety of actions, such as a fake security application~\cite{small}.
\item[\bf Toga] is a Trojan that can perform a variety of actions of the attacker's choice~\cite{toga}. 
\item[\bf VB] is a backdoor that enables an attacker to gain access to a computer~\cite{vb}. 
\item[\bf VBinject] is a generic description of malicious files that are obfuscated in a specific manner~\cite{vbinject}. 
\item[\bf Vobfus] is a worm that downloads malware and spreads through USB drives or other 
removable devices~\cite{vobfus}. 
\item[\bf Vundo] displays pop-up ads and may download files. It uses advanced techniques to 
defeat detection~\cite{vundo}.
\item[\bf Winwebsec] displays alerts that ask the user for money to 
fix supposed issues~\cite{winwebsec}.
\item[\bf Zbot] is installed through email and shares a user's personal information with attackers.
In addition, Zbot can disable a firewall~\cite{zbot}.
\end{description}

%\subsection{Features}

From each available malware sample, 
we extract the first~1000 mnemonic opcodes using 
the reversing tool Radare2 (also know as R2)~\cite{radare2}. 
We discard any malware executable 
that yields less than~1000 opcodes,
as well as a number of executables that were found to be corrupted.
The resulting opcode sequences, each of length~1000,
serve as the feature vectors for our machine learning experiments. 

Table~\ref{tab:samps} gives the number of samples (per family)
from which we successfully obtained opcode feature vectors. 
Note that our dataset contains a total of~9725 samples from the~21 malware families
and that the dataset is highly imbalanced---the number of samples per family 
varies from a low of~129 to a high of nearly~1000. 

\begin{table}[!htb]
\begin{center}
\caption{Type of each malware family}\label{tab:samps}
{%\footnotesize
\begin{tabular}{ccc|ccc}\midrule\midrule
\textbf{Index} & \textbf{Family} & \textbf{Samples} & \textbf{Index} & \textbf{Family}  & \textbf{Samples} \\
\midrule
 \z1 & Adload     & \z162    & 12 & Renos         & \z532 \\
 \z2 & Agent       & \z184    & 13 & Rimecud     & \z153 \\
 \z3 & Allaple      & \z986    & 14 & Small          & \z180 \\
 \z4 & BHO         & \z332    & 15 & Toga           & \z406 \\ 
 \z5 & Bifrose      & \z156    & 16 & VB              & \z346 \\
 \z6 & CeeInject  & \z873    & 17 & VBinject      & \z937 \\  
 \z7 & Cycbot      & \z597    & 18 & Vobfus        & \z929 \\             
 \z8 & FakeRean & \z553    & 19 & Vundo         & \z762 \\ 
 \z9 & Hotbar       & \z129    & 20 & Winwebsec & \z837 \\ 
 10  & Injector      & \z158    & 21 & Zbot            & \z303 \\ \cline{4-6} %\cmidrule{4-6}
 11 & OnLineGames & \z210 &  & $\mbox{Total}^{\vphantom{M^n}}$ & 9725 \\
\midrule\midrule 
\end{tabular}
}
\end{center}
\end{table}

\subsection{Metrics}

The metrics used to quantify the success of our experiments
are accuracy, balanced accuracy, precision, recall, and the F1 score.
Accuracy is simply the ratio of correct classifications to the total number of classifications.
In contrast, the balanced accuracy is the average accuracy per family.

Precision, which is also known as the positive predictive value, is the number of true positives
divided by the sum of the true positives and false positives. That is, the precision is the ratio
of samples classified as positives that are actually positive to all samples that are classified as positive. 
Recall, which is also known as the true positive rate or sensitivity, 
is the computed by dividing the number of true positives
by the number true positives plus the number of false negatives. That is, the
recall is the fraction of positive samples that are classified as such.
The F1 score is computed as
$$
  \mbox{F1} = 2\cdot\frac{\mbox{precision}\cdot\mbox{recall}}{\mbox{precision} + \mbox{recall}} ,
$$
which is the harmonic mean of the precision and recall.

\subsection{Software}

%Table~\ref{tab:exp_BB} summarizes our bagging and boosting experiments, 
%while Table~\ref{tab:exp_stack} summarizes our stacking experiments.
%Here, ``bagging'' simply means that we generate multiple models based
%on subsets of the features, while ``boosting'' means that we
%generate multiple models using the same data, but different 
%hyper-parameters. For example, bagged HMMs could be generated using 
%different subsets of the data, and boosted HMMs could be generated
%using different initial values for the HMMs. In each case, we select the
%best model, except for random forest and AdaBoost, which use
%multiple models.

The software packages used in our experiments include
\texttt{hmmlearn}~\cite{hmmlearn}, \texttt{XGBoost}~\cite{xgb}, \texttt{Keras}~\cite{keras}, 
and \texttt{TensorFlow}~\cite{tensorflow}, and \texttt{scikit-learn}~\cite{sklearn},
as indicated in Table~\ref{tab:tools}.
In addition, we use \texttt{Numpy}~\cite{numpy} 
for linear algebra and various tools available in the package \texttt{scikit-learn}
(also known as \texttt{sklearn}) for general
data processing. These packages are all widely used 
in machine learning.

\begin{table}[!htb]
\centering
\caption{Software used in experiments}\label{tab:tools}
\begin{tabular}{cc}\midrule\midrule
\textbf{Technique} & \textbf{Software} \\ \midrule
HMM &  \texttt{hmmlearn} \\
XGBoost &  \texttt{XGBoost} \\
AdaBoost &  \texttt{sklearn} \\
CNN &  \texttt{Keras}, \texttt{TensorFlow} \\
LSTM &  \texttt{Keras},  \texttt{TensorFlow} \\
Random Forest &  \texttt{sklearn} \\ \midrule\midrule
\end{tabular}
\end{table}

\subsection{Overview of Experiments}

For all of our experiments, we use opcode sequences of length~1000
as features. For CNNs, the sequences are interpreted as images. 

We consider three broad categories of experiments. First, we
apply ``standard'' machine learning techniques. These experiments,
serve as a baseline for comparison for our subsequent experiments.
Among other things, these standard experiments
show that the malware classification problem that we are dealing
with is challenging.

We also conduct bagging and boosting experiments based
on a subset of the techniques considered in our baseline
standard experiments. These results demonstrate that 
both bagging and boosting can provide some improvement
over our baseline techniques.

Finally, we consider a set of stacking experiments, where we restrict our 
attention to simple voting schemes, all of which are based on 
architectures previously considered in this paper. Although these 
are very basic stacking architectures, they
clearly show the potential benefit of stacking multiple techniques.

\subsection{Standard Techniques}

For our ``standard'' techniques, we test several machine learning
methods that are typically used individually. Specifically, we consider
hidden Markov models (HMM), convolutional neural networks (CNN),
random forest, and long short-term memory (LSTM). The parameters that we have tested
in each of these cases are listed in Table~\ref{tab:parmsStandard}, with those that gave the
best results in boldface.

\begin{table}[!htb]
\caption{Parameters for standard techniques}\label{tab:parmsStandard}
\centering
\begin{tabular}{c|ll}\midrule\midrule
\textbf{Technique} & \textbf{Parameters} & \textbf{Values tested}\\ \midrule
\multirow{3}{*}{HMM} & \texttt{n\_components} & [1,2,5,\textbf{10}] \\
 & \texttt{n\_iter} & [50,100,\textbf{200},300,500]\\
 & \texttt{tol} & [0.01,0.5]\\ \midrule
\multirow{3}{*}{CNN} & \texttt{learning\_rate} & [\textbf{0.001},0.0001] \\
 & \texttt{batch\_size} & [\textbf{32},64,128]\\
 & \texttt{epochs} & [50,75,\textbf{100}\\ 
 \midrule
\multirow{5}{*}{Random Forest} & \texttt{n\_estimators} & [100,200,300,500,\textbf{800}]\\
 & \texttt{min\_samples\_split} & [\textbf{2},5,10,15,20]\\
 & \texttt{min\_samples\_leaf} & [\textbf{1},2,5,10,15]\\ 
 & \texttt{max\_features} & [\textbf{auto},sqrt,$\mbox{log}_2$]\\
 & \texttt{max\_depth} & [30,\textbf{40},50,60,70,80]\\ 
 \midrule
\multirow{5}{*}{LSTM} & \texttt{layers} & [\textbf{1},3]\\
 & \texttt{directional} & [uni-dir,\textbf{bi-dir}]\\
 & \texttt{learning\_rate} & [\textbf{0.01}]\\
 & \texttt{batch\_size} & [\textbf{1},16,32]\\
 & \texttt{epochs} & [\textbf{20}]\\
\midrule\midrule
\end{tabular}
\end{table}

From Table~\ref{tab:parmsStandard}, we note that
a significant number of parameter combinations were tested in each case.
For example, in the case of our random forest model, we tested
$$
  5^3\cdot 3\cdot 6 = 2250
$$
different combinations of parameters.

The confusion matrices for all of the experiments in this section
can be found
in the Appendix in Figure~\ref{fig:conf_2}~(a) through Figure~\ref{fig:conf_2}~(d).
We present the results of all of these experiments---in terms of the
metrics discussed previously (i.e., accuracy, balanced accuracy, precision, recall, 
and F1 score)---in Section~\ref{sect:discuss}, below.

%\subsection{Additional Bagging and Boosting Experiments}
\subsection{Bagging Experiments}

Recall from our discussion above, that we use the term bagging to mean 
a multi-model approach where the individual models are trained
with the same technique and essentially the same parameters, but different
subsets of the data or features. In contrast, we use boosting to refer to
multi-model cases where the data and features are essentially the same
and the models are of the same type, with the model parameters varied.

We will use AdaBoost and XGBoost results to serve as representative examples
of boosting. We also consider bagging experiments (in the sense described in
the previous paragraph) involving each of the HMM, CNN, and LSTM architectures.
The results of these three distinct bagging experiments---in the form of confusion matrices---are
given in Figure~\ref{fig:conf_1} in the Appendix. In terms of the metrics discussed above,
the results of these experiments are summarized in Section~\ref{sect:discuss}, below.

%%%%% ????? A table summarizing these experiments would be good ?????

\subsection{Boosting Experiments}

As representative examples of boosting techniques, we consider
AdaBoost and XGBoost. In each case, we experiment
with a variety of parameters as listed in Table~\ref{tab:parmsBoost}.
The parameter selection that yielded the best results
are highlighted in boldface.

\begin{table}[!htb]
\caption{Parameters for boosting techniques}\label{tab:parmsBoost}
\centering
\begin{tabular}{c|ll}\midrule\midrule
\textbf{Technique} & \textbf{Parameters} & \textbf{Values tested}\\ \midrule
\multirow{3}{*}{AdaBoost} & \texttt{n\_estimators} & [100,200,300,500,800,\textbf{1000}]\\
 & \texttt{learning\_rate} & [0.5,1.0,1.5,2.0]\\
 & \texttt{algorithm} & [\textbf{SAMME},SAMME.R]\\ 
 \midrule
\multirow{4}{*}{XGBoost} & \texttt{eta} & [0.05,0.1,0.2,\textbf{0.3},0.5]\\
 & \texttt{max\_depth} & [1,2,\textbf{3},4]\\
 & \texttt{objective} & [\textbf{multi:softprob},binary:logistic]\\
 & \texttt{steps} & [1,5,10,\textbf{20},50]\\ 
\midrule\midrule
\end{tabular}
\end{table}

Confusion matrices for these two boosting experiments are
given in Figure~\ref{fig:conf_2b} in the Appendix. 
The results of these experiments are summarized in Section~\ref{sect:discuss}, 
below, in terms of accuracy, balanced accuracy, and so on.

%For Bagging, we built X number of models with each model built using a subsample 
%(with replacement) of the training dataset, which had size of 60\%\ of the training dataset. 
%Afterwards, we implemented a voting system where each model had equal weights on the test set.
%In the event of a tie, we took into account the scoring/log-likehood (HMM)) 
%or the confidence percentage (CNN/LSTM) for each model of its top prediction to 
%deduce the final outcome.
%
%For Boosting, we built X number of models with each model built using a subsample 
%(with replacement) of the training dataset, which had size of 60\%\ of the training dataset. 
%For each subsequent model, the errors of the previous ensemble were automatically inputted 
%into the subsample dataset to train this particular model. Afterwards, we implemented a voting 
%system where each model had equal weights on the test set. In the event of a tie, we took into 
%account the scoring/log-likehood (HMM)) or the confidence percentage (CNN/LSTM) for each 
%model of its top prediction to deduce the final outcome.

%\subsubsection{Bagged HMM}

%\subsubsection{Boosted HMM}

%\subsubsection{Bagged CNN}

%\subsubsection{Boosted CNN}

%\subsubsection{Bagged LSTM}

%\subsubsection{Boosted LSTM}

\subsection{Voting Experiments}

Since there exists an essentially unlimited number of possible stacking architectures, we have
limited our attention to one of the simplest, namely, voting. These results serve as a lower
bound on the results that can be obtained with stacking architectures.

We consider six different stacking architectures. %, which we refer to as
%``all CNN'', 
%``all LSTM'',
%``all bagged neural networks'',
%``all boosted neural networks'',
%``all classic techniques'',
%``all neural networks''
%and
%``all models''.
These stacking experiments can be summarized as follows.
\begin{description}
\item[\bf CNN] consists of the plain and bagged CNN models discussed above.
The confusion matrix for this experiment is given in Figure~\ref{fig:conf_3}~(a).
\item[\bf LSTM] consists of the plain and bagged LSTM models discussed above. 
The confusion matrix for this experiment is given in Figure~\ref{fig:conf_3}~(b).
\item[\bf Bagged neural networks] combines our bagged CNN and bagged LSTM models.
The confusion matrix for this experiment is given in Figure~\ref{fig:conf_3}~(c).
%\item[\bf All boosted neural networks] consists of our previous boosted CNN and boosted LSTM models.
%The confusion matrix for this experiment is given in Figure~\ref{fig:conf_3}~(d).
\item[\bf Classic techniques] combines (via voting) all of the classic models considered
above, namely, HMM, bagged HMM, random forest, AdaBoost, and XGBoost.
The confusion matrix for this experiment is given in Figure~\ref{fig:conf_3}~(d).
\item[\bf All neural networks] consists of all of the CNN and LSTM models, bagged and plain.
The confusion matrix for this experiment is given in Figure~\ref{fig:conf_3}~(e).
\item[\bf All models] combines all of the classic and neural network models into one voting scheme.
The confusion matrix for this experiment is given in Figure~\ref{fig:conf_3}~(f).
\end{description}

In the next section, we present the results for each of the voting experiments 
discussed in this section in terms
of the our various metrics. These metrics enable us to directly compare
all of our experimental results.

\subsection{Discussion}\label{sect:discuss}

Table~\ref{tab:comp} summarizes the results of all of the experiments discussed above,
in term of the following metrics: accuracy, balanced accuracy, precision, recall,
and F1 score. These metrics have been introduced in Section~\ref{sect:dm}, above.

\begin{table}[!htb]
\caption{Comparison of experimental results}\label{tab:comp}
\centering
\resizebox{0.85\textwidth}{!}{%
\begin{tabular}{cc|ccccc}\midrule\midrule
\multirow{2}{*}{\textbf{Experiments}} & \multirow{2}{*}{\textbf{Case}} 
	& \multirow{2}{*}{\textbf{Accuracy}} & \textbf{Balanced} 
	& \multirow{2}{*}{\textbf{Precision}} & \multirow{2}{*}{\ \textbf{Recall}\ \ } 
	& \multirow{2}{*}{\textbf{F1 score}} \\
 & & & \textbf{accuracy} \\ \midrule
\multirow{4}{*}{Standard} &
HMM & 0.6717 & 0.6336 & 0.7325 & 0.6717 & 0.6848 \\
& CNN & 0.8211 & \textbf{0.7245} & \textbf{0.8364} & \textbf{0.8211} & 0.8104 \\
& Random Forest & 0.7549 & 0.6610 & 0.7545 & 0.7523 & 0.7448 \\
& LSTM & \textbf{0.8410} & 0.7185 & 0.7543 & 0.7185 & \textbf{0.8145} \\ \midrule
\multirow{3}{*}{Bagging} &
%Bagged HMM & 0.6820 & 0.6294 & 0.7389 & 0.6820 & 0.6905 \\
Bagged HMM & 0.7168 & 0.6462 & 0.7484 & 0.7168 & 0.7165 \\ % boosted
& Bagged CNN & \textbf{0.8910} & \textbf{0.8105} & \textbf{0.9032} & \textbf{0.8910} & \textbf{0.8838} \\
%& Bagged LSTM & 0.7969 & 0.6951 & 0.7890 & 0.7969 & 0.7847 \\ \midrule
& Bagged LSTM & 0.8602 & 0.7754 & 0.8571 & 0.8602 & 0.8549 \\% boosted
\midrule
\multirow{2}{*}{Boosting} &
AdaBoost & 0.5378 & 0.4060 & 0.5231 & 0.5378 & 0.5113 \\
& XGBoost & \textbf{0.7472} & \textbf{0.6636} & \textbf{0.7371} & \textbf{0.7472} & \textbf{0.7285} \\ \midrule
\multirow{6}{*}{Voting} &
Classic  & 0.8766 & 0.8079 & 0.8747 & 0.8766 & 0.8719 \\
& CNN & 0.9260 & 0.8705 & 0.9321 & 0.9260 & 0.9231 \\
& LSTM & 0.8560 & 0.7470 & 0.8511 & 0.8560 & 0.8408 \\
%& Bagged neural networks & 0.8905 & 0.8098 & 0.9012 & 0.8905 & 0.8838 \\
& Bagged neural networks & \fbox{\textbf{0.9337}} % boosted
	& \fbox{\textbf{0.8816}} & \fbox{\textbf{0.9384}} 
	& \fbox{\textbf{0.9337}} & \fbox{\textbf{0.9313}} \\
& All neural networks & 0.9208 & 0.8613 & 0.9284 & 0.9208 & 0.9171 \\
& All models & 0.9188 & 0.8573 & 0.9249 & 0.9188 & 0.9154 \\
\midrule\midrule
\end{tabular}}
\end{table}

In Table~\ref{tab:comp},
the best result for each type of experiment is in boldface, with the best results 
overall also being boxed. We see that a voting strategy based on all of the 
bagged neural network techniques gives us the best result for
each of the five statistics that we have computed. 

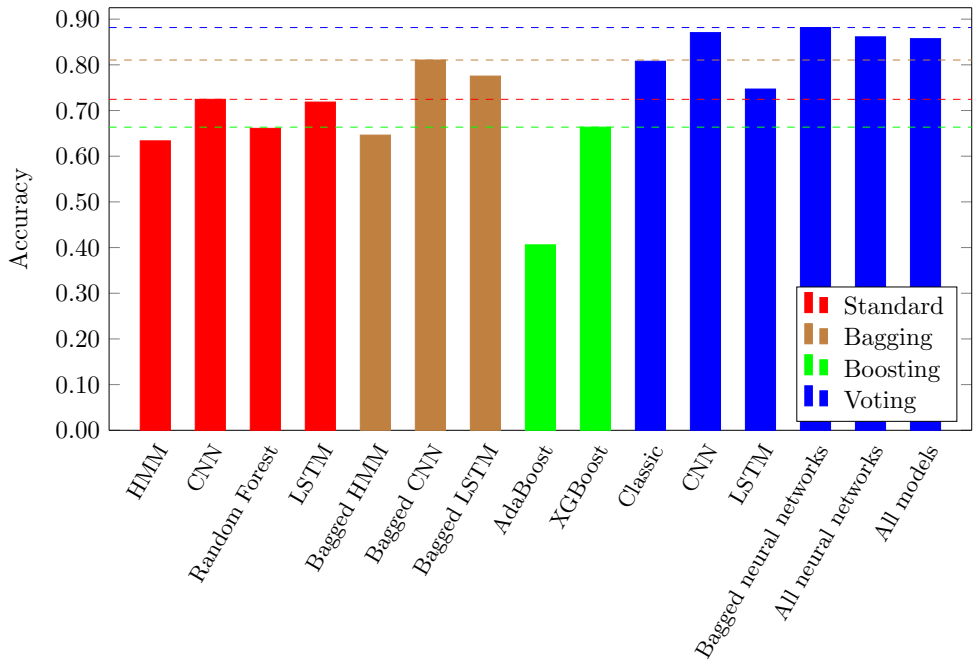
\begin{figure}[!htb]
	\centering
	\begin{tikzpicture}[scale=0.95, every node/.style={scale=1.0}]
\begin{axis}[%bar shift=0pt,
        width  = 0.85*\textwidth,
        height = 7.5cm,
        ymin=0.0,ymax=0.925,
        ytick={0.0,0.1,0.2,0.3,0.4,0.5,0.6,0.7,0.8,0.9},
        separate axis lines,
        major x tick style = transparent,
        ybar=5*\pgflinewidth,
        every axis plot/.append style={
          ybar,
          bar width=12.0pt,
          bar shift=0pt,
          fill
        },
        bar width=10.0pt,
%        ymajorgrids = true,
        ylabel = {Accuracy},
        xtick={1,2,3,4,5,6,7,8,9,10,11,12,13,14,15},
        x tick style={draw=none},
        xticklabels={
HMM,
CNN,
Random Forest,
LSTM,
Bagged HMM,
Bagged CNN,
Bagged LSTM,
AdaBoost,
XGBoost,
%Boosted HMM,
%Boosted CNN,
%Boosted LSTM,
Classic,
CNN,
LSTM,
Bagged neural networks,
%Voting Boosted ANN,
All neural networks,
All models
},
	y tick label style={
    		/pgf/number format/.cd,
   		fixed,
   		fixed zerofill,
    		precision=2},
%	yticklabel pos=right,
%        xtick = data,
        x tick label style={
        		rotate=60,
		font=\small,
		anchor=north east,
		inner sep=0mm},
%		font=\small},
%        scaled y ticks = false,
	%%%%% numbers on bars and rotated
%        nodes near coords,
%        every node near coord/.append style={rotate=90, 
%        								   anchor=west,
%								   font=\footnotesize,
%								   /pgf/number format/.cd,
%								   fixed,
%								   fixed zerofill,
%								   precision=4},
        %%%%%
%        enlarge x limits=0.03,
        enlarge x limits=0.06,
        legend cell align=left,
        legend style={
%                at={(1,1.05)},
%                anchor=south east,
%	        nodes={rotate=90},%%%%% rotate text in legend
%                at={(0.125,0)},
%                at={(0.125,0)},
%                at={(0.8775,0)},
                at={(0.89,0.02)},
                anchor=south,
                column sep=1ex
        },
]
\addplot[red]coordinates {(1,0.6336)};
\addplot[red,forget plot]coordinates {(2,0.7245)};
\addplot[red,forget plot]coordinates {(3,0.6610)};
\addplot[red,forget plot]coordinates {(4,0.7185)};
\addplot[brown]coordinates {(5,0.6462)};
\addplot[brown,forget plot]coordinates {(6,0.8105)};
\addplot[brown,forget plot]coordinates {(7,0.7754)};
\addplot[green]coordinates {(8,0.4060)};
\addplot[green,forget plot]coordinates {(9,0.6636)};
\addplot[blue]coordinates {(10,0.8079)};
\addplot[blue,forget plot]coordinates {(11,0.8705)};
\addplot[blue,forget plot]coordinates {(12,0.7470)};
\addplot[blue,forget plot]coordinates {(13,0.8816)};
\addplot[blue,forget plot]coordinates {(14,0.8613)};
\addplot[blue,forget plot]coordinates {(15,0.8573)};
\addlegendentry{Standard}
\addlegendentry{Bagging}
\addlegendentry{Boosting}
\addlegendentry{Voting}
\coordinate (A) at (axis cs:1,0.8816);
\coordinate (B) at (axis cs:1,0.6636);
\coordinate (C) at (axis cs:1,0.8105);
\coordinate (D) at (axis cs:1,0.7245);
\coordinate (O1) at (rel axis cs:0,0);
\coordinate (O2) at (rel axis cs:1,0);
\draw [blue,dashed] (A -| O1) -- (A -| O2);
\draw [green,dashed] (B -| O1) -- (B -| O2);
\draw [brown,dashed] (C -| O1) -- (C -| O2);
\draw [red,dashed] (D -| O1) -- (D -| O2);
\end{axis}
\end{tikzpicture}
	\caption{Balanced accuracy results}\label{fig:bar}
\end{figure}

Since our dataset is highly imbalanced, we consider the balanced
accuracy as the best measure of success.
The balanced accuracy results in Table~\ref{tab:comp}
are given in the form of a bar graph in Figure~\ref{fig:bar}.

Note that the results in Figure~\ref{fig:bar}
clearly show that stacking techniques
are beneficial, as compared to the corresponding
``standard'' techniques. Stacking not only yields the best
results, but it dominates in all categories. 
We note that five of the six stacking experiments perform better than any
of the standard, bagging, or boosting experiments. This is particularly
noteworthy since we only considered a simple stacking approach.
As a results,
our stacking experiments likely provide a poor lower bound on stacking
in general, and more advanced stacking techniques may
improve significantly over the results that we have obtained.

\section{Conclusion and Future Work}\label{sect:conc}

In this paper, we have attempted to impose some structure on the field of ensemble learning.
We showed that combination architectures can be classified as either bagging,
boosting, or in the more general case, stacking. We then provided experimental results
involving a challenging malware dataset to illustrate the potential benefits of
ensemble architectures. Our results clearly show that ensembles
improve on standard techniques, with respect to our specific dataset. 
Of course, in principle, we expect such combination
architectures to outperform standard techniques, but it is instructive to confirm this
empirically, and to show that the improvement can be substantial. These results make it
clear that there is a reason why complex stacking architectures win machine learning 
competitions.

However, stacking models are not without potential pitfalls. As the architectures
become more involved, training can become impractical. Furthermore, scoring can also
become prohibitively costly, especially if large numbers of
features are used in complex schemes involving extensive 
use of bagging or boosting.

For future work, it would be useful to quantify the tradeoff between accuracy and model
complexity. While stacking will generally improve results, marginal improvements in accuracy
that come at great additional cost in training and scoring are unlikely to be of any
value in real world applications. More concretely, future work involving additional
features would be very interesting, as it would allow for a more thorough analysis
of bagging, and it would enable us to draw firmer conclusions regarding the 
relative merits of bagging and boosting. Of course, more more complex classes of 
stacking techniques could be considered.

\bibliographystyle{plain}

\bibliography{Stamp-Mark.bib,references.bib,referencesMulti.bib}

\clearpage

\appendix

\section*{Appendix: Confusion Matrices}

%
% Plain
%

\begin{figure}[!htb]
	\centering
	\begin{tabular}{ccc}
	\resizebox{0.45\textwidth}{0.25\textheight}{%
	%\begin{tikzpicture}[scale=0.60,every node/.style={scale=0.9}]
%\begin{tikzpicture}[scale=0.55]
\begin{tikzpicture}[scale=0.60]
    \begin{axis}[%colorbar/width=2.5mm,
        width=20cm,
        height=20cm,
%        colormap={blackwhite}{gray(0cm)=(1); gray(1cm)=(0.5)},
%   colormap={bluewhite}{color=(white) color=(blue)},
%   colormap={bluewhite}{color=(white) rgb255=(0,191,255)},
    colormap={bluewhite}{color=(white) rgb255=(100,149,237)},
        xticklabels={Adload,Agent,Allaple,BHO,Bifrose,CeeInject,Cycbot,FakeRean,Hotbar,Injector,
        OnLineGames,Renos,Rimecud,Small,Toga,VB,VBinject,Vobfus,Vundo,Winwebsec,Zbot},
        xtick={0,...,20},
        xtick style={draw=none},
    xticklabel style={anchor=east,rotate=45,yshift=-5pt},
        yticklabels={Adload,Agent,Allaple,BHO,Bifrose,CeeInject,Cycbot,FakeRean,Hotbar,Injector,
        OnLineGames,Renos,Rimecud,Small,Toga,VB,VBinject,Vobfus,Vundo,Winwebsec,Zbot},
        ytick={0,...,20},
        ytick style={draw=none},
        enlargelimits=false,
        xticklabel style={font=\large},
        yticklabel style={font=\large},
        colorbar,
        colorbar style={
%           width=0.05*\pgfkeysvalueof{/pgfplots/parent axis width},%%% added this
%           height=0.5*\pgfkeysvalueof{/pgfplots/parent axis height},
%       plot graphics/node/.style={scale=1.33,anchor=south west,inner sep=0pt,}, %%% scale colorbar fill %%%
            ytick={0.0,0.2,0.4,0.6,0.8,1.0},
            yticklabels={0.0,0.2,0.4,0.6,0.8,1.0},
            yticklabel={\pgfmathprintnumber\tick},
            yticklabel style={font=\large,
                    /pgf/number format/fixed,
            /pgf/number format/precision=1}
        },
%        point meta min=0,
%        point meta max=100,
        point meta min=0.0,
        point meta max=1.0,
        nodes near coords={\pgfmathprintnumber\pgfplotspointmeta},
        % ---------------------------------------------------------------------
        % show `nodes near coords' but adapt the style so that values
        % above a threshold get another style
        % (adapted from <http://tex.stackexchange.com/a/141006/95441>)
        % #1: the THRESHOLD after which we switch to a special display.
        nodes near coords black white/.style={
            % define the style of the nodes with "small" values
            small value/.style={
                yshift=-7pt,
%                text=white,
                text=black,
                /pgf/number format/fixed,
                /pgf/number format/precision=3,
                /pgf/number format/zerofill
            },
            % define the style of the nodes with "large" values
            large value/.style={
                yshift=-7pt,
%                text=black,
                text=white,
                /pgf/number format/fixed,
                /pgf/number format/precision=3,
                /pgf/number format/zerofill
            },
            every node near coord/.style={
                check for zero/.code={
                    \pgfmathfloatifflags{\pgfplotspointmeta}{0}{
                        % If meta=0, make the node a coordinate
                        % (which doesn't have text)
                        \pgfkeys{/tikz/coordinate}
                    }{
                        \begingroup
                        % this group is merely to switch to FPU locally.
                        % Might be unnecessary, but who knows.
                        \pgfkeys{/pgf/fpu}
                        \pgfmathparse{\pgfplotspointmeta<#1}
                        \global\let\result=\pgfmathresult
                        \endgroup
                        %
                        % simplifies debugging:
                        %\show\result
                        %
                        \pgfmathfloatcreate{1}{1.0}{0}
                        \let\ONE=\pgfmathresult
                        \ifx\result\ONE
                            % AH: our condition 'y < #1' is met.
                            \pgfkeysalso{/pgfplots/small value}
                        \else
                            % ok, proceed as usual.
                            \pgfkeysalso{/pgfplots/large value}
                        \fi
                    }
                },
                check for zero,
            },
        },
        % asign a value to the new style which is the threshold at which
        % the two style `small value' or `large value' are used
%        nodes near coords black white=50,
        nodes near coords black white=0.5,
        % -----------------------------------------------------------------
    ]
        \addplot[
            matrix plot,
            mesh/cols=21,
            point meta=explicit,draw=gray
        ] table [meta=C] {
            x y C
0 0 0.829
1 0 0.098
2 0 0.0
3 0 0.049
4 0 0.0
5 0 0.0
6 0 0.0
7 0 0.0
8 0 0.0
9 0 0.0
10 0 0.0
11 0 0.0
12 0 0.0
13 0 0.024
14 0 0.0
15 0 0.0
16 0 0.0
17 0 0.0
18 0 0.0
19 0 0.0
20 0 0.0
0 1 0.0
1 1 0.348
2 1 0.0
3 1 0.022
4 1 0.022
5 1 0.022
6 1 0.0
7 1 0.022
8 1 0.0
9 1 0.022
10 1 0.109
11 1 0.0
12 1 0.022
13 1 0.152
14 1 0.0
15 1 0.043
16 1 0.13
17 1 0.0
18 1 0.065
19 1 0.022
20 1 0.0
0 2 0.0
1 2 0.0
2 2 1.0
3 2 0.0
4 2 0.0
5 2 0.0
6 2 0.0
7 2 0.0
8 2 0.0
9 2 0.0
10 2 0.0
11 2 0.0
12 2 0.0
13 2 0.0
14 2 0.0
15 2 0.0
16 2 0.0
17 2 0.0
18 2 0.0
19 2 0.0
20 2 0.0
0 3 0.0
1 3 0.06
2 3 0.0
3 3 0.867
4 3 0.012
5 3 0.0
6 3 0.0
7 3 0.0
8 3 0.0
9 3 0.0
10 3 0.036
11 3 0.0
12 3 0.0
13 3 0.012
14 3 0.0
15 3 0.0
16 3 0.0
17 3 0.0
18 3 0.012
19 3 0.0
20 3 0.0
0 4 0.0
1 4 0.0
2 4 0.0
3 4 0.026
4 4 0.436
5 4 0.026
6 4 0.0
7 4 0.051
8 4 0.0
9 4 0.026
10 4 0.128
11 4 0.0
12 4 0.077
13 4 0.0
14 4 0.051
15 4 0.051
16 4 0.077
17 4 0.051
18 4 0.0
19 4 0.0
20 4 0.0
0 5 0.0
1 5 0.018
2 5 0.0
3 5 0.005
4 5 0.091
5 5 0.723
6 5 0.0
7 5 0.009
8 5 0.0
9 5 0.014
10 5 0.059
11 5 0.0
12 5 0.0
13 5 0.036
14 5 0.009
15 5 0.005
16 5 0.0
17 5 0.0
18 5 0.009
19 5 0.005
20 5 0.018
0 6 0.0
1 6 0.007
2 6 0.0
3 6 0.0
4 6 0.0
5 6 0.0
6 6 0.839
7 6 0.007
8 6 0.0
9 6 0.0
10 6 0.007
11 6 0.007
12 6 0.0
13 6 0.0
14 6 0.0
15 6 0.0
16 6 0.0
17 6 0.0
18 6 0.0
19 6 0.094
20 6 0.04
0 7 0.0
1 7 0.014
2 7 0.0
3 7 0.0
4 7 0.007
5 7 0.0
6 7 0.0
7 7 0.741
8 7 0.0
9 7 0.0
10 7 0.029
11 7 0.0
12 7 0.0
13 7 0.05
14 7 0.029
15 7 0.036
16 7 0.0
17 7 0.0
18 7 0.022
19 7 0.072
20 7 0.0
0 8 0.0
1 8 0.0
2 8 0.0
3 8 0.0
4 8 0.0
5 8 0.0
6 8 0.0
7 8 0.0
8 8 0.969
9 8 0.0
10 8 0.0
11 8 0.0
12 8 0.0
13 8 0.031
14 8 0.0
15 8 0.0
16 8 0.0
17 8 0.0
18 8 0.0
19 8 0.0
20 8 0.0
0 9 0.05
1 9 0.1
2 9 0.0
3 9 0.025
4 9 0.1
5 9 0.025
6 9 0.0
7 9 0.025
8 9 0.0
9 9 0.075
10 9 0.0
11 9 0.0
12 9 0.05
13 9 0.175
14 9 0.05
15 9 0.025
16 9 0.1
17 9 0.025
18 9 0.0
19 9 0.0
20 9 0.175
0 10 0.074
1 10 0.056
2 10 0.0
3 10 0.037
4 10 0.0
5 10 0.0
6 10 0.0
7 10 0.019
8 10 0.0
9 10 0.019
10 10 0.704
11 10 0.0
12 10 0.0
13 10 0.074
14 10 0.0
15 10 0.0
16 10 0.0
17 10 0.0
18 10 0.0
19 10 0.019
20 10 0.0
0 11 0.0
1 11 0.015
2 11 0.0
3 11 0.0
4 11 0.008
5 11 0.0
6 11 0.0
7 11 0.015
8 11 0.0
9 11 0.015
10 11 0.023
11 11 0.818
12 11 0.0
13 11 0.023
14 11 0.008
15 11 0.008
16 11 0.015
17 11 0.0
18 11 0.008
19 11 0.0
20 11 0.045
0 12 0.0
1 12 0.077
2 12 0.0
3 12 0.0
4 12 0.0
5 12 0.0
6 12 0.0
7 12 0.0
8 12 0.0
9 12 0.0
10 12 0.026
11 12 0.0
12 12 0.846
13 12 0.0
14 12 0.0
15 12 0.0
16 12 0.026
17 12 0.0
18 12 0.026
19 12 0.0
20 12 0.0
0 13 0.043
1 13 0.022
2 13 0.0
3 13 0.022
4 13 0.065
5 13 0.0
6 13 0.0
7 13 0.087
8 13 0.0
9 13 0.0
10 13 0.196
11 13 0.0
12 13 0.0
13 13 0.326
14 13 0.065
15 13 0.152
16 13 0.0
17 13 0.0
18 13 0.0
19 13 0.0
20 13 0.022
0 14 0.04
1 14 0.0
2 14 0.0
3 14 0.01
4 14 0.208
5 14 0.0
6 14 0.0
7 14 0.03
8 14 0.0
9 14 0.03
10 14 0.03
11 14 0.0
12 14 0.02
13 14 0.03
14 14 0.465
15 14 0.04
16 14 0.079
17 14 0.02
18 14 0.0
19 14 0.0
20 14 0.0
0 15 0.023
1 15 0.023
2 15 0.0
3 15 0.0
4 15 0.011
5 15 0.0
6 15 0.0
7 15 0.011
8 15 0.0
9 15 0.0
10 15 0.011
11 15 0.0
12 15 0.0
13 15 0.023
14 15 0.023
15 15 0.529
16 15 0.092
17 15 0.241
18 15 0.0
19 15 0.0
20 15 0.011
0 16 0.004
1 16 0.021
2 16 0.0
3 16 0.0
4 16 0.204
5 16 0.0
6 16 0.0
7 16 0.021
8 16 0.0
9 16 0.043
10 16 0.021
11 16 0.0
12 16 0.0
13 16 0.047
14 16 0.055
15 16 0.106
16 16 0.328
17 16 0.132
18 16 0.0
19 16 0.009
20 16 0.009
0 17 0.0
1 17 0.009
2 17 0.0
3 17 0.0
4 17 0.0
5 17 0.0
6 17 0.0
7 17 0.004
8 17 0.0
9 17 0.0
10 17 0.0
11 17 0.0
12 17 0.0
13 17 0.013
14 17 0.0
15 17 0.009
16 17 0.017
17 17 0.948
18 17 0.0
19 17 0.0
20 17 0.0
0 18 0.005
1 18 0.073
2 18 0.0
3 18 0.005
4 18 0.068
5 18 0.047
6 18 0.0
7 18 0.016
8 18 0.0
9 18 0.0
10 18 0.105
11 18 0.0
12 18 0.005
13 18 0.052
14 18 0.105
15 18 0.0
16 18 0.0
17 18 0.0
18 18 0.492
19 18 0.021
20 18 0.005
0 19 0.01
1 19 0.048
2 19 0.0
3 19 0.0
4 19 0.019
5 19 0.138
6 19 0.0
7 19 0.033
8 19 0.0
9 19 0.0
10 19 0.005
11 19 0.01
12 19 0.0
13 19 0.048
14 19 0.024
15 19 0.0
16 19 0.0
17 19 0.0
18 19 0.005
19 19 0.562
20 19 0.1
0 20 0.013
1 20 0.053
2 20 0.0
3 20 0.013
4 20 0.026
5 20 0.026
6 20 0.0
7 20 0.039
8 20 0.0
9 20 0.0
10 20 0.026
11 20 0.026
12 20 0.053
13 20 0.039
14 20 0.145
15 20 0.013
16 20 0.013
17 20 0.013
18 20 0.013
19 20 0.026
20 20 0.461
        };
    \end{axis}
\end{tikzpicture}
%
%\caption{I'm confused~5!}\label{tab:CM5}
%\end{figure*}
	}
& & 
	\resizebox{0.45\textwidth}{0.25\textheight}{%
	%\begin{tikzpicture}[scale=0.60,every node/.style={scale=0.9}]
%\begin{tikzpicture}[scale=0.55]
\begin{tikzpicture}[scale=0.60]
    \begin{axis}[%colorbar/width=2.5mm,
        width=20cm,
        height=20cm,
%        colormap={blackwhite}{gray(0cm)=(1); gray(1cm)=(0.5)},
%   colormap={bluewhite}{color=(white) color=(blue)},
%   colormap={bluewhite}{color=(white) rgb255=(0,191,255)},
    colormap={bluewhite}{color=(white) rgb255=(100,149,237)},
        xticklabels={Adload,Agent,Allaple,BHO,Bifrose,CeeInject,Cycbot,FakeRean,Hotbar,Injector,
        OnLineGames,Renos,Rimecud,Small,Toga,VB,VBinject,Vobfus,Vundo,Winwebsec,Zbot},
        xtick={0,...,20},
        xtick style={draw=none},
    xticklabel style={anchor=east,rotate=45,yshift=-5pt},
        yticklabels={Adload,Agent,Allaple,BHO,Bifrose,CeeInject,Cycbot,FakeRean,Hotbar,Injector,
        OnLineGames,Renos,Rimecud,Small,Toga,VB,VBinject,Vobfus,Vundo,Winwebsec,Zbot},
        ytick={0,...,20},
        ytick style={draw=none},
        enlargelimits=false,
        xticklabel style={font=\large},
        yticklabel style={font=\large},
        colorbar,
        colorbar style={
%           width=0.05*\pgfkeysvalueof{/pgfplots/parent axis width},%%% added this
%           height=0.5*\pgfkeysvalueof{/pgfplots/parent axis height},
%       plot graphics/node/.style={scale=1.33,anchor=south west,inner sep=0pt,}, %%% scale colorbar fill %%%
            ytick={0.0,0.2,0.4,0.6,0.8,1.0},
            yticklabels={0.0,0.2,0.4,0.6,0.8,1.0},
            yticklabel={\pgfmathprintnumber\tick},
            yticklabel style={font=\large,
                    /pgf/number format/fixed,
            /pgf/number format/precision=1}
        },
%        point meta min=0,
%        point meta max=100,
        point meta min=0.0,
        point meta max=1.0,
        nodes near coords={\pgfmathprintnumber\pgfplotspointmeta},
        % ---------------------------------------------------------------------
        % show `nodes near coords' but adapt the style so that values
        % above a threshold get another style
        % (adapted from <http://tex.stackexchange.com/a/141006/95441>)
        % #1: the THRESHOLD after which we switch to a special display.
        nodes near coords black white/.style={
            % define the style of the nodes with "small" values
            small value/.style={
                yshift=-7pt,
%                text=white,
                text=black,
                /pgf/number format/fixed,
                /pgf/number format/precision=3,
                /pgf/number format/zerofill
            },
            % define the style of the nodes with "large" values
            large value/.style={
                yshift=-7pt,
%                text=black,
                text=white,
                /pgf/number format/fixed,
                /pgf/number format/precision=3,
                /pgf/number format/zerofill
            },
            every node near coord/.style={
                check for zero/.code={
                    \pgfmathfloatifflags{\pgfplotspointmeta}{0}{
                        % If meta=0, make the node a coordinate
                        % (which doesn't have text)
                        \pgfkeys{/tikz/coordinate}
                    }{
                        \begingroup
                        % this group is merely to switch to FPU locally.
                        % Might be unnecessary, but who knows.
                        \pgfkeys{/pgf/fpu}
                        \pgfmathparse{\pgfplotspointmeta<#1}
                        \global\let\result=\pgfmathresult
                        \endgroup
                        %
                        % simplifies debugging:
                        %\show\result
                        %
                        \pgfmathfloatcreate{1}{1.0}{0}
                        \let\ONE=\pgfmathresult
                        \ifx\result\ONE
                            % AH: our condition 'y < #1' is met.
                            \pgfkeysalso{/pgfplots/small value}
                        \else
                            % ok, proceed as usual.
                            \pgfkeysalso{/pgfplots/large value}
                        \fi
                    }
                },
                check for zero,
            },
        },
        % asign a value to the new style which is the threshold at which
        % the two style `small value' or `large value' are used
%        nodes near coords black white=50,
        nodes near coords black white=0.5,
        % -----------------------------------------------------------------
    ]
        \addplot[
            matrix plot,
            mesh/cols=21,
            point meta=explicit,draw=gray
        ] table [meta=C] {
            x y C
0 0 0.881
1 0 0.0
2 0 0.0
3 0 0.119
4 0 0.0
5 0 0.0
6 0 0.0
7 0 0.0
8 0 0.0
9 0 0.0
10 0 0.0
11 0 0.0
12 0 0.0
13 0 0.0
14 0 0.0
15 0 0.0
16 0 0.0
17 0 0.0
18 0 0.0
19 0 0.0
20 0 0.0
0 1 0.0
1 1 0.486
2 1 0.0
3 1 0.0
4 1 0.0
5 1 0.0
6 1 0.0
7 1 0.114
8 1 0.0
9 1 0.0
10 1 0.029
11 1 0.029
12 1 0.0
13 1 0.057
14 1 0.029
15 1 0.0
16 1 0.086
17 1 0.029
18 1 0.057
19 1 0.086
20 1 0.0
0 2 0.0
1 2 0.0
2 2 1.0
3 2 0.0
4 2 0.0
5 2 0.0
6 2 0.0
7 2 0.0
8 2 0.0
9 2 0.0
10 2 0.0
11 2 0.0
12 2 0.0
13 2 0.0
14 2 0.0
15 2 0.0
16 2 0.0
17 2 0.0
18 2 0.0
19 2 0.0
20 2 0.0
0 3 0.042
1 3 0.0
2 3 0.0
3 3 0.887
4 3 0.0
5 3 0.014
6 3 0.0
7 3 0.0
8 3 0.0
9 3 0.0
10 3 0.0
11 3 0.0
12 3 0.0
13 3 0.0
14 3 0.0
15 3 0.0
16 3 0.0
17 3 0.0
18 3 0.028
19 3 0.014
20 3 0.014
0 4 0.0
1 4 0.0
2 4 0.0
3 4 0.0
4 4 0.105
5 4 0.158
6 4 0.0
7 4 0.105
8 4 0.0
9 4 0.026
10 4 0.0
11 4 0.132
12 4 0.0
13 4 0.0
14 4 0.0
15 4 0.0
16 4 0.395
17 4 0.053
18 4 0.026
19 4 0.0
20 4 0.0
0 5 0.0
1 5 0.0
2 5 0.0
3 5 0.006
4 5 0.0
5 5 0.899
6 5 0.006
7 5 0.013
8 5 0.0
9 5 0.0
10 5 0.0
11 5 0.032
12 5 0.0
13 5 0.0
14 5 0.013
15 5 0.0
16 5 0.0
17 5 0.0
18 5 0.006
19 5 0.019
20 5 0.006
0 6 0.0
1 6 0.0
2 6 0.0
3 6 0.0
4 6 0.0
5 6 0.0
6 6 0.953
7 6 0.031
8 6 0.0
9 6 0.0
10 6 0.0
11 6 0.0
12 6 0.0
13 6 0.0
14 6 0.0
15 6 0.0
16 6 0.0
17 6 0.0
18 6 0.0
19 6 0.016
20 6 0.0
0 7 0.0
1 7 0.0
2 7 0.0
3 7 0.0
4 7 0.0
5 7 0.011
6 7 0.0
7 7 0.874
8 7 0.0
9 7 0.0
10 7 0.0
11 7 0.011
12 7 0.0
13 7 0.0
14 7 0.0
15 7 0.0
16 7 0.021
17 7 0.0
18 7 0.021
19 7 0.063
20 7 0.0
0 8 0.0
1 8 0.0
2 8 0.0
3 8 0.0
4 8 0.0
5 8 0.0
6 8 0.0
7 8 0.038
8 8 0.962
9 8 0.0
10 8 0.0
11 8 0.0
12 8 0.0
13 8 0.0
14 8 0.0
15 8 0.0
16 8 0.0
17 8 0.0
18 8 0.0
19 8 0.0
20 8 0.0
0 9 0.0
1 9 0.032
2 9 0.0
3 9 0.0
4 9 0.0
5 9 0.161
6 9 0.0
7 9 0.097
8 9 0.0
9 9 0.226
10 9 0.0
11 9 0.129
12 9 0.0
13 9 0.0
14 9 0.065
15 9 0.032
16 9 0.161
17 9 0.0
18 9 0.065
19 9 0.032
20 9 0.0
0 10 0.0
1 10 0.0
2 10 0.0
3 10 0.0
4 10 0.0
5 10 0.0
6 10 0.021
7 10 0.043
8 10 0.0
9 10 0.0
10 10 0.745
11 10 0.021
12 10 0.0
13 10 0.021
14 10 0.0
15 10 0.0
16 10 0.043
17 10 0.0
18 10 0.064
19 10 0.043
20 10 0.0
0 11 0.0
1 11 0.0
2 11 0.0
3 11 0.01
4 11 0.0
5 11 0.0
6 11 0.01
7 11 0.019
8 11 0.0
9 11 0.0
10 11 0.0
11 11 0.886
12 11 0.0
13 11 0.0
14 11 0.01
15 11 0.0
16 11 0.019
17 11 0.038
18 11 0.01
19 11 0.0
20 11 0.0
0 12 0.0
1 12 0.0
2 12 0.0
3 12 0.0
4 12 0.0
5 12 0.0
6 12 0.0
7 12 0.088
8 12 0.0
9 12 0.0
10 12 0.0
11 12 0.0
12 12 0.853
13 12 0.0
14 12 0.0
15 12 0.0
16 12 0.0
17 12 0.0
18 12 0.0
19 12 0.059
20 12 0.0
0 13 0.0
1 13 0.029
2 13 0.0
3 13 0.029
4 13 0.0
5 13 0.029
6 13 0.0
7 13 0.029
8 13 0.0
9 13 0.0
10 13 0.0
11 13 0.057
12 13 0.0
13 13 0.429
14 13 0.0
15 13 0.0
16 13 0.086
17 13 0.0
18 13 0.057
19 13 0.257
20 13 0.0
0 14 0.0
1 14 0.0
2 14 0.0
3 14 0.025
4 14 0.013
5 14 0.013
6 14 0.0
7 14 0.114
8 14 0.0
9 14 0.0
10 14 0.0
11 14 0.025
12 14 0.0
13 14 0.0
14 14 0.443
15 14 0.013
16 14 0.241
17 14 0.013
18 14 0.038
19 14 0.051
20 14 0.013
0 15 0.0
1 15 0.0
2 15 0.0
3 15 0.0
4 15 0.0
5 15 0.0
6 15 0.0
7 15 0.034
8 15 0.0
9 15 0.017
10 15 0.0
11 15 0.017
12 15 0.0
13 15 0.0
14 15 0.0
15 15 0.483
16 15 0.397
17 15 0.034
18 15 0.0
19 15 0.017
20 15 0.0
0 16 0.0
1 16 0.0
2 16 0.0
3 16 0.0
4 16 0.0
5 16 0.006
6 16 0.0
7 16 0.045
8 16 0.0
9 16 0.0
10 16 0.0
11 16 0.022
12 16 0.006
13 16 0.0
14 16 0.034
15 16 0.011
16 16 0.815
17 16 0.0
18 16 0.011
19 16 0.039
20 16 0.011
0 17 0.0
1 17 0.0
2 17 0.0
3 17 0.0
4 17 0.0
5 17 0.0
6 17 0.005
7 17 0.0
8 17 0.0
9 17 0.0
10 17 0.0
11 17 0.005
12 17 0.0
13 17 0.0
14 17 0.0
15 17 0.0
16 17 0.074
17 17 0.905
18 17 0.0
19 17 0.011
20 17 0.0
0 18 0.0
1 18 0.0
2 18 0.006
3 18 0.006
4 18 0.0
5 18 0.0
6 18 0.0
7 18 0.036
8 18 0.0
9 18 0.0
10 18 0.0
11 18 0.018
12 18 0.0
13 18 0.0
14 18 0.0
15 18 0.0
16 18 0.0
17 18 0.0
18 18 0.916
19 18 0.018
20 18 0.0
0 19 0.0
1 19 0.006
2 19 0.0
3 19 0.006
4 19 0.0
5 19 0.0
6 19 0.0
7 19 0.024
8 19 0.0
9 19 0.0
10 19 0.0
11 19 0.006
12 19 0.0
13 19 0.0
14 19 0.0
15 19 0.0
16 19 0.012
17 19 0.0
18 19 0.006
19 19 0.941
20 19 0.0
0 20 0.0
1 20 0.0
2 20 0.0
3 20 0.0
4 20 0.0
5 20 0.0
6 20 0.0
7 20 0.036
8 20 0.0
9 20 0.0
10 20 0.018
11 20 0.182
12 20 0.018
13 20 0.0
14 20 0.036
15 20 0.0
16 20 0.073
17 20 0.0
18 20 0.036
19 20 0.073
20 20 0.527
        };
    \end{axis}
\end{tikzpicture}
%
%\caption{I'm confused~5!}\label{tab:CM5}
%\end{figure*}
	}
\\
(a) HMM
& & 
(b) CNN
\\ \\
	\resizebox{0.45\textwidth}{0.25\textheight}{%
	%\begin{tikzpicture}[scale=0.60,every node/.style={scale=0.9}]
%\begin{tikzpicture}[scale=0.55]
\begin{tikzpicture}[scale=0.60]
    \begin{axis}[%colorbar/width=2.5mm,
        width=20cm,
        height=20cm,
%        colormap={blackwhite}{gray(0cm)=(1); gray(1cm)=(0.5)},
%   colormap={bluewhite}{color=(white) color=(blue)},
%   colormap={bluewhite}{color=(white) rgb255=(0,191,255)},
    colormap={bluewhite}{color=(white) rgb255=(100,149,237)},
        xticklabels={Adload,Agent,Allaple,BHO,Bifrose,CeeInject,Cycbot,FakeRean,Hotbar,Injector,
        OnLineGames,Renos,Rimecud,Small,Toga,VB,VBinject,Vobfus,Vundo,Winwebsec,Zbot},
        xtick={0,...,20},
        xtick style={draw=none},
    xticklabel style={anchor=east,rotate=45,yshift=-5pt},
        yticklabels={Adload,Agent,Allaple,BHO,Bifrose,CeeInject,Cycbot,FakeRean,Hotbar,Injector,
        OnLineGames,Renos,Rimecud,Small,Toga,VB,VBinject,Vobfus,Vundo,Winwebsec,Zbot},
        ytick={0,...,20},
        ytick style={draw=none},
        enlargelimits=false,
        xticklabel style={font=\large},
        yticklabel style={font=\large},
        colorbar,
        colorbar style={
%           width=0.05*\pgfkeysvalueof{/pgfplots/parent axis width},%%% added this
%           height=0.5*\pgfkeysvalueof{/pgfplots/parent axis height},
%       plot graphics/node/.style={scale=1.33,anchor=south west,inner sep=0pt,}, %%% scale colorbar fill %%%
            ytick={0.0,0.2,0.4,0.6,0.8,1.0},
            yticklabels={0.0,0.2,0.4,0.6,0.8,1.0},
            yticklabel={\pgfmathprintnumber\tick},
            yticklabel style={font=\large,
                    /pgf/number format/fixed,
            /pgf/number format/precision=1}
        },
%        point meta min=0,
%        point meta max=100,
        point meta min=0.0,
        point meta max=1.0,
        nodes near coords={\pgfmathprintnumber\pgfplotspointmeta},
        % ---------------------------------------------------------------------
        % show `nodes near coords' but adapt the style so that values
        % above a threshold get another style
        % (adapted from <http://tex.stackexchange.com/a/141006/95441>)
        % #1: the THRESHOLD after which we switch to a special display.
        nodes near coords black white/.style={
            % define the style of the nodes with "small" values
            small value/.style={
                yshift=-7pt,
%                text=white,
                text=black,
                /pgf/number format/fixed,
                /pgf/number format/precision=3,
                /pgf/number format/zerofill
            },
            % define the style of the nodes with "large" values
            large value/.style={
                yshift=-7pt,
%                text=black,
                text=white,
                /pgf/number format/fixed,
                /pgf/number format/precision=3,
                /pgf/number format/zerofill
            },
            every node near coord/.style={
                check for zero/.code={
                    \pgfmathfloatifflags{\pgfplotspointmeta}{0}{
                        % If meta=0, make the node a coordinate
                        % (which doesn't have text)
                        \pgfkeys{/tikz/coordinate}
                    }{
                        \begingroup
                        % this group is merely to switch to FPU locally.
                        % Might be unnecessary, but who knows.
                        \pgfkeys{/pgf/fpu}
                        \pgfmathparse{\pgfplotspointmeta<#1}
                        \global\let\result=\pgfmathresult
                        \endgroup
                        %
                        % simplifies debugging:
                        %\show\result
                        %
                        \pgfmathfloatcreate{1}{1.0}{0}
                        \let\ONE=\pgfmathresult
                        \ifx\result\ONE
                            % AH: our condition 'y < #1' is met.
                            \pgfkeysalso{/pgfplots/small value}
                        \else
                            % ok, proceed as usual.
                            \pgfkeysalso{/pgfplots/large value}
                        \fi
                    }
                },
                check for zero,
            },
        },
        % asign a value to the new style which is the threshold at which
        % the two style `small value' or `large value' are used
%        nodes near coords black white=50,
        nodes near coords black white=0.5,
        % -----------------------------------------------------------------
    ]
        \addplot[
            matrix plot,
            mesh/cols=21,
            point meta=explicit,draw=gray
        ] table [meta=C] {
            x y C
0 0 0.929
1 0 0.0
2 0 0.024
3 0 0.024
4 0 0.0
5 0 0.0
6 0 0.0
7 0 0.0
8 0 0.0
9 0 0.0
10 0 0.0
11 0 0.024
12 0 0.0
13 0 0.0
14 0 0.0
15 0 0.0
16 0 0.0
17 0 0.0
18 0 0.0
19 0 0.0
20 0 0.0
0 1 0.0
1 1 0.486
2 1 0.057
3 1 0.0
4 1 0.0
5 1 0.0
6 1 0.0
7 1 0.114
8 1 0.0
9 1 0.086
10 1 0.0
11 1 0.029
12 1 0.0
13 1 0.029
14 1 0.0
15 1 0.029
16 1 0.086
17 1 0.029
18 1 0.0
19 1 0.057
20 1 0.0
0 2 0.0
1 2 0.0
2 2 1.0
3 2 0.0
4 2 0.0
5 2 0.0
6 2 0.0
7 2 0.0
8 2 0.0
9 2 0.0
10 2 0.0
11 2 0.0
12 2 0.0
13 2 0.0
14 2 0.0
15 2 0.0
16 2 0.0
17 2 0.0
18 2 0.0
19 2 0.0
20 2 0.0
0 3 0.042
1 3 0.0
2 3 0.014
3 3 0.915
4 3 0.0
5 3 0.0
6 3 0.0
7 3 0.014
8 3 0.0
9 3 0.0
10 3 0.0
11 3 0.0
12 3 0.0
13 3 0.0
14 3 0.0
15 3 0.0
16 3 0.0
17 3 0.0
18 3 0.014
19 3 0.0
20 3 0.0
0 4 0.0
1 4 0.026
2 4 0.079
3 4 0.0
4 4 0.553
5 4 0.053
6 4 0.026
7 4 0.026
8 4 0.0
9 4 0.053
10 4 0.0
11 4 0.0
12 4 0.0
13 4 0.0
14 4 0.0
15 4 0.0
16 4 0.184
17 4 0.0
18 4 0.0
19 4 0.0
20 4 0.0
0 5 0.0
1 5 0.0
2 5 0.038
3 5 0.0
4 5 0.006
5 5 0.905
6 5 0.0
7 5 0.006
8 5 0.0
9 5 0.006
10 5 0.0
11 5 0.013
12 5 0.0
13 5 0.0
14 5 0.0
15 5 0.0
16 5 0.006
17 5 0.0
18 5 0.0
19 5 0.013
20 5 0.006
0 6 0.0
1 6 0.0
2 6 0.0
3 6 0.0
4 6 0.0
5 6 0.0
6 6 1.0
7 6 0.0
8 6 0.0
9 6 0.0
10 6 0.0
11 6 0.0
12 6 0.0
13 6 0.0
14 6 0.0
15 6 0.0
16 6 0.0
17 6 0.0
18 6 0.0
19 6 0.0
20 6 0.0
0 7 0.0
1 7 0.0
2 7 0.042
3 7 0.0
4 7 0.0
5 7 0.0
6 7 0.042
7 7 0.874
8 7 0.0
9 7 0.0
10 7 0.0
11 7 0.021
12 7 0.0
13 7 0.0
14 7 0.0
15 7 0.0
16 7 0.011
17 7 0.0
18 7 0.0
19 7 0.011
20 7 0.0
0 8 0.0
1 8 0.0
2 8 0.0
3 8 0.0
4 8 0.0
5 8 0.0
6 8 0.0
7 8 0.0
8 8 1.0
9 8 0.0
10 8 0.0
11 8 0.0
12 8 0.0
13 8 0.0
14 8 0.0
15 8 0.0
16 8 0.0
17 8 0.0
18 8 0.0
19 8 0.0
20 8 0.0
0 9 0.0
1 9 0.0
2 9 0.065
3 9 0.0
4 9 0.0
5 9 0.097
6 9 0.0
7 9 0.0
8 9 0.0
9 9 0.677
10 9 0.0
11 9 0.032
12 9 0.0
13 9 0.0
14 9 0.0
15 9 0.0
16 9 0.097
17 9 0.0
18 9 0.032
19 9 0.0
20 9 0.0
0 10 0.0
1 10 0.0
2 10 0.064
3 10 0.0
4 10 0.0
5 10 0.0
6 10 0.0
7 10 0.0
8 10 0.0
9 10 0.021
10 10 0.766
11 10 0.064
12 10 0.0
13 10 0.0
14 10 0.0
15 10 0.0
16 10 0.0
17 10 0.0
18 10 0.064
19 10 0.021
20 10 0.0
0 11 0.0
1 11 0.0
2 11 0.067
3 11 0.0
4 11 0.0
5 11 0.0
6 11 0.01
7 11 0.0
8 11 0.0
9 11 0.0
10 11 0.0
11 11 0.914
12 11 0.0
13 11 0.0
14 11 0.0
15 11 0.0
16 11 0.0
17 11 0.0
18 11 0.0
19 11 0.0
20 11 0.01
0 12 0.0
1 12 0.0
2 12 0.118
3 12 0.0
4 12 0.0
5 12 0.0
6 12 0.0
7 12 0.0
8 12 0.0
9 12 0.0
10 12 0.0
11 12 0.029
12 12 0.824
13 12 0.0
14 12 0.0
15 12 0.0
16 12 0.0
17 12 0.0
18 12 0.0
19 12 0.029
20 12 0.0
0 13 0.0
1 13 0.029
2 13 0.114
3 13 0.0
4 13 0.0
5 13 0.057
6 13 0.0
7 13 0.0
8 13 0.0
9 13 0.029
10 13 0.0
11 13 0.086
12 13 0.0
13 13 0.571
14 13 0.0
15 13 0.0
16 13 0.057
17 13 0.0
18 13 0.0
19 13 0.057
20 13 0.0
0 14 0.0
1 14 0.013
2 14 0.177
3 14 0.0
4 14 0.025
5 14 0.0
6 14 0.025
7 14 0.013
8 14 0.0
9 14 0.013
10 14 0.0
11 14 0.025
12 14 0.0
13 14 0.0
14 14 0.519
15 14 0.0
16 14 0.152
17 14 0.025
18 14 0.0
19 14 0.013
20 14 0.0
0 15 0.0
1 15 0.034
2 15 0.034
3 15 0.0
4 15 0.0
5 15 0.0
6 15 0.017
7 15 0.0
8 15 0.0
9 15 0.0
10 15 0.0
11 15 0.0
12 15 0.0
13 15 0.0
14 15 0.0
15 15 0.517
16 15 0.259
17 15 0.103
18 15 0.0
19 15 0.034
20 15 0.0
0 16 0.0
1 16 0.0
2 16 0.022
3 16 0.0
4 16 0.011
5 16 0.0
6 16 0.0
7 16 0.011
8 16 0.0
9 16 0.0
10 16 0.006
11 16 0.006
12 16 0.006
13 16 0.0
14 16 0.006
15 16 0.0
16 16 0.854
17 16 0.073
18 16 0.0
19 16 0.006
20 16 0.0
0 17 0.0
1 17 0.0
2 17 0.005
3 17 0.0
4 17 0.0
5 17 0.005
6 17 0.0
7 17 0.0
8 17 0.0
9 17 0.0
10 17 0.0
11 17 0.0
12 17 0.0
13 17 0.0
14 17 0.0
15 17 0.0
16 17 0.047
17 17 0.942
18 17 0.0
19 17 0.0
20 17 0.0
0 18 0.0
1 18 0.0
2 18 0.006
3 18 0.0
4 18 0.0
5 18 0.0
6 18 0.006
7 18 0.03
8 18 0.0
9 18 0.0
10 18 0.0
11 18 0.012
12 18 0.0
13 18 0.0
14 18 0.0
15 18 0.0
16 18 0.0
17 18 0.0
18 18 0.934
19 18 0.006
20 18 0.006
0 19 0.0
1 19 0.0
2 19 0.035
3 19 0.0
4 19 0.0
5 19 0.0
6 19 0.018
7 19 0.006
8 19 0.0
9 19 0.0
10 19 0.0
11 19 0.006
12 19 0.0
13 19 0.0
14 19 0.0
15 19 0.0
16 19 0.0
17 19 0.0
18 19 0.0
19 19 0.935
20 19 0.0
0 20 0.0
1 20 0.0
2 20 0.145
3 20 0.0
4 20 0.0
5 20 0.0
6 20 0.018
7 20 0.018
8 20 0.0
9 20 0.018
10 20 0.0
11 20 0.036
12 20 0.0
13 20 0.0
14 20 0.0
15 20 0.0
16 20 0.018
17 20 0.0
18 20 0.0
19 20 0.055
20 20 0.691
        };
    \end{axis}
\end{tikzpicture}
%
%\caption{I'm confused~5!}\label{tab:CM5}
%\end{figure*}
	}

& & 
	\resizebox{0.45\textwidth}{0.25\textheight}{%
	%\begin{tikzpicture}[scale=0.60,every node/.style={scale=0.9}]
%\begin{tikzpicture}[scale=0.55]
\begin{tikzpicture}[scale=0.60]
    \begin{axis}[%colorbar/width=2.5mm,
        width=20cm,
        height=20cm,
%        colormap={blackwhite}{gray(0cm)=(1); gray(1cm)=(0.5)},
%   colormap={bluewhite}{color=(white) color=(blue)},
%   colormap={bluewhite}{color=(white) rgb255=(0,191,255)},
    colormap={bluewhite}{color=(white) rgb255=(100,149,237)},
        xticklabels={Adload,Agent,Allaple,BHO,Bifrose,CeeInject,Cycbot,FakeRean,Hotbar,Injector,
        OnLineGames,Renos,Rimecud,Small,Toga,VB,VBinject,Vobfus,Vundo,Winwebsec,Zbot},
        xtick={0,...,20},
        xtick style={draw=none},
    xticklabel style={anchor=east,rotate=45,yshift=-5pt},
        yticklabels={Adload,Agent,Allaple,BHO,Bifrose,CeeInject,Cycbot,FakeRean,Hotbar,Injector,
        OnLineGames,Renos,Rimecud,Small,Toga,VB,VBinject,Vobfus,Vundo,Winwebsec,Zbot},
        ytick={0,...,20},
        ytick style={draw=none},
        enlargelimits=false,
        xticklabel style={font=\large},
        yticklabel style={font=\large},
        colorbar,
        colorbar style={
%           width=0.05*\pgfkeysvalueof{/pgfplots/parent axis width},%%% added this
%           height=0.5*\pgfkeysvalueof{/pgfplots/parent axis height},
%       plot graphics/node/.style={scale=1.33,anchor=south west,inner sep=0pt,}, %%% scale colorbar fill %%%
            ytick={0.0,0.2,0.4,0.6,0.8,1.0},
            yticklabels={0.0,0.2,0.4,0.6,0.8,1.0},
            yticklabel={\pgfmathprintnumber\tick},
            yticklabel style={font=\large,
                    /pgf/number format/fixed,
            /pgf/number format/precision=1}
        },
%        point meta min=0,
%        point meta max=100,
        point meta min=0.0,
        point meta max=1.0,
        nodes near coords={\pgfmathprintnumber\pgfplotspointmeta},
        % ---------------------------------------------------------------------
        % show `nodes near coords' but adapt the style so that values
        % above a threshold get another style
        % (adapted from <http://tex.stackexchange.com/a/141006/95441>)
        % #1: the THRESHOLD after which we switch to a special display.
        nodes near coords black white/.style={
            % define the style of the nodes with "small" values
            small value/.style={
                yshift=-7pt,
%                text=white,
                text=black,
                /pgf/number format/fixed,
                /pgf/number format/precision=3,
                /pgf/number format/zerofill
            },
            % define the style of the nodes with "large" values
            large value/.style={
                yshift=-7pt,
%                text=black,
                text=white,
                /pgf/number format/fixed,
                /pgf/number format/precision=3,
                /pgf/number format/zerofill
            },
            every node near coord/.style={
                check for zero/.code={
                    \pgfmathfloatifflags{\pgfplotspointmeta}{0}{
                        % If meta=0, make the node a coordinate
                        % (which doesn't have text)
                        \pgfkeys{/tikz/coordinate}
                    }{
                        \begingroup
                        % this group is merely to switch to FPU locally.
                        % Might be unnecessary, but who knows.
                        \pgfkeys{/pgf/fpu}
                        \pgfmathparse{\pgfplotspointmeta<#1}
                        \global\let\result=\pgfmathresult
                        \endgroup
                        %
                        % simplifies debugging:
                        %\show\result
                        %
                        \pgfmathfloatcreate{1}{1.0}{0}
                        \let\ONE=\pgfmathresult
                        \ifx\result\ONE
                            % AH: our condition 'y < #1' is met.
                            \pgfkeysalso{/pgfplots/small value}
                        \else
                            % ok, proceed as usual.
                            \pgfkeysalso{/pgfplots/large value}
                        \fi
                    }
                },
                check for zero,
            },
        },
        % asign a value to the new style which is the threshold at which
        % the two style `small value' or `large value' are used
%        nodes near coords black white=50,
        nodes near coords black white=0.5,
        % -----------------------------------------------------------------
    ]
        \addplot[
            matrix plot,
            mesh/cols=21,
            point meta=explicit,draw=gray
        ] table [meta=C] {
            x y C
0 0 0.81
1 0 0.0
2 0 0.0
3 0 0.119
4 0 0.0
5 0 0.0
6 0 0.0
7 0 0.0
8 0 0.0
9 0 0.0
10 0 0.0
11 0 0.024
12 0 0.0
13 0 0.024
14 0 0.0
15 0 0.0
16 0 0.024
17 0 0.0
18 0 0.0
19 0 0.0
20 0 0.0
0 1 0.0
1 1 0.171
2 1 0.0
3 1 0.057
4 1 0.0
5 1 0.029
6 1 0.0
7 1 0.114
8 1 0.0
9 1 0.0
10 1 0.057
11 1 0.086
12 1 0.0
13 1 0.0
14 1 0.029
15 1 0.114
16 1 0.0
17 1 0.2
18 1 0.114
19 1 0.029
20 1 0.0
0 2 0.0
1 2 0.0
2 2 1.0
3 2 0.0
4 2 0.0
5 2 0.0
6 2 0.0
7 2 0.0
8 2 0.0
9 2 0.0
10 2 0.0
11 2 0.0
12 2 0.0
13 2 0.0
14 2 0.0
15 2 0.0
16 2 0.0
17 2 0.0
18 2 0.0
19 2 0.0
20 2 0.0
0 3 0.0
1 3 0.0
2 3 0.0
3 3 0.944
4 3 0.0
5 3 0.0
6 3 0.0
7 3 0.0
8 3 0.0
9 3 0.0
10 3 0.0
11 3 0.0
12 3 0.0
13 3 0.0
14 3 0.0
15 3 0.0
16 3 0.014
17 3 0.0
18 3 0.042
19 3 0.0
20 3 0.0
0 4 0.0
1 4 0.0
2 4 0.053
3 4 0.0
4 4 0.0
5 4 0.105
6 4 0.0
7 4 0.053
8 4 0.0
9 4 0.0
10 4 0.0
11 4 0.026
12 4 0.0
13 4 0.026
14 4 0.105
15 4 0.026
16 4 0.237
17 4 0.132
18 4 0.132
19 4 0.079
20 4 0.026
0 5 0.0
1 5 0.006
2 5 0.0
3 5 0.019
4 5 0.0
5 5 0.829
6 5 0.0
7 5 0.0
8 5 0.0
9 5 0.006
10 5 0.0
11 5 0.032
12 5 0.0
13 5 0.006
14 5 0.006
15 5 0.0
16 5 0.006
17 5 0.07
18 5 0.006
19 5 0.013
20 5 0.0
0 6 0.0
1 6 0.0
2 6 0.0
3 6 0.0
4 6 0.0
5 6 0.0
6 6 0.977
7 6 0.0
8 6 0.0
9 6 0.0
10 6 0.0
11 6 0.0
12 6 0.0
13 6 0.0
14 6 0.0
15 6 0.0
16 6 0.0
17 6 0.0
18 6 0.0
19 6 0.023
20 6 0.0
0 7 0.0
1 7 0.011
2 7 0.0
3 7 0.0
4 7 0.0
5 7 0.0
6 7 0.063
7 7 0.642
8 7 0.0
9 7 0.0
10 7 0.0
11 7 0.011
12 7 0.0
13 7 0.011
14 7 0.0
15 7 0.0
16 7 0.0
17 7 0.011
18 7 0.095
19 7 0.095
20 7 0.063
0 8 0.0
1 8 0.0
2 8 0.0
3 8 0.0
4 8 0.0
5 8 0.0
6 8 0.0
7 8 0.0
8 8 0.962
9 8 0.0
10 8 0.0
11 8 0.038
12 8 0.0
13 8 0.0
14 8 0.0
15 8 0.0
16 8 0.0
17 8 0.0
18 8 0.0
19 8 0.0
20 8 0.0
0 9 0.0
1 9 0.032
2 9 0.0
3 9 0.032
4 9 0.0
5 9 0.097
6 9 0.0
7 9 0.032
8 9 0.0
9 9 0.0
10 9 0.032
11 9 0.097
12 9 0.0
13 9 0.0
14 9 0.161
15 9 0.032
16 9 0.0
17 9 0.323
18 9 0.065
19 9 0.097
20 9 0.0
0 10 0.0
1 10 0.043
2 10 0.021
3 10 0.021
4 10 0.0
5 10 0.021
6 10 0.0
7 10 0.043
8 10 0.0
9 10 0.0
10 10 0.553
11 10 0.0
12 10 0.0
13 10 0.021
14 10 0.021
15 10 0.0
16 10 0.0
17 10 0.0
18 10 0.191
19 10 0.043
20 10 0.021
0 11 0.0
1 11 0.01
2 11 0.01
3 11 0.01
4 11 0.0
5 11 0.01
6 11 0.01
7 11 0.0
8 11 0.0
9 11 0.0
10 11 0.01
11 11 0.81
12 11 0.0
13 11 0.01
14 11 0.038
15 11 0.0
16 11 0.029
17 11 0.0
18 11 0.01
19 11 0.029
20 11 0.019
0 12 0.0
1 12 0.0
2 12 0.0
3 12 0.0
4 12 0.0
5 12 0.0
6 12 0.059
7 12 0.0
8 12 0.0
9 12 0.0
10 12 0.029
11 12 0.0
12 12 0.706
13 12 0.0
14 12 0.088
15 12 0.0
16 12 0.0
17 12 0.0
18 12 0.0
19 12 0.118
20 12 0.0
0 13 0.0
1 13 0.029
2 13 0.029
3 13 0.0
4 13 0.0
5 13 0.0
6 13 0.0
7 13 0.029
8 13 0.0
9 13 0.0
10 13 0.029
11 13 0.0
12 13 0.0
13 13 0.029
14 13 0.086
15 13 0.0
16 13 0.086
17 13 0.143
18 13 0.229
19 13 0.314
20 13 0.0
0 14 0.0
1 14 0.013
2 14 0.0
3 14 0.013
4 14 0.0
5 14 0.0
6 14 0.038
7 14 0.0
8 14 0.0
9 14 0.0
10 14 0.025
11 14 0.0
12 14 0.0
13 14 0.025
14 14 0.278
15 14 0.013
16 14 0.291
17 14 0.114
18 14 0.063
19 14 0.038
20 14 0.089
0 15 0.0
1 15 0.0
2 15 0.0
3 15 0.017
4 15 0.0
5 15 0.0
6 15 0.0
7 15 0.0
8 15 0.0
9 15 0.0
10 15 0.0
11 15 0.0
12 15 0.0
13 15 0.0
14 15 0.017
15 15 0.362
16 15 0.293
17 15 0.241
18 15 0.017
19 15 0.034
20 15 0.017
0 16 0.0
1 16 0.006
2 16 0.006
3 16 0.006
4 16 0.0
5 16 0.006
6 16 0.006
7 16 0.006
8 16 0.0
9 16 0.0
10 16 0.006
11 16 0.017
12 16 0.006
13 16 0.0
14 16 0.045
15 16 0.146
16 16 0.393
17 16 0.326
18 16 0.017
19 16 0.011
20 16 0.0
0 17 0.0
1 17 0.0
2 17 0.0
3 17 0.0
4 17 0.0
5 17 0.0
6 17 0.0
7 17 0.005
8 17 0.0
9 17 0.0
10 17 0.011
11 17 0.0
12 17 0.0
13 17 0.0
14 17 0.0
15 17 0.042
16 17 0.021
17 17 0.916
18 17 0.0
19 17 0.005
20 17 0.0
0 18 0.0
1 18 0.0
2 18 0.006
3 18 0.006
4 18 0.0
5 18 0.006
6 18 0.018
7 18 0.012
8 18 0.0
9 18 0.0
10 18 0.0
11 18 0.006
12 18 0.0
13 18 0.0
14 18 0.0
15 18 0.0
16 18 0.0
17 18 0.0
18 18 0.94
19 18 0.006
20 18 0.0
0 19 0.0
1 19 0.0
2 19 0.006
3 19 0.029
4 19 0.0
5 19 0.006
6 19 0.024
7 19 0.024
8 19 0.0
9 19 0.0
10 19 0.0
11 19 0.018
12 19 0.0
13 19 0.006
14 19 0.006
15 19 0.0
16 19 0.0
17 19 0.012
18 19 0.006
19 19 0.865
20 19 0.0
0 20 0.0
1 20 0.0
2 20 0.0
3 20 0.036
4 20 0.0
5 20 0.018
6 20 0.018
7 20 0.0
8 20 0.0
9 20 0.0
10 20 0.0
11 20 0.145
12 20 0.018
13 20 0.0
14 20 0.127
15 20 0.0
16 20 0.018
17 20 0.109
18 20 0.073
19 20 0.091
20 20 0.345
        };
    \end{axis}
\end{tikzpicture}
%
%\caption{I'm confused~5!}\label{tab:CM5}
%\end{figure*}
	}
\\
(c) Random Forest
& & 
(d) LSTM
	\end{tabular}
	\caption{Confusion matrices for standard techniques}
	\label{fig:conf_2}
\end{figure}

%
% Bagging
%

\begin{figure}[!htb]
	\centering
	\begin{tabular}{ccc}
	\resizebox{0.45\textwidth}{0.25\textheight}{%
	%\begin{tikzpicture}[scale=0.60,every node/.style={scale=0.9}]
%\begin{tikzpicture}[scale=0.55]
\begin{tikzpicture}[scale=0.60]
    \begin{axis}[%colorbar/width=2.5mm,
        width=20cm,
        height=20cm,
%        colormap={blackwhite}{gray(0cm)=(1); gray(1cm)=(0.5)},
%   colormap={bluewhite}{color=(white) color=(blue)},
%   colormap={bluewhite}{color=(white) rgb255=(0,191,255)},
    colormap={bluewhite}{color=(white) rgb255=(100,149,237)},
        xticklabels={Adload,Agent,Allaple,BHO,Bifrose,CeeInject,Cycbot,FakeRean,Hotbar,Injector,
        OnLineGames,Renos,Rimecud,Small,Toga,VB,VBinject,Vobfus,Vundo,Winwebsec,Zbot},
        xtick={0,...,20},
        xtick style={draw=none},
    xticklabel style={anchor=east,rotate=45,yshift=-5pt},
        yticklabels={Adload,Agent,Allaple,BHO,Bifrose,CeeInject,Cycbot,FakeRean,Hotbar,Injector,
        OnLineGames,Renos,Rimecud,Small,Toga,VB,VBinject,Vobfus,Vundo,Winwebsec,Zbot},
        ytick={0,...,20},
        ytick style={draw=none},
        enlargelimits=false,
        xticklabel style={font=\large},
        yticklabel style={font=\large},
        colorbar,
        colorbar style={
%           width=0.05*\pgfkeysvalueof{/pgfplots/parent axis width},%%% added this
%           height=0.5*\pgfkeysvalueof{/pgfplots/parent axis height},
%       plot graphics/node/.style={scale=1.33,anchor=south west,inner sep=0pt,}, %%% scale colorbar fill %%%
            ytick={0.0,0.2,0.4,0.6,0.8,1.0},
            yticklabels={0.0,0.2,0.4,0.6,0.8,1.0},
            yticklabel={\pgfmathprintnumber\tick},
            yticklabel style={font=\large,
                    /pgf/number format/fixed,
            /pgf/number format/precision=1}
        },
%        point meta min=0,
%        point meta max=100,
        point meta min=0.0,
        point meta max=1.0,
        nodes near coords={\pgfmathprintnumber\pgfplotspointmeta},
        % ---------------------------------------------------------------------
        % show `nodes near coords' but adapt the style so that values
        % above a threshold get another style
        % (adapted from <http://tex.stackexchange.com/a/141006/95441>)
        % #1: the THRESHOLD after which we switch to a special display.
        nodes near coords black white/.style={
            % define the style of the nodes with "small" values
            small value/.style={
                yshift=-7pt,
%                text=white,
                text=black,
                /pgf/number format/fixed,
                /pgf/number format/precision=3,
                /pgf/number format/zerofill
            },
            % define the style of the nodes with "large" values
            large value/.style={
                yshift=-7pt,
%                text=black,
                text=white,
                /pgf/number format/fixed,
                /pgf/number format/precision=3,
                /pgf/number format/zerofill
            },
            every node near coord/.style={
                check for zero/.code={
                    \pgfmathfloatifflags{\pgfplotspointmeta}{0}{
                        % If meta=0, make the node a coordinate
                        % (which doesn't have text)
                        \pgfkeys{/tikz/coordinate}
                    }{
                        \begingroup
                        % this group is merely to switch to FPU locally.
                        % Might be unnecessary, but who knows.
                        \pgfkeys{/pgf/fpu}
                        \pgfmathparse{\pgfplotspointmeta<#1}
                        \global\let\result=\pgfmathresult
                        \endgroup
                        %
                        % simplifies debugging:
                        %\show\result
                        %
                        \pgfmathfloatcreate{1}{1.0}{0}
                        \let\ONE=\pgfmathresult
                        \ifx\result\ONE
                            % AH: our condition 'y < #1' is met.
                            \pgfkeysalso{/pgfplots/small value}
                        \else
                            % ok, proceed as usual.
                            \pgfkeysalso{/pgfplots/large value}
                        \fi
                    }
                },
                check for zero,
            },
        },
        % asign a value to the new style which is the threshold at which
        % the two style `small value' or `large value' are used
%        nodes near coords black white=50,
        nodes near coords black white=0.5,
        % -----------------------------------------------------------------
    ]
        \addplot[
            matrix plot,
            mesh/cols=21,
            point meta=explicit,draw=gray
        ] table [meta=C] {
            x y C
0 0 0.146
1 0 0.0
2 0 0.0
3 0 0.829
4 0 0.0
5 0 0.0
6 0 0.0
7 0 0.0
8 0 0.0
9 0 0.0
10 0 0.0
11 0 0.0
12 0 0.0
13 0 0.0
14 0 0.024
15 0 0.0
16 0 0.0
17 0 0.0
18 0 0.0
19 0 0.0
20 0 0.0
0 1 0.022
1 1 0.304
2 1 0.0
3 1 0.0
4 1 0.0
5 1 0.022
6 1 0.0
7 1 0.0
8 1 0.022
9 1 0.065
10 1 0.13
11 1 0.0
12 1 0.022
13 1 0.043
14 1 0.022
15 1 0.022
16 1 0.0
17 1 0.152
18 1 0.152
19 1 0.022
20 1 0.0
0 2 0.0
1 2 0.0
2 2 1.0
3 2 0.0
4 2 0.0
5 2 0.0
6 2 0.0
7 2 0.0
8 2 0.0
9 2 0.0
10 2 0.0
11 2 0.0
12 2 0.0
13 2 0.0
14 2 0.0
15 2 0.0
16 2 0.0
17 2 0.0
18 2 0.0
19 2 0.0
20 2 0.0
0 3 0.06
1 3 0.012
2 3 0.0
3 3 0.831
4 3 0.0
5 3 0.0
6 3 0.0
7 3 0.0
8 3 0.0
9 3 0.0
10 3 0.024
11 3 0.0
12 3 0.0
13 3 0.0
14 3 0.0
15 3 0.0
16 3 0.0
17 3 0.0
18 3 0.072
19 3 0.0
20 3 0.0
0 4 0.0
1 4 0.0
2 4 0.0
3 4 0.0
4 4 0.41
5 4 0.051
6 4 0.0
7 4 0.051
8 4 0.0
9 4 0.026
10 4 0.103
11 4 0.0
12 4 0.051
13 4 0.026
14 4 0.051
15 4 0.0
16 4 0.128
17 4 0.051
18 4 0.051
19 4 0.0
20 4 0.0
0 5 0.0
1 5 0.05
2 5 0.0
3 5 0.009
4 5 0.064
5 5 0.764
6 5 0.0
7 5 0.005
8 5 0.0
9 5 0.032
10 5 0.018
11 5 0.0
12 5 0.0
13 5 0.005
14 5 0.014
15 5 0.0
16 5 0.005
17 5 0.0
18 5 0.023
19 5 0.005
20 5 0.009
0 6 0.0
1 6 0.0
2 6 0.0
3 6 0.0
4 6 0.0
5 6 0.0
6 6 0.872
7 6 0.034
8 6 0.0
9 6 0.0
10 6 0.0
11 6 0.0
12 6 0.0
13 6 0.0
14 6 0.0
15 6 0.0
16 6 0.0
17 6 0.0
18 6 0.007
19 6 0.087
20 6 0.0
0 7 0.0
1 7 0.007
2 7 0.0
3 7 0.0
4 7 0.0
5 7 0.0
6 7 0.0
7 7 0.777
8 7 0.0
9 7 0.007
10 7 0.014
11 7 0.0
12 7 0.007
13 7 0.029
14 7 0.029
15 7 0.036
16 7 0.0
17 7 0.0
18 7 0.036
19 7 0.05
20 7 0.007
0 8 0.0
1 8 0.0
2 8 0.0
3 8 0.0
4 8 0.0
5 8 0.0
6 8 0.0
7 8 0.0
8 8 0.969
9 8 0.0
10 8 0.0
11 8 0.0
12 8 0.0
13 8 0.031
14 8 0.0
15 8 0.0
16 8 0.0
17 8 0.0
18 8 0.0
19 8 0.0
20 8 0.0
0 9 0.0
1 9 0.15
2 9 0.0
3 9 0.1
4 9 0.025
5 9 0.075
6 9 0.0
7 9 0.025
8 9 0.0
9 9 0.225
10 9 0.0
11 9 0.025
12 9 0.025
13 9 0.025
14 9 0.05
15 9 0.05
16 9 0.075
17 9 0.025
18 9 0.025
19 9 0.0
20 9 0.1
0 10 0.0
1 10 0.111
2 10 0.0
3 10 0.056
4 10 0.0
5 10 0.0
6 10 0.0
7 10 0.0
8 10 0.0
9 10 0.0
10 10 0.759
11 10 0.0
12 10 0.0
13 10 0.019
14 10 0.0
15 10 0.0
16 10 0.037
17 10 0.0
18 10 0.0
19 10 0.0
20 10 0.019
0 11 0.0
1 11 0.008
2 11 0.0
3 11 0.0
4 11 0.0
5 11 0.0
6 11 0.0
7 11 0.008
8 11 0.0
9 11 0.0
10 11 0.008
11 11 0.886
12 11 0.0
13 11 0.015
14 11 0.0
15 11 0.0
16 11 0.008
17 11 0.0
18 11 0.053
19 11 0.0
20 11 0.015
0 12 0.0
1 12 0.0
2 12 0.0
3 12 0.0
4 12 0.0
5 12 0.0
6 12 0.0
7 12 0.0
8 12 0.0
9 12 0.0
10 12 0.0
11 12 0.0
12 12 0.872
13 12 0.026
14 12 0.0
15 12 0.0
16 12 0.026
17 12 0.0
18 12 0.026
19 12 0.0
20 12 0.051
0 13 0.0
1 13 0.152
2 13 0.0
3 13 0.043
4 13 0.043
5 13 0.0
6 13 0.0
7 13 0.043
8 13 0.0
9 13 0.043
10 13 0.065
11 13 0.0
12 13 0.0
13 13 0.348
14 13 0.065
15 13 0.087
16 13 0.0
17 13 0.0
18 13 0.065
19 13 0.0
20 13 0.043
0 14 0.0
1 14 0.0
2 14 0.0
3 14 0.01
4 14 0.139
5 14 0.0
6 14 0.0
7 14 0.03
8 14 0.0
9 14 0.02
10 14 0.03
11 14 0.0
12 14 0.02
13 14 0.03
14 14 0.495
15 14 0.099
16 14 0.03
17 14 0.02
18 14 0.05
19 14 0.01
20 14 0.02
0 15 0.0
1 15 0.0
2 15 0.0
3 15 0.023
4 15 0.0
5 15 0.0
6 15 0.0
7 15 0.0
8 15 0.0
9 15 0.011
10 15 0.011
11 15 0.0
12 15 0.0
13 15 0.034
14 15 0.023
15 15 0.586
16 15 0.069
17 15 0.23
18 15 0.0
19 15 0.0
20 15 0.011
0 16 0.009
1 16 0.026
2 16 0.0
3 16 0.0
4 16 0.14
5 16 0.0
6 16 0.0
7 16 0.017
8 16 0.0
9 16 0.081
10 16 0.013
11 16 0.0
12 16 0.0
13 16 0.051
14 16 0.06
15 16 0.123
16 16 0.285
17 16 0.132
18 16 0.047
19 16 0.0
20 16 0.017
0 17 0.0
1 17 0.013
2 17 0.0
3 17 0.0
4 17 0.0
5 17 0.0
6 17 0.0
7 17 0.0
8 17 0.0
9 17 0.0
10 17 0.0
11 17 0.0
12 17 0.0
13 17 0.0
14 17 0.0
15 17 0.021
16 17 0.004
17 17 0.957
18 17 0.0
19 17 0.0
20 17 0.004
0 18 0.0
1 18 0.026
2 18 0.0
3 18 0.01
4 18 0.005
5 18 0.005
6 18 0.0
7 18 0.01
8 18 0.0
9 18 0.0
10 18 0.01
11 18 0.005
12 18 0.01
13 18 0.01
14 18 0.0
15 18 0.0
16 18 0.0
17 18 0.0
18 18 0.869
19 18 0.01
20 18 0.026
0 19 0.024
1 19 0.005
2 19 0.0
3 19 0.0
4 19 0.005
5 19 0.0
6 19 0.0
7 19 0.119
8 19 0.0
9 19 0.095
10 19 0.01
11 19 0.005
12 19 0.0
13 19 0.005
14 19 0.01
15 19 0.0
16 19 0.0
17 19 0.0
18 19 0.01
19 19 0.7
20 19 0.014
0 20 0.0
1 20 0.053
2 20 0.0
3 20 0.013
4 20 0.0
5 20 0.013
6 20 0.0
7 20 0.026
8 20 0.0
9 20 0.026
10 20 0.013
11 20 0.013
12 20 0.039
13 20 0.066
14 20 0.092
15 20 0.013
16 20 0.013
17 20 0.013
18 20 0.053
19 20 0.039
20 20 0.513
        };
    \end{axis}
\end{tikzpicture}
%
%\caption{I'm confused~5!}\label{tab:CM5}
%\end{figure*}
	}
& &
	\resizebox{0.45\textwidth}{0.25\textheight}{%
	%\begin{tikzpicture}[scale=0.60,every node/.style={scale=0.9}]
%\begin{tikzpicture}[scale=0.55]
\begin{tikzpicture}[scale=0.60]
    \begin{axis}[%colorbar/width=2.5mm,
        width=20cm,
        height=20cm,
%        colormap={blackwhite}{gray(0cm)=(1); gray(1cm)=(0.5)},
%   colormap={bluewhite}{color=(white) color=(blue)},
%   colormap={bluewhite}{color=(white) rgb255=(0,191,255)},
    colormap={bluewhite}{color=(white) rgb255=(100,149,237)},
        xticklabels={Adload,Agent,Allaple,BHO,Bifrose,CeeInject,Cycbot,FakeRean,Hotbar,Injector,
        OnLineGames,Renos,Rimecud,Small,Toga,VB,VBinject,Vobfus,Vundo,Winwebsec,Zbot},
        xtick={0,...,20},
        xtick style={draw=none},
    xticklabel style={anchor=east,rotate=45,yshift=-5pt},
        yticklabels={Adload,Agent,Allaple,BHO,Bifrose,CeeInject,Cycbot,FakeRean,Hotbar,Injector,
        OnLineGames,Renos,Rimecud,Small,Toga,VB,VBinject,Vobfus,Vundo,Winwebsec,Zbot},
        ytick={0,...,20},
        ytick style={draw=none},
        enlargelimits=false,
        xticklabel style={font=\large},
        yticklabel style={font=\large},
        colorbar,
        colorbar style={
%           width=0.05*\pgfkeysvalueof{/pgfplots/parent axis width},%%% added this
%           height=0.5*\pgfkeysvalueof{/pgfplots/parent axis height},
%       plot graphics/node/.style={scale=1.33,anchor=south west,inner sep=0pt,}, %%% scale colorbar fill %%%
            ytick={0.0,0.2,0.4,0.6,0.8,1.0},
            yticklabels={0.0,0.2,0.4,0.6,0.8,1.0},
            yticklabel={\pgfmathprintnumber\tick},
            yticklabel style={font=\large,
                    /pgf/number format/fixed,
            /pgf/number format/precision=1}
        },
%        point meta min=0,
%        point meta max=100,
        point meta min=0.0,
        point meta max=1.0,
        nodes near coords={\pgfmathprintnumber\pgfplotspointmeta},
        % ---------------------------------------------------------------------
        % show `nodes near coords' but adapt the style so that values
        % above a threshold get another style
        % (adapted from <http://tex.stackexchange.com/a/141006/95441>)
        % #1: the THRESHOLD after which we switch to a special display.
        nodes near coords black white/.style={
            % define the style of the nodes with "small" values
            small value/.style={
                yshift=-7pt,
%                text=white,
                text=black,
                /pgf/number format/fixed,
                /pgf/number format/precision=3,
                /pgf/number format/zerofill
            },
            % define the style of the nodes with "large" values
            large value/.style={
                yshift=-7pt,
%                text=black,
                text=white,
                /pgf/number format/fixed,
                /pgf/number format/precision=3,
                /pgf/number format/zerofill
            },
            every node near coord/.style={
                check for zero/.code={
                    \pgfmathfloatifflags{\pgfplotspointmeta}{0}{
                        % If meta=0, make the node a coordinate
                        % (which doesn't have text)
                        \pgfkeys{/tikz/coordinate}
                    }{
                        \begingroup
                        % this group is merely to switch to FPU locally.
                        % Might be unnecessary, but who knows.
                        \pgfkeys{/pgf/fpu}
                        \pgfmathparse{\pgfplotspointmeta<#1}
                        \global\let\result=\pgfmathresult
                        \endgroup
                        %
                        % simplifies debugging:
                        %\show\result
                        %
                        \pgfmathfloatcreate{1}{1.0}{0}
                        \let\ONE=\pgfmathresult
                        \ifx\result\ONE
                            % AH: our condition 'y < #1' is met.
                            \pgfkeysalso{/pgfplots/small value}
                        \else
                            % ok, proceed as usual.
                            \pgfkeysalso{/pgfplots/large value}
                        \fi
                    }
                },
                check for zero,
            },
        },
        % asign a value to the new style which is the threshold at which
        % the two style `small value' or `large value' are used
%        nodes near coords black white=50,
        nodes near coords black white=0.5,
        % -----------------------------------------------------------------
    ]
        \addplot[
            matrix plot,
            mesh/cols=21,
            point meta=explicit,draw=gray
        ] table [meta=C] {
            x y C
0 0 0.952
1 0 0.0
2 0 0.0
3 0 0.024
4 0 0.0
5 0 0.0
6 0 0.0
7 0 0.0
8 0 0.0
9 0 0.0
10 0 0.0
11 0 0.0
12 0 0.0
13 0 0.0
14 0 0.0
15 0 0.0
16 0 0.024
17 0 0.0
18 0 0.0
19 0 0.0
20 0 0.0
0 1 0.0
1 1 0.629
2 1 0.0
3 1 0.0
4 1 0.0
5 1 0.0
6 1 0.0
7 1 0.086
8 1 0.0
9 1 0.0
10 1 0.029
11 1 0.057
12 1 0.0
13 1 0.057
14 1 0.0
15 1 0.0
16 1 0.086
17 1 0.0
18 1 0.057
19 1 0.0
20 1 0.0
0 2 0.0
1 2 0.0
2 2 1.0
3 2 0.0
4 2 0.0
5 2 0.0
6 2 0.0
7 2 0.0
8 2 0.0
9 2 0.0
10 2 0.0
11 2 0.0
12 2 0.0
13 2 0.0
14 2 0.0
15 2 0.0
16 2 0.0
17 2 0.0
18 2 0.0
19 2 0.0
20 2 0.0
0 3 0.042
1 3 0.0
2 3 0.0
3 3 0.915
4 3 0.0
5 3 0.0
6 3 0.0
7 3 0.0
8 3 0.0
9 3 0.0
10 3 0.0
11 3 0.0
12 3 0.0
13 3 0.0
14 3 0.0
15 3 0.0
16 3 0.014
17 3 0.0
18 3 0.014
19 3 0.014
20 3 0.0
0 4 0.0
1 4 0.053
2 4 0.0
3 4 0.0
4 4 0.132
5 4 0.184
6 4 0.026
7 4 0.026
8 4 0.0
9 4 0.0
10 4 0.0
11 4 0.053
12 4 0.0
13 4 0.0
14 4 0.079
15 4 0.0
16 4 0.421
17 4 0.0
18 4 0.0
19 4 0.026
20 4 0.0
0 5 0.0
1 5 0.0
2 5 0.0
3 5 0.0
4 5 0.0
5 5 0.924
6 5 0.0
7 5 0.0
8 5 0.0
9 5 0.0
10 5 0.0
11 5 0.019
12 5 0.0
13 5 0.0
14 5 0.006
15 5 0.0
16 5 0.019
17 5 0.0
18 5 0.006
19 5 0.019
20 5 0.006
0 6 0.0
1 6 0.0
2 6 0.0
3 6 0.0
4 6 0.0
5 6 0.0
6 6 0.992
7 6 0.0
8 6 0.0
9 6 0.0
10 6 0.0
11 6 0.0
12 6 0.0
13 6 0.0
14 6 0.0
15 6 0.0
16 6 0.008
17 6 0.0
18 6 0.0
19 6 0.0
20 6 0.0
0 7 0.0
1 7 0.0
2 7 0.0
3 7 0.0
4 7 0.0
5 7 0.0
6 7 0.0
7 7 0.958
8 7 0.0
9 7 0.0
10 7 0.011
11 7 0.0
12 7 0.0
13 7 0.0
14 7 0.011
15 7 0.0
16 7 0.0
17 7 0.0
18 7 0.0
19 7 0.021
20 7 0.0
0 8 0.0
1 8 0.0
2 8 0.0
3 8 0.0
4 8 0.0
5 8 0.0
6 8 0.0
7 8 0.038
8 8 0.962
9 8 0.0
10 8 0.0
11 8 0.0
12 8 0.0
13 8 0.0
14 8 0.0
15 8 0.0
16 8 0.0
17 8 0.0
18 8 0.0
19 8 0.0
20 8 0.0
0 9 0.0
1 9 0.0
2 9 0.0
3 9 0.0
4 9 0.0
5 9 0.065
6 9 0.0
7 9 0.032
8 9 0.0
9 9 0.355
10 9 0.032
11 9 0.065
12 9 0.0
13 9 0.0
14 9 0.032
15 9 0.0
16 9 0.29
17 9 0.0
18 9 0.097
19 9 0.032
20 9 0.0
0 10 0.0
1 10 0.021
2 10 0.0
3 10 0.0
4 10 0.0
5 10 0.0
6 10 0.0
7 10 0.021
8 10 0.0
9 10 0.0
10 10 0.915
11 10 0.0
12 10 0.0
13 10 0.0
14 10 0.0
15 10 0.0
16 10 0.0
17 10 0.0
18 10 0.021
19 10 0.021
20 10 0.0
0 11 0.0
1 11 0.0
2 11 0.0
3 11 0.01
4 11 0.0
5 11 0.0
6 11 0.01
7 11 0.019
8 11 0.0
9 11 0.0
10 11 0.0
11 11 0.943
12 11 0.0
13 11 0.0
14 11 0.01
15 11 0.0
16 11 0.01
17 11 0.0
18 11 0.0
19 11 0.0
20 11 0.0
0 12 0.0
1 12 0.0
2 12 0.0
3 12 0.0
4 12 0.0
5 12 0.0
6 12 0.0
7 12 0.118
8 12 0.0
9 12 0.0
10 12 0.0
11 12 0.0
12 12 0.824
13 12 0.0
14 12 0.029
15 12 0.0
16 12 0.0
17 12 0.0
18 12 0.0
19 12 0.029
20 12 0.0
0 13 0.0
1 13 0.086
2 13 0.0
3 13 0.0
4 13 0.0
5 13 0.0
6 13 0.0
7 13 0.029
8 13 0.0
9 13 0.0
10 13 0.0
11 13 0.0
12 13 0.0
13 13 0.571
14 13 0.0
15 13 0.0
16 13 0.143
17 13 0.0
18 13 0.057
19 13 0.029
20 13 0.086
0 14 0.0
1 14 0.0
2 14 0.0
3 14 0.0
4 14 0.0
5 14 0.013
6 14 0.0
7 14 0.025
8 14 0.0
9 14 0.0
10 14 0.0
11 14 0.013
12 14 0.0
13 14 0.0
14 14 0.747
15 14 0.0
16 14 0.19
17 14 0.0
18 14 0.0
19 14 0.0
20 14 0.013
0 15 0.0
1 15 0.017
2 15 0.0
3 15 0.0
4 15 0.0
5 15 0.0
6 15 0.0
7 15 0.034
8 15 0.0
9 15 0.0
10 15 0.0
11 15 0.017
12 15 0.0
13 15 0.0
14 15 0.0
15 15 0.638
16 15 0.259
17 15 0.0
18 15 0.0
19 15 0.034
20 15 0.0
0 16 0.0
1 16 0.0
2 16 0.0
3 16 0.0
4 16 0.0
5 16 0.0
6 16 0.0
7 16 0.006
8 16 0.0
9 16 0.0
10 16 0.0
11 16 0.0
12 16 0.0
13 16 0.0
14 16 0.011
15 16 0.006
16 16 0.938
17 16 0.0
18 16 0.006
19 16 0.022
20 16 0.011
0 17 0.0
1 17 0.0
2 17 0.0
3 17 0.0
4 17 0.0
5 17 0.0
6 17 0.005
7 17 0.0
8 17 0.0
9 17 0.0
10 17 0.0
11 17 0.0
12 17 0.0
13 17 0.005
14 17 0.005
15 17 0.0
16 17 0.026
17 17 0.958
18 17 0.0
19 17 0.0
20 17 0.0
0 18 0.0
1 18 0.0
2 18 0.0
3 18 0.006
4 18 0.0
5 18 0.0
6 18 0.0
7 18 0.036
8 18 0.0
9 18 0.006
10 18 0.0
11 18 0.0
12 18 0.0
13 18 0.0
14 18 0.0
15 18 0.0
16 18 0.006
17 18 0.0
18 18 0.922
19 18 0.018
20 18 0.006
0 19 0.0
1 19 0.0
2 19 0.0
3 19 0.0
4 19 0.0
5 19 0.006
6 19 0.0
7 19 0.0
8 19 0.0
9 19 0.0
10 19 0.0
11 19 0.006
12 19 0.0
13 19 0.0
14 19 0.006
15 19 0.0
16 19 0.0
17 19 0.0
18 19 0.0
19 19 0.982
20 19 0.0
0 20 0.0
1 20 0.0
2 20 0.0
3 20 0.0
4 20 0.0
5 20 0.018
6 20 0.0
7 20 0.109
8 20 0.0
9 20 0.0
10 20 0.0
11 20 0.036
12 20 0.0
13 20 0.0
14 20 0.036
15 20 0.0
16 20 0.036
17 20 0.0
18 20 0.0
19 20 0.0
20 20 0.764
        };
    \end{axis}
\end{tikzpicture}
%
%\caption{I'm confused~5!}\label{tab:CM5}
%\end{figure*}
	}
\\
(a) Bagged HMM
& & 
(b) Bagged CNN
\\ \\
\multicolumn{3}{c}{
	\resizebox{0.45\textwidth}{0.25\textheight}{%
	%\begin{tikzpicture}[scale=0.60,every node/.style={scale=0.9}]
%\begin{tikzpicture}[scale=0.55]
\begin{tikzpicture}[scale=0.60]
    \begin{axis}[%colorbar/width=2.5mm,
        width=20cm,
        height=20cm,
%        colormap={blackwhite}{gray(0cm)=(1); gray(1cm)=(0.5)},
%   colormap={bluewhite}{color=(white) color=(blue)},
%   colormap={bluewhite}{color=(white) rgb255=(0,191,255)},
    colormap={bluewhite}{color=(white) rgb255=(100,149,237)},
        xticklabels={Adload,Agent,Allaple,BHO,Bifrose,CeeInject,Cycbot,FakeRean,Hotbar,Injector,
        OnLineGames,Renos,Rimecud,Small,Toga,VB,VBinject,Vobfus,Vundo,Winwebsec,Zbot},
        xtick={0,...,20},
        xtick style={draw=none},
    xticklabel style={anchor=east,rotate=45,yshift=-5pt},
        yticklabels={Adload,Agent,Allaple,BHO,Bifrose,CeeInject,Cycbot,FakeRean,Hotbar,Injector,
        OnLineGames,Renos,Rimecud,Small,Toga,VB,VBinject,Vobfus,Vundo,Winwebsec,Zbot},
        ytick={0,...,20},
        ytick style={draw=none},
        enlargelimits=false,
        xticklabel style={font=\large},
        yticklabel style={font=\large},
        colorbar,
        colorbar style={
%           width=0.05*\pgfkeysvalueof{/pgfplots/parent axis width},%%% added this
%           height=0.5*\pgfkeysvalueof{/pgfplots/parent axis height},
%       plot graphics/node/.style={scale=1.33,anchor=south west,inner sep=0pt,}, %%% scale colorbar fill %%%
            ytick={0.0,0.2,0.4,0.6,0.8,1.0},
            yticklabels={0.0,0.2,0.4,0.6,0.8,1.0},
            yticklabel={\pgfmathprintnumber\tick},
            yticklabel style={font=\large,
                    /pgf/number format/fixed,
            /pgf/number format/precision=1}
        },
%        point meta min=0,
%        point meta max=100,
        point meta min=0.0,
        point meta max=1.0,
        nodes near coords={\pgfmathprintnumber\pgfplotspointmeta},
        % ---------------------------------------------------------------------
        % show `nodes near coords' but adapt the style so that values
        % above a threshold get another style
        % (adapted from <http://tex.stackexchange.com/a/141006/95441>)
        % #1: the THRESHOLD after which we switch to a special display.
        nodes near coords black white/.style={
            % define the style of the nodes with "small" values
            small value/.style={
                yshift=-7pt,
%                text=white,
                text=black,
                /pgf/number format/fixed,
                /pgf/number format/precision=3,
                /pgf/number format/zerofill
            },
            % define the style of the nodes with "large" values
            large value/.style={
                yshift=-7pt,
%                text=black,
                text=white,
                /pgf/number format/fixed,
                /pgf/number format/precision=3,
                /pgf/number format/zerofill
            },
            every node near coord/.style={
                check for zero/.code={
                    \pgfmathfloatifflags{\pgfplotspointmeta}{0}{
                        % If meta=0, make the node a coordinate
                        % (which doesn't have text)
                        \pgfkeys{/tikz/coordinate}
                    }{
                        \begingroup
                        % this group is merely to switch to FPU locally.
                        % Might be unnecessary, but who knows.
                        \pgfkeys{/pgf/fpu}
                        \pgfmathparse{\pgfplotspointmeta<#1}
                        \global\let\result=\pgfmathresult
                        \endgroup
                        %
                        % simplifies debugging:
                        %\show\result
                        %
                        \pgfmathfloatcreate{1}{1.0}{0}
                        \let\ONE=\pgfmathresult
                        \ifx\result\ONE
                            % AH: our condition 'y < #1' is met.
                            \pgfkeysalso{/pgfplots/small value}
                        \else
                            % ok, proceed as usual.
                            \pgfkeysalso{/pgfplots/large value}
                        \fi
                    }
                },
                check for zero,
            },
        },
        % asign a value to the new style which is the threshold at which
        % the two style `small value' or `large value' are used
%        nodes near coords black white=50,
        nodes near coords black white=0.5,
        % -----------------------------------------------------------------
    ]
        \addplot[
            matrix plot,
            mesh/cols=21,
            point meta=explicit,draw=gray
        ] table [meta=C] {
            x y C
0 0 0.833
1 0 0.0
2 0 0.0
3 0 0.119
4 0 0.0
5 0 0.0
6 0 0.0
7 0 0.0
8 0 0.0
9 0 0.0
10 0 0.0
11 0 0.024
12 0 0.0
13 0 0.0
14 0 0.024
15 0 0.0
16 0 0.0
17 0 0.0
18 0 0.0
19 0 0.0
20 0 0.0
0 1 0.0
1 1 0.514
2 1 0.0
3 1 0.029
4 1 0.0
5 1 0.086
6 1 0.0
7 1 0.0
8 1 0.0
9 1 0.0
10 1 0.086
11 1 0.086
12 1 0.0
13 1 0.057
14 1 0.029
15 1 0.029
16 1 0.029
17 1 0.0
18 1 0.029
19 1 0.029
20 1 0.0
0 2 0.0
1 2 0.0
2 2 1.0
3 2 0.0
4 2 0.0
5 2 0.0
6 2 0.0
7 2 0.0
8 2 0.0
9 2 0.0
10 2 0.0
11 2 0.0
12 2 0.0
13 2 0.0
14 2 0.0
15 2 0.0
16 2 0.0
17 2 0.0
18 2 0.0
19 2 0.0
20 2 0.0
0 3 0.0
1 3 0.0
2 3 0.0
3 3 0.944
4 3 0.0
5 3 0.0
6 3 0.0
7 3 0.0
8 3 0.0
9 3 0.0
10 3 0.0
11 3 0.0
12 3 0.0
13 3 0.014
14 3 0.0
15 3 0.0
16 3 0.014
17 3 0.0
18 3 0.028
19 3 0.0
20 3 0.0
0 4 0.0
1 4 0.026
2 4 0.0
3 4 0.0
4 4 0.158
5 4 0.158
6 4 0.026
7 4 0.026
8 4 0.0
9 4 0.026
10 4 0.0
11 4 0.053
12 4 0.0
13 4 0.0
14 4 0.053
15 4 0.026
16 4 0.368
17 4 0.0
18 4 0.053
19 4 0.026
20 4 0.0
0 5 0.0
1 5 0.006
2 5 0.0
3 5 0.013
4 5 0.0
5 5 0.918
6 5 0.0
7 5 0.0
8 5 0.0
9 5 0.006
10 5 0.006
11 5 0.019
12 5 0.0
13 5 0.0
14 5 0.0
15 5 0.0
16 5 0.0
17 5 0.0
18 5 0.006
19 5 0.025
20 5 0.0
0 6 0.0
1 6 0.0
2 6 0.0
3 6 0.0
4 6 0.0
5 6 0.0
6 6 1.0
7 6 0.0
8 6 0.0
9 6 0.0
10 6 0.0
11 6 0.0
12 6 0.0
13 6 0.0
14 6 0.0
15 6 0.0
16 6 0.0
17 6 0.0
18 6 0.0
19 6 0.0
20 6 0.0
0 7 0.0
1 7 0.0
2 7 0.0
3 7 0.0
4 7 0.0
5 7 0.0
6 7 0.011
7 7 0.916
8 7 0.0
9 7 0.0
10 7 0.0
11 7 0.011
12 7 0.011
13 7 0.0
14 7 0.021
15 7 0.0
16 7 0.0
17 7 0.0
18 7 0.011
19 7 0.021
20 7 0.0
0 8 0.0
1 8 0.0
2 8 0.0
3 8 0.0
4 8 0.0
5 8 0.0
6 8 0.0
7 8 0.0
8 8 0.962
9 8 0.0
10 8 0.0
11 8 0.038
12 8 0.0
13 8 0.0
14 8 0.0
15 8 0.0
16 8 0.0
17 8 0.0
18 8 0.0
19 8 0.0
20 8 0.0
0 9 0.0
1 9 0.032
2 9 0.0
3 9 0.0
4 9 0.0
5 9 0.29
6 9 0.0
7 9 0.0
8 9 0.0
9 9 0.161
10 9 0.032
11 9 0.161
12 9 0.0
13 9 0.032
14 9 0.0
15 9 0.032
16 9 0.097
17 9 0.032
18 9 0.032
19 9 0.065
20 9 0.032
0 10 0.0
1 10 0.064
2 10 0.0
3 10 0.0
4 10 0.0
5 10 0.0
6 10 0.0
7 10 0.0
8 10 0.0
9 10 0.021
10 10 0.809
11 10 0.021
12 10 0.0
13 10 0.0
14 10 0.0
15 10 0.0
16 10 0.0
17 10 0.0
18 10 0.064
19 10 0.021
20 10 0.0
0 11 0.0
1 11 0.0
2 11 0.0
3 11 0.01
4 11 0.0
5 11 0.0
6 11 0.0
7 11 0.0
8 11 0.0
9 11 0.0
10 11 0.0
11 11 0.933
12 11 0.0
13 11 0.0
14 11 0.029
15 11 0.0
16 11 0.0
17 11 0.0
18 11 0.0
19 11 0.01
20 11 0.019
0 12 0.0
1 12 0.0
2 12 0.0
3 12 0.0
4 12 0.0
5 12 0.0
6 12 0.029
7 12 0.0
8 12 0.0
9 12 0.0
10 12 0.0
11 12 0.0
12 12 0.912
13 12 0.0
14 12 0.029
15 12 0.0
16 12 0.0
17 12 0.0
18 12 0.0
19 12 0.029
20 12 0.0
0 13 0.0
1 13 0.143
2 13 0.0
3 13 0.029
4 13 0.029
5 13 0.029
6 13 0.0
7 13 0.029
8 13 0.0
9 13 0.057
10 13 0.029
11 13 0.0
12 13 0.0
13 13 0.457
14 13 0.029
15 13 0.0
16 13 0.029
17 13 0.0
18 13 0.114
19 13 0.029
20 13 0.0
0 14 0.0
1 14 0.013
2 14 0.0
3 14 0.025
4 14 0.0
5 14 0.025
6 14 0.013
7 14 0.025
8 14 0.0
9 14 0.0
10 14 0.0
11 14 0.0
12 14 0.0
13 14 0.025
14 14 0.494
15 14 0.0
16 14 0.342
17 14 0.0
18 14 0.0
19 14 0.025
20 14 0.013
0 15 0.0
1 15 0.0
2 15 0.0
3 15 0.0
4 15 0.0
5 15 0.0
6 15 0.0
7 15 0.017
8 15 0.0
9 15 0.017
10 15 0.0
11 15 0.0
12 15 0.0
13 15 0.0
14 15 0.0
15 15 0.69
16 15 0.155
17 15 0.121
18 15 0.0
19 15 0.0
20 15 0.0
0 16 0.0
1 16 0.0
2 16 0.0
3 16 0.0
4 16 0.011
5 16 0.006
6 16 0.006
7 16 0.0
8 16 0.0
9 16 0.0
10 16 0.006
11 16 0.011
12 16 0.006
13 16 0.0
14 16 0.006
15 16 0.062
16 16 0.86
17 16 0.017
18 16 0.006
19 16 0.006
20 16 0.0
0 17 0.0
1 17 0.016
2 17 0.0
3 17 0.0
4 17 0.0
5 17 0.005
6 17 0.0
7 17 0.0
8 17 0.0
9 17 0.0
10 17 0.0
11 17 0.0
12 17 0.0
13 17 0.0
14 17 0.0
15 17 0.0
16 17 0.011
17 17 0.968
18 17 0.0
19 17 0.0
20 17 0.0
0 18 0.0
1 18 0.0
2 18 0.0
3 18 0.006
4 18 0.0
5 18 0.006
6 18 0.0
7 18 0.006
8 18 0.0
9 18 0.0
10 18 0.0
11 18 0.006
12 18 0.0
13 18 0.0
14 18 0.0
15 18 0.0
16 18 0.006
17 18 0.0
18 18 0.952
19 18 0.012
20 18 0.006
0 19 0.0
1 19 0.0
2 19 0.0
3 19 0.0
4 19 0.0
5 19 0.018
6 19 0.006
7 19 0.006
8 19 0.0
9 19 0.0
10 19 0.0
11 19 0.0
12 19 0.0
13 19 0.0
14 19 0.0
15 19 0.0
16 19 0.006
17 19 0.0
18 19 0.006
19 19 0.947
20 19 0.012
0 20 0.0
1 20 0.018
2 20 0.0
3 20 0.0
4 20 0.0
5 20 0.127
6 20 0.0
7 20 0.055
8 20 0.0
9 20 0.0
10 20 0.0
11 20 0.036
12 20 0.018
13 20 0.018
14 20 0.055
15 20 0.018
16 20 0.018
17 20 0.0
18 20 0.036
19 20 0.036
20 20 0.564
        };
    \end{axis}
\end{tikzpicture}
%
%\caption{I'm confused~5!}\label{tab:CM5}
%\end{figure*}
	}
}
\\
\multicolumn{3}{c}{(c) Bagged LSTM}
	\end{tabular}
	\caption{Confusion matrices for bagging experiments}
	\label{fig:conf_1}
\end{figure}

%
% Boosting
%

\begin{figure}[!htb]
	\centering
	\begin{tabular}{ccc}
	\resizebox{0.45\textwidth}{0.25\textheight}{%
	%\begin{tikzpicture}[scale=0.60,every node/.style={scale=0.9}]
%\begin{tikzpicture}[scale=0.55]
\begin{tikzpicture}[scale=0.60]
    \begin{axis}[%colorbar/width=2.5mm,
        width=20cm,
        height=20cm,
%        colormap={blackwhite}{gray(0cm)=(1); gray(1cm)=(0.5)},
%   colormap={bluewhite}{color=(white) color=(blue)},
%   colormap={bluewhite}{color=(white) rgb255=(0,191,255)},
    colormap={bluewhite}{color=(white) rgb255=(100,149,237)},
        xticklabels={Adload,Agent,Allaple,BHO,Bifrose,CeeInject,Cycbot,FakeRean,Hotbar,Injector,
        OnLineGames,Renos,Rimecud,Small,Toga,VB,VBinject,Vobfus,Vundo,Winwebsec,Zbot},
        xtick={0,...,20},
        xtick style={draw=none},
    xticklabel style={anchor=east,rotate=45,yshift=-5pt},
        yticklabels={Adload,Agent,Allaple,BHO,Bifrose,CeeInject,Cycbot,FakeRean,Hotbar,Injector,
        OnLineGames,Renos,Rimecud,Small,Toga,VB,VBinject,Vobfus,Vundo,Winwebsec,Zbot},
        ytick={0,...,20},
        ytick style={draw=none},
        enlargelimits=false,
        xticklabel style={font=\large},
        yticklabel style={font=\large},
        colorbar,
        colorbar style={
%           width=0.05*\pgfkeysvalueof{/pgfplots/parent axis width},%%% added this
%           height=0.5*\pgfkeysvalueof{/pgfplots/parent axis height},
%       plot graphics/node/.style={scale=1.33,anchor=south west,inner sep=0pt,}, %%% scale colorbar fill %%%
            ytick={0.0,0.2,0.4,0.6,0.8,1.0},
            yticklabels={0.0,0.2,0.4,0.6,0.8,1.0},
            yticklabel={\pgfmathprintnumber\tick},
            yticklabel style={font=\large,
                    /pgf/number format/fixed,
            /pgf/number format/precision=1}
        },
%        point meta min=0,
%        point meta max=100,
        point meta min=0.0,
        point meta max=1.0,
        nodes near coords={\pgfmathprintnumber\pgfplotspointmeta},
        % ---------------------------------------------------------------------
        % show `nodes near coords' but adapt the style so that values
        % above a threshold get another style
        % (adapted from <http://tex.stackexchange.com/a/141006/95441>)
        % #1: the THRESHOLD after which we switch to a special display.
        nodes near coords black white/.style={
            % define the style of the nodes with "small" values
            small value/.style={
                yshift=-7pt,
%                text=white,
                text=black,
                /pgf/number format/fixed,
                /pgf/number format/precision=3,
                /pgf/number format/zerofill
            },
            % define the style of the nodes with "large" values
            large value/.style={
                yshift=-7pt,
%                text=black,
                text=white,
                /pgf/number format/fixed,
                /pgf/number format/precision=3,
                /pgf/number format/zerofill
            },
            every node near coord/.style={
                check for zero/.code={
                    \pgfmathfloatifflags{\pgfplotspointmeta}{0}{
                        % If meta=0, make the node a coordinate
                        % (which doesn't have text)
                        \pgfkeys{/tikz/coordinate}
                    }{
                        \begingroup
                        % this group is merely to switch to FPU locally.
                        % Might be unnecessary, but who knows.
                        \pgfkeys{/pgf/fpu}
                        \pgfmathparse{\pgfplotspointmeta<#1}
                        \global\let\result=\pgfmathresult
                        \endgroup
                        %
                        % simplifies debugging:
                        %\show\result
                        %
                        \pgfmathfloatcreate{1}{1.0}{0}
                        \let\ONE=\pgfmathresult
                        \ifx\result\ONE
                            % AH: our condition 'y < #1' is met.
                            \pgfkeysalso{/pgfplots/small value}
                        \else
                            % ok, proceed as usual.
                            \pgfkeysalso{/pgfplots/large value}
                        \fi
                    }
                },
                check for zero,
            },
        },
        % asign a value to the new style which is the threshold at which
        % the two style `small value' or `large value' are used
%        nodes near coords black white=50,
        nodes near coords black white=0.5,
        % -----------------------------------------------------------------
    ]
        \addplot[
            matrix plot,
            mesh/cols=21,
            point meta=explicit,draw=gray
        ] table [meta=C] {
            x y C
0 0 0.231
1 0 0.0
2 0 0.0
3 0 0.077
4 0 0.0
5 0 0.0
6 0 0.0
7 0 0.0
8 0 0.0
9 0 0.0
10 0 0.0
11 0 0.0
12 0 0.0
13 0 0.0
14 0 0.0
15 0 0.0
16 0 0.0
17 0 0.038
18 0 0.0
19 0 0.615
20 0 0.038
0 1 0.0
1 1 0.0
2 1 0.0
3 1 0.0
4 1 0.0
5 1 0.158
6 1 0.0
7 1 0.053
8 1 0.0
9 1 0.0
10 1 0.0
11 1 0.053
12 1 0.0
13 1 0.0
14 1 0.0
15 1 0.0
16 1 0.158
17 1 0.053
18 1 0.0
19 1 0.526
20 1 0.0
0 2 0.035
1 2 0.0
2 2 0.894
3 2 0.0
4 2 0.0
5 2 0.0
6 2 0.0
7 2 0.0
8 2 0.0
9 2 0.0
10 2 0.0
11 2 0.035
12 2 0.0
13 2 0.0
14 2 0.024
15 2 0.0
16 2 0.0
17 2 0.0
18 2 0.0
19 2 0.012
20 2 0.0
0 3 0.0
1 3 0.0
2 3 0.0
3 3 0.688
4 3 0.0
5 3 0.0
6 3 0.0
7 3 0.031
8 3 0.0
9 3 0.0
10 3 0.0
11 3 0.0
12 3 0.0
13 3 0.0
14 3 0.0
15 3 0.0
16 3 0.0
17 3 0.0
18 3 0.062
19 3 0.219
20 3 0.0
0 4 0.0
1 4 0.0
2 4 0.062
3 4 0.125
4 4 0.0
5 4 0.188
6 4 0.0
7 4 0.0
8 4 0.0
9 4 0.0
10 4 0.062
11 4 0.0
12 4 0.0
13 4 0.0
14 4 0.188
15 4 0.062
16 4 0.188
17 4 0.062
18 4 0.062
19 4 0.0
20 4 0.0
0 5 0.0
1 5 0.0
2 5 0.0
3 5 0.011
4 5 0.0
5 5 0.659
6 5 0.0
7 5 0.022
8 5 0.0
9 5 0.0
10 5 0.0
11 5 0.044
12 5 0.0
13 5 0.0
14 5 0.0
15 5 0.0
16 5 0.011
17 5 0.011
18 5 0.143
19 5 0.099
20 5 0.0
0 6 0.0
1 6 0.0
2 6 0.0
3 6 0.0
4 6 0.0
5 6 0.0
6 6 0.47
7 6 0.045
8 6 0.0
9 6 0.0
10 6 0.0
11 6 0.0
12 6 0.0
13 6 0.0
14 6 0.288
15 6 0.0
16 6 0.015
17 6 0.0
18 6 0.03
19 6 0.152
20 6 0.0
0 7 0.0
1 7 0.0
2 7 0.0
3 7 0.0
4 7 0.0
5 7 0.0
6 7 0.024
7 7 0.381
8 7 0.0
9 7 0.0
10 7 0.0
11 7 0.024
12 7 0.0
13 7 0.0
14 7 0.095
15 7 0.0
16 7 0.071
17 7 0.0
18 7 0.024
19 7 0.381
20 7 0.0
0 8 0.0
1 8 0.0
2 8 0.0
3 8 0.0
4 8 0.0
5 8 0.385
6 8 0.0
7 8 0.0
8 8 0.615
9 8 0.0
10 8 0.0
11 8 0.0
12 8 0.0
13 8 0.0
14 8 0.0
15 8 0.0
16 8 0.0
17 8 0.0
18 8 0.0
19 8 0.0
20 8 0.0
0 9 0.0
1 9 0.091
2 9 0.0
3 9 0.0
4 9 0.0
5 9 0.091
6 9 0.0
7 9 0.091
8 9 0.0
9 9 0.0
10 9 0.0
11 9 0.0
12 9 0.0
13 9 0.0
14 9 0.0
15 9 0.091
16 9 0.182
17 9 0.0
18 9 0.091
19 9 0.273
20 9 0.091
0 10 0.0
1 10 0.0
2 10 0.0
3 10 0.038
4 10 0.0
5 10 0.077
6 10 0.0
7 10 0.0
8 10 0.0
9 10 0.0
10 10 0.346
11 10 0.038
12 10 0.0
13 10 0.0
14 10 0.154
15 10 0.0
16 10 0.0
17 10 0.0
18 10 0.115
19 10 0.231
20 10 0.0
0 11 0.0
1 11 0.0
2 11 0.125
3 11 0.018
4 11 0.0
5 11 0.018
6 11 0.0
7 11 0.018
8 11 0.0
9 11 0.0
10 11 0.018
11 11 0.429
12 11 0.0
13 11 0.0
14 11 0.036
15 11 0.0
16 11 0.054
17 11 0.0
18 11 0.018
19 11 0.268
20 11 0.0
0 12 0.0
1 12 0.0
2 12 0.077
3 12 0.0
4 12 0.0
5 12 0.0
6 12 0.0
7 12 0.0
8 12 0.0
9 12 0.0
10 12 0.0
11 12 0.0
12 12 0.538
13 12 0.0
14 12 0.231
15 12 0.0
16 12 0.0
17 12 0.0
18 12 0.0
19 12 0.154
20 12 0.0
0 13 0.0
1 13 0.0
2 13 0.105
3 13 0.0
4 13 0.0
5 13 0.0
6 13 0.0
7 13 0.0
8 13 0.0
9 13 0.0
10 13 0.0
11 13 0.158
12 13 0.0
13 13 0.053
14 13 0.0
15 13 0.0
16 13 0.158
17 13 0.0
18 13 0.105
19 13 0.421
20 13 0.0
0 14 0.0
1 14 0.0
2 14 0.0
3 14 0.0
4 14 0.0
5 14 0.0
6 14 0.023
7 14 0.023
8 14 0.0
9 14 0.0
10 14 0.0
11 14 0.047
12 14 0.0
13 14 0.0
14 14 0.372
15 14 0.0
16 14 0.349
17 14 0.047
18 14 0.0
19 14 0.14
20 14 0.0
0 15 0.0
1 15 0.0
2 15 0.0
3 15 0.0
4 15 0.0
5 15 0.0
6 15 0.0
7 15 0.0
8 15 0.0
9 15 0.0
10 15 0.0
11 15 0.0
12 15 0.0
13 15 0.0
14 15 0.036
15 15 0.143
16 15 0.571
17 15 0.143
18 15 0.0
19 15 0.107
20 15 0.0
0 16 0.0
1 16 0.0
2 16 0.011
3 16 0.011
4 16 0.0
5 16 0.0
6 16 0.022
7 16 0.033
8 16 0.0
9 16 0.0
10 16 0.0
11 16 0.011
12 16 0.011
13 16 0.0
14 16 0.011
15 16 0.033
16 16 0.6
17 16 0.244
18 16 0.011
19 16 0.0
20 16 0.0
0 17 0.0
1 17 0.019
2 17 0.0
3 17 0.0
4 17 0.0
5 17 0.0
6 17 0.0
7 17 0.0
8 17 0.0
9 17 0.0
10 17 0.0
11 17 0.0
12 17 0.0
13 17 0.0
14 17 0.01
15 17 0.058
16 17 0.442
17 17 0.452
18 17 0.0
19 17 0.019
20 17 0.0
0 18 0.0
1 18 0.0
2 18 0.0
3 18 0.0
4 18 0.0
5 18 0.0
6 18 0.0
7 18 0.046
8 18 0.0
9 18 0.0
10 18 0.0
11 18 0.0
12 18 0.0
13 18 0.0
14 18 0.0
15 18 0.0
16 18 0.023
17 18 0.0
18 18 0.874
19 18 0.057
20 18 0.0
0 19 0.0
1 19 0.0
2 19 0.037
3 19 0.012
4 19 0.0
5 19 0.012
6 19 0.012
7 19 0.025
8 19 0.0
9 19 0.0
10 19 0.012
11 19 0.012
12 19 0.0
13 19 0.0
14 19 0.025
15 19 0.0
16 19 0.049
17 19 0.0
18 19 0.222
19 19 0.58
20 19 0.0
0 20 0.0
1 20 0.0
2 20 0.04
3 20 0.0
4 20 0.0
5 20 0.0
6 20 0.08
7 20 0.04
8 20 0.0
9 20 0.0
10 20 0.0
11 20 0.04
12 20 0.04
13 20 0.0
14 20 0.28
15 20 0.0
16 20 0.08
17 20 0.08
18 20 0.08
19 20 0.2
20 20 0.04
        };
    \end{axis}
\end{tikzpicture}
%
%\caption{I'm confused~5!}\label{tab:CM5}
%\end{figure*}
	}
& & 
	\resizebox{0.45\textwidth}{0.25\textheight}{%
	%\begin{tikzpicture}[scale=0.60,every node/.style={scale=0.9}]
%\begin{tikzpicture}[scale=0.55]
\begin{tikzpicture}[scale=0.60]
    \begin{axis}[%colorbar/width=2.5mm,
        width=20cm,
        height=20cm,
%        colormap={blackwhite}{gray(0cm)=(1); gray(1cm)=(0.5)},
%   colormap={bluewhite}{color=(white) color=(blue)},
%   colormap={bluewhite}{color=(white) rgb255=(0,191,255)},
    colormap={bluewhite}{color=(white) rgb255=(100,149,237)},
        xticklabels={Adload,Agent,Allaple,BHO,Bifrose,CeeInject,Cycbot,FakeRean,Hotbar,Injector,
        OnLineGames,Renos,Rimecud,Small,Toga,VB,VBinject,Vobfus,Vundo,Winwebsec,Zbot},
        xtick={0,...,20},
        xtick style={draw=none},
    xticklabel style={anchor=east,rotate=45,yshift=-5pt},
        yticklabels={Adload,Agent,Allaple,BHO,Bifrose,CeeInject,Cycbot,FakeRean,Hotbar,Injector,
        OnLineGames,Renos,Rimecud,Small,Toga,VB,VBinject,Vobfus,Vundo,Winwebsec,Zbot},
        ytick={0,...,20},
        ytick style={draw=none},
        enlargelimits=false,
        xticklabel style={font=\large},
        yticklabel style={font=\large},
        colorbar,
        colorbar style={
%           width=0.05*\pgfkeysvalueof{/pgfplots/parent axis width},%%% added this
%           height=0.5*\pgfkeysvalueof{/pgfplots/parent axis height},
%       plot graphics/node/.style={scale=1.33,anchor=south west,inner sep=0pt,}, %%% scale colorbar fill %%%
            ytick={0.0,0.2,0.4,0.6,0.8,1.0},
            yticklabels={0.0,0.2,0.4,0.6,0.8,1.0},
            yticklabel={\pgfmathprintnumber\tick},
            yticklabel style={font=\large,
                    /pgf/number format/fixed,
            /pgf/number format/precision=1}
        },
%        point meta min=0,
%        point meta max=100,
        point meta min=0.0,
        point meta max=1.0,
        nodes near coords={\pgfmathprintnumber\pgfplotspointmeta},
        % ---------------------------------------------------------------------
        % show `nodes near coords' but adapt the style so that values
        % above a threshold get another style
        % (adapted from <http://tex.stackexchange.com/a/141006/95441>)
        % #1: the THRESHOLD after which we switch to a special display.
        nodes near coords black white/.style={
            % define the style of the nodes with "small" values
            small value/.style={
                yshift=-7pt,
%                text=white,
                text=black,
                /pgf/number format/fixed,
                /pgf/number format/precision=3,
                /pgf/number format/zerofill
            },
            % define the style of the nodes with "large" values
            large value/.style={
                yshift=-7pt,
%                text=black,
                text=white,
                /pgf/number format/fixed,
                /pgf/number format/precision=3,
                /pgf/number format/zerofill
            },
            every node near coord/.style={
                check for zero/.code={
                    \pgfmathfloatifflags{\pgfplotspointmeta}{0}{
                        % If meta=0, make the node a coordinate
                        % (which doesn't have text)
                        \pgfkeys{/tikz/coordinate}
                    }{
                        \begingroup
                        % this group is merely to switch to FPU locally.
                        % Might be unnecessary, but who knows.
                        \pgfkeys{/pgf/fpu}
                        \pgfmathparse{\pgfplotspointmeta<#1}
                        \global\let\result=\pgfmathresult
                        \endgroup
                        %
                        % simplifies debugging:
                        %\show\result
                        %
                        \pgfmathfloatcreate{1}{1.0}{0}
                        \let\ONE=\pgfmathresult
                        \ifx\result\ONE
                            % AH: our condition 'y < #1' is met.
                            \pgfkeysalso{/pgfplots/small value}
                        \else
                            % ok, proceed as usual.
                            \pgfkeysalso{/pgfplots/large value}
                        \fi
                    }
                },
                check for zero,
            },
        },
        % asign a value to the new style which is the threshold at which
        % the two style `small value' or `large value' are used
%        nodes near coords black white=50,
        nodes near coords black white=0.5,
        % -----------------------------------------------------------------
    ]
        \addplot[
            matrix plot,
            mesh/cols=21,
            point meta=explicit,draw=gray
        ] table [meta=C] {
            x y C
0 0 0.923
1 0 0.0
2 0 0.0
3 0 0.038
4 0 0.0
5 0 0.0
6 0 0.0
7 0 0.0
8 0 0.0
9 0 0.0
10 0 0.0
11 0 0.038
12 0 0.0
13 0 0.0
14 0 0.0
15 0 0.0
16 0 0.0
17 0 0.0
18 0 0.0
19 0 0.0
20 0 0.0
0 1 0.0
1 1 0.211
2 1 0.0
3 1 0.053
4 1 0.0
5 1 0.053
6 1 0.0
7 1 0.158
8 1 0.0
9 1 0.053
10 1 0.053
11 1 0.053
12 1 0.0
13 1 0.0
14 1 0.0
15 1 0.0
16 1 0.105
17 1 0.053
18 1 0.053
19 1 0.158
20 1 0.0
0 2 0.0
1 2 0.0
2 2 0.976
3 2 0.0
4 2 0.0
5 2 0.0
6 2 0.0
7 2 0.0
8 2 0.0
9 2 0.0
10 2 0.0
11 2 0.012
12 2 0.0
13 2 0.0
14 2 0.0
15 2 0.0
16 2 0.0
17 2 0.0
18 2 0.0
19 2 0.012
20 2 0.0
0 3 0.062
1 3 0.0
2 3 0.0
3 3 0.875
4 3 0.0
5 3 0.0
6 3 0.0
7 3 0.0
8 3 0.0
9 3 0.0
10 3 0.0
11 3 0.0
12 3 0.0
13 3 0.0
14 3 0.0
15 3 0.0
16 3 0.0
17 3 0.0
18 3 0.031
19 3 0.031
20 3 0.0
0 4 0.0
1 4 0.0
2 4 0.125
3 4 0.062
4 4 0.188
5 4 0.062
6 4 0.0
7 4 0.0
8 4 0.0
9 4 0.062
10 4 0.0
11 4 0.0
12 4 0.0
13 4 0.0
14 4 0.062
15 4 0.0
16 4 0.312
17 4 0.062
18 4 0.0
19 4 0.062
20 4 0.0
0 5 0.0
1 5 0.022
2 5 0.0
3 5 0.0
4 5 0.0
5 5 0.846
6 5 0.0
7 5 0.011
8 5 0.0
9 5 0.011
10 5 0.011
11 5 0.033
12 5 0.0
13 5 0.011
14 5 0.022
15 5 0.0
16 5 0.0
17 5 0.0
18 5 0.011
19 5 0.022
20 5 0.0
0 6 0.0
1 6 0.0
2 6 0.0
3 6 0.0
4 6 0.0
5 6 0.0
6 6 0.985
7 6 0.0
8 6 0.0
9 6 0.0
10 6 0.0
11 6 0.0
12 6 0.0
13 6 0.0
14 6 0.0
15 6 0.0
16 6 0.0
17 6 0.0
18 6 0.0
19 6 0.015
20 6 0.0
0 7 0.0
1 7 0.0
2 7 0.0
3 7 0.0
4 7 0.0
5 7 0.024
6 7 0.071
7 7 0.738
8 7 0.0
9 7 0.0
10 7 0.0
11 7 0.024
12 7 0.0
13 7 0.0
14 7 0.024
15 7 0.0
16 7 0.095
17 7 0.0
18 7 0.0
19 7 0.0
20 7 0.024
0 8 0.0
1 8 0.0
2 8 0.0
3 8 0.0
4 8 0.0
5 8 0.0
6 8 0.0
7 8 0.0
8 8 1.0
9 8 0.0
10 8 0.0
11 8 0.0
12 8 0.0
13 8 0.0
14 8 0.0
15 8 0.0
16 8 0.0
17 8 0.0
18 8 0.0
19 8 0.0
20 8 0.0
0 9 0.0
1 9 0.091
2 9 0.0
3 9 0.0
4 9 0.0
5 9 0.091
6 9 0.0
7 9 0.0
8 9 0.0
9 9 0.182
10 9 0.091
11 9 0.091
12 9 0.0
13 9 0.0
14 9 0.0
15 9 0.0
16 9 0.273
17 9 0.091
18 9 0.0
19 9 0.091
20 9 0.0
0 10 0.0
1 10 0.0
2 10 0.038
3 10 0.0
4 10 0.0
5 10 0.0
6 10 0.0
7 10 0.038
8 10 0.0
9 10 0.038
10 10 0.577
11 10 0.077
12 10 0.0
13 10 0.0
14 10 0.0
15 10 0.0
16 10 0.0
17 10 0.0
18 10 0.115
19 10 0.115
20 10 0.0
0 11 0.0
1 11 0.0
2 11 0.018
3 11 0.018
4 11 0.0
5 11 0.0
6 11 0.0
7 11 0.018
8 11 0.0
9 11 0.0
10 11 0.0
11 11 0.857
12 11 0.0
13 11 0.0
14 11 0.0
15 11 0.0
16 11 0.018
17 11 0.0
18 11 0.0
19 11 0.054
20 11 0.018
0 12 0.0
1 12 0.0
2 12 0.0
3 12 0.0
4 12 0.0
5 12 0.0
6 12 0.0
7 12 0.077
8 12 0.0
9 12 0.0
10 12 0.077
11 12 0.0
12 12 0.615
13 12 0.0
14 12 0.0
15 12 0.0
16 12 0.0
17 12 0.0
18 12 0.0
19 12 0.154
20 12 0.077
0 13 0.0
1 13 0.105
2 13 0.105
3 13 0.0
4 13 0.0
5 13 0.105
6 13 0.0
7 13 0.0
8 13 0.0
9 13 0.105
10 13 0.0
11 13 0.105
12 13 0.0
13 13 0.158
14 13 0.0
15 13 0.0
16 13 0.105
17 13 0.0
18 13 0.0
19 13 0.158
20 13 0.053
0 14 0.0
1 14 0.0
2 14 0.0
3 14 0.0
4 14 0.0
5 14 0.047
6 14 0.023
7 14 0.023
8 14 0.0
9 14 0.0
10 14 0.0
11 14 0.023
12 14 0.0
13 14 0.0
14 14 0.302
15 14 0.0
16 14 0.372
17 14 0.047
18 14 0.0
19 14 0.116
20 14 0.047
0 15 0.0
1 15 0.0
2 15 0.0
3 15 0.0
4 15 0.0
5 15 0.0
6 15 0.036
7 15 0.0
8 15 0.0
9 15 0.0
10 15 0.0
11 15 0.0
12 15 0.0
13 15 0.0
14 15 0.0
15 15 0.321
16 15 0.464
17 15 0.107
18 15 0.0
19 15 0.071
20 15 0.0
0 16 0.0
1 16 0.0
2 16 0.0
3 16 0.0
4 16 0.022
5 16 0.0
6 16 0.0
7 16 0.0
8 16 0.0
9 16 0.0
10 16 0.0
11 16 0.033
12 16 0.0
13 16 0.0
14 16 0.022
15 16 0.056
16 16 0.7
17 16 0.122
18 16 0.011
19 16 0.033
20 16 0.0
0 17 0.0
1 17 0.01
2 17 0.0
3 17 0.0
4 17 0.0
5 17 0.019
6 17 0.0
7 17 0.0
8 17 0.0
9 17 0.01
10 17 0.0
11 17 0.0
12 17 0.0
13 17 0.0
14 17 0.0
15 17 0.019
16 17 0.106
17 17 0.837
18 17 0.0
19 17 0.0
20 17 0.0
0 18 0.0
1 18 0.0
2 18 0.0
3 18 0.0
4 18 0.011
5 18 0.0
6 18 0.0
7 18 0.034
8 18 0.0
9 18 0.0
10 18 0.0
11 18 0.011
12 18 0.0
13 18 0.0
14 18 0.0
15 18 0.0
16 18 0.0
17 18 0.011
18 18 0.885
19 18 0.034
20 18 0.011
0 19 0.0
1 19 0.0
2 19 0.0
3 19 0.0
4 19 0.0
5 19 0.0
6 19 0.025
7 19 0.049
8 19 0.0
9 19 0.0
10 19 0.012
11 19 0.012
12 19 0.0
13 19 0.0
14 19 0.012
15 19 0.0
16 19 0.0
17 19 0.0
18 19 0.0
19 19 0.877
20 19 0.012
0 20 0.0
1 20 0.0
2 20 0.0
3 20 0.0
4 20 0.0
5 20 0.12
6 20 0.04
7 20 0.0
8 20 0.0
9 20 0.0
10 20 0.0
11 20 0.12
12 20 0.04
13 20 0.0
14 20 0.2
15 20 0.0
16 20 0.0
17 20 0.04
18 20 0.04
19 20 0.16
20 20 0.24
        };
    \end{axis}
\end{tikzpicture}
%
%\caption{I'm confused~5!}\label{tab:CM5}
%\end{figure*}
	}
\\
(a) AdaBoost
& & 
(b) XGBoost
	\end{tabular}
	\caption{Confusion matrices for boosting techniques}
	\label{fig:conf_2b}
\end{figure}

%
% Voting ensembles
%

\begin{figure}[!htb]
	\centering
	\begin{tabular}{ccc}
	\resizebox{0.45\textwidth}{0.25\textheight}{%
	%\begin{tikzpicture}[scale=0.60,every node/.style={scale=0.9}]
%\begin{tikzpicture}[scale=0.55]
\begin{tikzpicture}[scale=0.60]
    \begin{axis}[%colorbar/width=2.5mm,
        width=20cm,
        height=20cm,
%        colormap={blackwhite}{gray(0cm)=(1); gray(1cm)=(0.5)},
%   colormap={bluewhite}{color=(white) color=(blue)},
%   colormap={bluewhite}{color=(white) rgb255=(0,191,255)},
    colormap={bluewhite}{color=(white) rgb255=(100,149,237)},
        xticklabels={Adload,Agent,Allaple,BHO,Bifrose,CeeInject,Cycbot,FakeRean,Hotbar,Injector,
        OnLineGames,Renos,Rimecud,Small,Toga,VB,VBinject,Vobfus,Vundo,Winwebsec,Zbot},
        xtick={0,...,20},
        xtick style={draw=none},
    xticklabel style={anchor=east,rotate=45,yshift=-5pt},
        yticklabels={Adload,Agent,Allaple,BHO,Bifrose,CeeInject,Cycbot,FakeRean,Hotbar,Injector,
        OnLineGames,Renos,Rimecud,Small,Toga,VB,VBinject,Vobfus,Vundo,Winwebsec,Zbot},
        ytick={0,...,20},
        ytick style={draw=none},
        enlargelimits=false,
        xticklabel style={font=\large},
        yticklabel style={font=\large},
        colorbar,
        colorbar style={
%           width=0.05*\pgfkeysvalueof{/pgfplots/parent axis width},%%% added this
%           height=0.5*\pgfkeysvalueof{/pgfplots/parent axis height},
%       plot graphics/node/.style={scale=1.33,anchor=south west,inner sep=0pt,}, %%% scale colorbar fill %%%
            ytick={0.0,0.2,0.4,0.6,0.8,1.0},
            yticklabels={0.0,0.2,0.4,0.6,0.8,1.0},
            yticklabel={\pgfmathprintnumber\tick},
            yticklabel style={font=\large,
                    /pgf/number format/fixed,
            /pgf/number format/precision=1}
        },
%        point meta min=0,
%        point meta max=100,
        point meta min=0.0,
        point meta max=1.0,
        nodes near coords={\pgfmathprintnumber\pgfplotspointmeta},
        % ---------------------------------------------------------------------
        % show `nodes near coords' but adapt the style so that values
        % above a threshold get another style
        % (adapted from <http://tex.stackexchange.com/a/141006/95441>)
        % #1: the THRESHOLD after which we switch to a special display.
        nodes near coords black white/.style={
            % define the style of the nodes with "small" values
            small value/.style={
                yshift=-7pt,
%                text=white,
                text=black,
                /pgf/number format/fixed,
                /pgf/number format/precision=3,
                /pgf/number format/zerofill
            },
            % define the style of the nodes with "large" values
            large value/.style={
                yshift=-7pt,
%                text=black,
                text=white,
                /pgf/number format/fixed,
                /pgf/number format/precision=3,
                /pgf/number format/zerofill
            },
            every node near coord/.style={
                check for zero/.code={
                    \pgfmathfloatifflags{\pgfplotspointmeta}{0}{
                        % If meta=0, make the node a coordinate
                        % (which doesn't have text)
                        \pgfkeys{/tikz/coordinate}
                    }{
                        \begingroup
                        % this group is merely to switch to FPU locally.
                        % Might be unnecessary, but who knows.
                        \pgfkeys{/pgf/fpu}
                        \pgfmathparse{\pgfplotspointmeta<#1}
                        \global\let\result=\pgfmathresult
                        \endgroup
                        %
                        % simplifies debugging:
                        %\show\result
                        %
                        \pgfmathfloatcreate{1}{1.0}{0}
                        \let\ONE=\pgfmathresult
                        \ifx\result\ONE
                            % AH: our condition 'y < #1' is met.
                            \pgfkeysalso{/pgfplots/small value}
                        \else
                            % ok, proceed as usual.
                            \pgfkeysalso{/pgfplots/large value}
                        \fi
                    }
                },
                check for zero,
            },
        },
        % asign a value to the new style which is the threshold at which
        % the two style `small value' or `large value' are used
%        nodes near coords black white=50,
        nodes near coords black white=0.5,
        % -----------------------------------------------------------------
    ]
        \addplot[
            matrix plot,
            mesh/cols=21,
            point meta=explicit,draw=gray
        ] table [meta=C] {
            x y C
0 0 0.976
1 0 0.0
2 0 0.0
3 0 0.024
4 0 0.0
5 0 0.0
6 0 0.0
7 0 0.0
8 0 0.0
9 0 0.0
10 0 0.0
11 0 0.0
12 0 0.0
13 0 0.0
14 0 0.0
15 0 0.0
16 0 0.0
17 0 0.0
18 0 0.0
19 0 0.0
20 0 0.0
0 1 0.0
1 1 0.743
2 1 0.0
3 1 0.0
4 1 0.0
5 1 0.0
6 1 0.0
7 1 0.057
8 1 0.0
9 1 0.0
10 1 0.0
11 1 0.029
12 1 0.0
13 1 0.029
14 1 0.0
15 1 0.0
16 1 0.057
17 1 0.029
18 1 0.057
19 1 0.0
20 1 0.0
0 2 0.0
1 2 0.0
2 2 1.0
3 2 0.0
4 2 0.0
5 2 0.0
6 2 0.0
7 2 0.0
8 2 0.0
9 2 0.0
10 2 0.0
11 2 0.0
12 2 0.0
13 2 0.0
14 2 0.0
15 2 0.0
16 2 0.0
17 2 0.0
18 2 0.0
19 2 0.0
20 2 0.0
0 3 0.042
1 3 0.0
2 3 0.0
3 3 0.915
4 3 0.0
5 3 0.0
6 3 0.0
7 3 0.0
8 3 0.0
9 3 0.0
10 3 0.0
11 3 0.0
12 3 0.0
13 3 0.0
14 3 0.0
15 3 0.0
16 3 0.014
17 3 0.0
18 3 0.028
19 3 0.0
20 3 0.0
0 4 0.0
1 4 0.0
2 4 0.0
3 4 0.0
4 4 0.368
5 4 0.158
6 4 0.026
7 4 0.0
8 4 0.0
9 4 0.026
10 4 0.0
11 4 0.0
12 4 0.0
13 4 0.0
14 4 0.026
15 4 0.0
16 4 0.368
17 4 0.0
18 4 0.0
19 4 0.026
20 4 0.0
0 5 0.0
1 5 0.0
2 5 0.0
3 5 0.0
4 5 0.0
5 5 0.949
6 5 0.0
7 5 0.0
8 5 0.0
9 5 0.0
10 5 0.0
11 5 0.019
12 5 0.0
13 5 0.0
14 5 0.0
15 5 0.0
16 5 0.013
17 5 0.0
18 5 0.013
19 5 0.0
20 5 0.006
0 6 0.0
1 6 0.0
2 6 0.0
3 6 0.0
4 6 0.0
5 6 0.0
6 6 0.984
7 6 0.008
8 6 0.0
9 6 0.0
10 6 0.0
11 6 0.0
12 6 0.0
13 6 0.0
14 6 0.0
15 6 0.0
16 6 0.0
17 6 0.0
18 6 0.0
19 6 0.008
20 6 0.0
0 7 0.0
1 7 0.0
2 7 0.0
3 7 0.0
4 7 0.0
5 7 0.0
6 7 0.0
7 7 0.989
8 7 0.0
9 7 0.0
10 7 0.0
11 7 0.0
12 7 0.0
13 7 0.0
14 7 0.011
15 7 0.0
16 7 0.0
17 7 0.0
18 7 0.0
19 7 0.0
20 7 0.0
0 8 0.0
1 8 0.0
2 8 0.0
3 8 0.0
4 8 0.0
5 8 0.0
6 8 0.0
7 8 0.0
8 8 1.0
9 8 0.0
10 8 0.0
11 8 0.0
12 8 0.0
13 8 0.0
14 8 0.0
15 8 0.0
16 8 0.0
17 8 0.0
18 8 0.0
19 8 0.0
20 8 0.0
0 9 0.0
1 9 0.0
2 9 0.0
3 9 0.0
4 9 0.032
5 9 0.161
6 9 0.0
7 9 0.0
8 9 0.0
9 9 0.548
10 9 0.0
11 9 0.0
12 9 0.0
13 9 0.0
14 9 0.0
15 9 0.0
16 9 0.161
17 9 0.0
18 9 0.065
19 9 0.032
20 9 0.0
0 10 0.0
1 10 0.0
2 10 0.0
3 10 0.0
4 10 0.0
5 10 0.0
6 10 0.0
7 10 0.043
8 10 0.0
9 10 0.0
10 10 0.936
11 10 0.0
12 10 0.0
13 10 0.0
14 10 0.0
15 10 0.0
16 10 0.0
17 10 0.0
18 10 0.021
19 10 0.0
20 10 0.0
0 11 0.0
1 11 0.0
2 11 0.0
3 11 0.01
4 11 0.0
5 11 0.0
6 11 0.0
7 11 0.0
8 11 0.0
9 11 0.0
10 11 0.0
11 11 0.99
12 11 0.0
13 11 0.0
14 11 0.0
15 11 0.0
16 11 0.0
17 11 0.0
18 11 0.0
19 11 0.0
20 11 0.0
0 12 0.0
1 12 0.0
2 12 0.0
3 12 0.0
4 12 0.0
5 12 0.0
6 12 0.0
7 12 0.029
8 12 0.0
9 12 0.0
10 12 0.0
11 12 0.0
12 12 0.941
13 12 0.0
14 12 0.0
15 12 0.0
16 12 0.029
17 12 0.0
18 12 0.0
19 12 0.0
20 12 0.0
0 13 0.0
1 13 0.029
2 13 0.0
3 13 0.0
4 13 0.0
5 13 0.0
6 13 0.0
7 13 0.0
8 13 0.0
9 13 0.0
10 13 0.029
11 13 0.0
12 13 0.0
13 13 0.714
14 13 0.0
15 13 0.0
16 13 0.114
17 13 0.0
18 13 0.029
19 13 0.086
20 13 0.0
0 14 0.0
1 14 0.0
2 14 0.0
3 14 0.0
4 14 0.013
5 14 0.0
6 14 0.0
7 14 0.013
8 14 0.0
9 14 0.0
10 14 0.0
11 14 0.0
12 14 0.0
13 14 0.0
14 14 0.759
15 14 0.0
16 14 0.215
17 14 0.0
18 14 0.0
19 14 0.0
20 14 0.0
0 15 0.0
1 15 0.0
2 15 0.0
3 15 0.0
4 15 0.0
5 15 0.0
6 15 0.0
7 15 0.052
8 15 0.0
9 15 0.017
10 15 0.0
11 15 0.0
12 15 0.0
13 15 0.0
14 15 0.0
15 15 0.724
16 15 0.19
17 15 0.017
18 15 0.0
19 15 0.0
20 15 0.0
0 16 0.0
1 16 0.0
2 16 0.0
3 16 0.0
4 16 0.0
5 16 0.0
6 16 0.0
7 16 0.006
8 16 0.0
9 16 0.0
10 16 0.006
11 16 0.0
12 16 0.0
13 16 0.0
14 16 0.006
15 16 0.0
16 16 0.955
17 16 0.0
18 16 0.006
19 16 0.011
20 16 0.011
0 17 0.0
1 17 0.0
2 17 0.0
3 17 0.0
4 17 0.0
5 17 0.0
6 17 0.0
7 17 0.0
8 17 0.0
9 17 0.0
10 17 0.0
11 17 0.0
12 17 0.0
13 17 0.0
14 17 0.0
15 17 0.0
16 17 0.011
17 17 0.989
18 17 0.0
19 17 0.0
20 17 0.0
0 18 0.0
1 18 0.0
2 18 0.0
3 18 0.0
4 18 0.0
5 18 0.0
6 18 0.0
7 18 0.018
8 18 0.0
9 18 0.0
10 18 0.0
11 18 0.0
12 18 0.0
13 18 0.0
14 18 0.0
15 18 0.0
16 18 0.0
17 18 0.0
18 18 0.958
19 18 0.018
20 18 0.006
0 19 0.0
1 19 0.0
2 19 0.0
3 19 0.0
4 19 0.0
5 19 0.006
6 19 0.0
7 19 0.0
8 19 0.0
9 19 0.0
10 19 0.0
11 19 0.0
12 19 0.0
13 19 0.0
14 19 0.006
15 19 0.0
16 19 0.0
17 19 0.0
18 19 0.0
19 19 0.982
20 19 0.006
0 20 0.0
1 20 0.0
2 20 0.0
3 20 0.0
4 20 0.0
5 20 0.0
6 20 0.0
7 20 0.055
8 20 0.0
9 20 0.0
10 20 0.0
11 20 0.0
12 20 0.0
13 20 0.0
14 20 0.036
15 20 0.0
16 20 0.055
17 20 0.0
18 20 0.0
19 20 0.0
20 20 0.855
        };
    \end{axis}
\end{tikzpicture}
%
%\caption{I'm confused~5!}\label{tab:CM5}
%\end{figure*}
	}
& & 
	\resizebox{0.45\textwidth}{0.25\textheight}{%
	%\begin{tikzpicture}[scale=0.60,every node/.style={scale=0.9}]
%\begin{tikzpicture}[scale=0.55]
\begin{tikzpicture}[scale=0.60]
    \begin{axis}[%colorbar/width=2.5mm,
        width=20cm,
        height=20cm,
%        colormap={blackwhite}{gray(0cm)=(1); gray(1cm)=(0.5)},
%   colormap={bluewhite}{color=(white) color=(blue)},
%   colormap={bluewhite}{color=(white) rgb255=(0,191,255)},
    colormap={bluewhite}{color=(white) rgb255=(100,149,237)},
        xticklabels={Adload,Agent,Allaple,BHO,Bifrose,CeeInject,Cycbot,FakeRean,Hotbar,Injector,
        OnLineGames,Renos,Rimecud,Small,Toga,VB,VBinject,Vobfus,Vundo,Winwebsec,Zbot},
        xtick={0,...,20},
        xtick style={draw=none},
    xticklabel style={anchor=east,rotate=45,yshift=-5pt},
        yticklabels={Adload,Agent,Allaple,BHO,Bifrose,CeeInject,Cycbot,FakeRean,Hotbar,Injector,
        OnLineGames,Renos,Rimecud,Small,Toga,VB,VBinject,Vobfus,Vundo,Winwebsec,Zbot},
        ytick={0,...,20},
        ytick style={draw=none},
        enlargelimits=false,
        xticklabel style={font=\large},
        yticklabel style={font=\large},
        colorbar,
        colorbar style={
%           width=0.05*\pgfkeysvalueof{/pgfplots/parent axis width},%%% added this
%           height=0.5*\pgfkeysvalueof{/pgfplots/parent axis height},
%       plot graphics/node/.style={scale=1.33,anchor=south west,inner sep=0pt,}, %%% scale colorbar fill %%%
            ytick={0.0,0.2,0.4,0.6,0.8,1.0},
            yticklabels={0.0,0.2,0.4,0.6,0.8,1.0},
            yticklabel={\pgfmathprintnumber\tick},
            yticklabel style={font=\large,
                    /pgf/number format/fixed,
            /pgf/number format/precision=1}
        },
%        point meta min=0,
%        point meta max=100,
        point meta min=0.0,
        point meta max=1.0,
        nodes near coords={\pgfmathprintnumber\pgfplotspointmeta},
        % ---------------------------------------------------------------------
        % show `nodes near coords' but adapt the style so that values
        % above a threshold get another style
        % (adapted from <http://tex.stackexchange.com/a/141006/95441>)
        % #1: the THRESHOLD after which we switch to a special display.
        nodes near coords black white/.style={
            % define the style of the nodes with "small" values
            small value/.style={
                yshift=-7pt,
%                text=white,
                text=black,
                /pgf/number format/fixed,
                /pgf/number format/precision=3,
                /pgf/number format/zerofill
            },
            % define the style of the nodes with "large" values
            large value/.style={
                yshift=-7pt,
%                text=black,
                text=white,
                /pgf/number format/fixed,
                /pgf/number format/precision=3,
                /pgf/number format/zerofill
            },
            every node near coord/.style={
                check for zero/.code={
                    \pgfmathfloatifflags{\pgfplotspointmeta}{0}{
                        % If meta=0, make the node a coordinate
                        % (which doesn't have text)
                        \pgfkeys{/tikz/coordinate}
                    }{
                        \begingroup
                        % this group is merely to switch to FPU locally.
                        % Might be unnecessary, but who knows.
                        \pgfkeys{/pgf/fpu}
                        \pgfmathparse{\pgfplotspointmeta<#1}
                        \global\let\result=\pgfmathresult
                        \endgroup
                        %
                        % simplifies debugging:
                        %\show\result
                        %
                        \pgfmathfloatcreate{1}{1.0}{0}
                        \let\ONE=\pgfmathresult
                        \ifx\result\ONE
                            % AH: our condition 'y < #1' is met.
                            \pgfkeysalso{/pgfplots/small value}
                        \else
                            % ok, proceed as usual.
                            \pgfkeysalso{/pgfplots/large value}
                        \fi
                    }
                },
                check for zero,
            },
        },
        % asign a value to the new style which is the threshold at which
        % the two style `small value' or `large value' are used
%        nodes near coords black white=50,
        nodes near coords black white=0.5,
        % -----------------------------------------------------------------
    ]
        \addplot[
            matrix plot,
            mesh/cols=21,
            point meta=explicit,draw=gray
        ] table [meta=C] {
            x y C
0 0 0.833
1 0 0.0
2 0 0.0
3 0 0.119
4 0 0.0
5 0 0.0
6 0 0.0
7 0 0.0
8 0 0.0
9 0 0.0
10 0 0.0
11 0 0.024
12 0 0.0
13 0 0.0
14 0 0.024
15 0 0.0
16 0 0.0
17 0 0.0
18 0 0.0
19 0 0.0
20 0 0.0
0 1 0.0
1 1 0.514
2 1 0.0
3 1 0.029
4 1 0.0
5 1 0.057
6 1 0.0
7 1 0.057
8 1 0.0
9 1 0.0
10 1 0.0
11 1 0.086
12 1 0.0
13 1 0.086
14 1 0.057
15 1 0.0
16 1 0.0
17 1 0.057
18 1 0.029
19 1 0.029
20 1 0.0
0 2 0.0
1 2 0.0
2 2 1.0
3 2 0.0
4 2 0.0
5 2 0.0
6 2 0.0
7 2 0.0
8 2 0.0
9 2 0.0
10 2 0.0
11 2 0.0
12 2 0.0
13 2 0.0
14 2 0.0
15 2 0.0
16 2 0.0
17 2 0.0
18 2 0.0
19 2 0.0
20 2 0.0
0 3 0.0
1 3 0.0
2 3 0.0
3 3 0.944
4 3 0.0
5 3 0.0
6 3 0.0
7 3 0.0
8 3 0.0
9 3 0.0
10 3 0.0
11 3 0.0
12 3 0.0
13 3 0.014
14 3 0.0
15 3 0.0
16 3 0.014
17 3 0.0
18 3 0.028
19 3 0.0
20 3 0.0
0 4 0.0
1 4 0.026
2 4 0.0
3 4 0.0
4 4 0.079
5 4 0.184
6 4 0.0
7 4 0.026
8 4 0.0
9 4 0.026
10 4 0.0
11 4 0.053
12 4 0.0
13 4 0.0
14 4 0.079
15 4 0.026
16 4 0.368
17 4 0.0
18 4 0.053
19 4 0.079
20 4 0.0
0 5 0.0
1 5 0.013
2 5 0.0
3 5 0.013
4 5 0.0
5 5 0.911
6 5 0.0
7 5 0.0
8 5 0.0
9 5 0.013
10 5 0.0
11 5 0.025
12 5 0.0
13 5 0.0
14 5 0.0
15 5 0.0
16 5 0.006
17 5 0.0
18 5 0.0
19 5 0.019
20 5 0.0
0 6 0.0
1 6 0.0
2 6 0.0
3 6 0.0
4 6 0.0
5 6 0.0
6 6 0.992
7 6 0.0
8 6 0.0
9 6 0.0
10 6 0.0
11 6 0.0
12 6 0.0
13 6 0.0
14 6 0.0
15 6 0.0
16 6 0.0
17 6 0.0
18 6 0.0
19 6 0.008
20 6 0.0
0 7 0.0
1 7 0.0
2 7 0.0
3 7 0.0
4 7 0.0
5 7 0.0
6 7 0.021
7 7 0.926
8 7 0.0
9 7 0.0
10 7 0.0
11 7 0.0
12 7 0.0
13 7 0.0
14 7 0.0
15 7 0.0
16 7 0.021
17 7 0.0
18 7 0.011
19 7 0.021
20 7 0.0
0 8 0.0
1 8 0.0
2 8 0.0
3 8 0.0
4 8 0.0
5 8 0.0
6 8 0.0
7 8 0.0
8 8 0.962
9 8 0.0
10 8 0.0
11 8 0.038
12 8 0.0
13 8 0.0
14 8 0.0
15 8 0.0
16 8 0.0
17 8 0.0
18 8 0.0
19 8 0.0
20 8 0.0
0 9 0.0
1 9 0.032
2 9 0.0
3 9 0.0
4 9 0.0
5 9 0.323
6 9 0.0
7 9 0.032
8 9 0.0
9 9 0.097
10 9 0.032
11 9 0.097
12 9 0.0
13 9 0.032
14 9 0.0
15 9 0.032
16 9 0.129
17 9 0.032
18 9 0.032
19 9 0.129
20 9 0.0
0 10 0.0
1 10 0.064
2 10 0.0
3 10 0.0
4 10 0.0
5 10 0.0
6 10 0.0
7 10 0.043
8 10 0.0
9 10 0.0
10 10 0.745
11 10 0.0
12 10 0.0
13 10 0.021
14 10 0.0
15 10 0.0
16 10 0.0
17 10 0.0
18 10 0.085
19 10 0.043
20 10 0.0
0 11 0.0
1 11 0.0
2 11 0.0
3 11 0.01
4 11 0.0
5 11 0.0
6 11 0.01
7 11 0.0
8 11 0.0
9 11 0.0
10 11 0.0
11 11 0.933
12 11 0.0
13 11 0.0
14 11 0.01
15 11 0.0
16 11 0.0
17 11 0.0
18 11 0.0
19 11 0.029
20 11 0.01
0 12 0.0
1 12 0.0
2 12 0.0
3 12 0.0
4 12 0.0
5 12 0.0
6 12 0.029
7 12 0.0
8 12 0.0
9 12 0.0
10 12 0.029
11 12 0.0
12 12 0.912
13 12 0.0
14 12 0.029
15 12 0.0
16 12 0.0
17 12 0.0
18 12 0.0
19 12 0.0
20 12 0.0
0 13 0.0
1 13 0.143
2 13 0.029
3 13 0.0
4 13 0.029
5 13 0.0
6 13 0.0
7 13 0.0
8 13 0.0
9 13 0.057
10 13 0.029
11 13 0.0
12 13 0.0
13 13 0.429
14 13 0.057
15 13 0.0
16 13 0.057
17 13 0.0
18 13 0.057
19 13 0.114
20 13 0.0
0 14 0.0
1 14 0.025
2 14 0.0
3 14 0.0
4 14 0.0
5 14 0.013
6 14 0.038
7 14 0.025
8 14 0.0
9 14 0.0
10 14 0.0
11 14 0.0
12 14 0.013
13 14 0.025
14 14 0.443
15 14 0.0
16 14 0.38
17 14 0.0
18 14 0.013
19 14 0.013
20 14 0.013
0 15 0.0
1 15 0.0
2 15 0.0
3 15 0.0
4 15 0.0
5 15 0.0
6 15 0.0
7 15 0.052
8 15 0.0
9 15 0.017
10 15 0.0
11 15 0.0
12 15 0.0
13 15 0.0
14 15 0.0
15 15 0.603
16 15 0.19
17 15 0.121
18 15 0.0
19 15 0.017
20 15 0.0
0 16 0.0
1 16 0.006
2 16 0.0
3 16 0.0
4 16 0.0
5 16 0.006
6 16 0.0
7 16 0.0
8 16 0.0
9 16 0.0
10 16 0.0
11 16 0.006
12 16 0.006
13 16 0.0
14 16 0.006
15 16 0.017
16 16 0.933
17 16 0.006
18 16 0.011
19 16 0.006
20 16 0.0
0 17 0.0
1 17 0.016
2 17 0.0
3 17 0.0
4 17 0.0
5 17 0.005
6 17 0.0
7 17 0.0
8 17 0.0
9 17 0.0
10 17 0.0
11 17 0.0
12 17 0.0
13 17 0.0
14 17 0.0
15 17 0.0
16 17 0.011
17 17 0.963
18 17 0.0
19 17 0.005
20 17 0.0
0 18 0.0
1 18 0.0
2 18 0.0
3 18 0.006
4 18 0.0
5 18 0.0
6 18 0.0
7 18 0.006
8 18 0.0
9 18 0.0
10 18 0.0
11 18 0.012
12 18 0.0
13 18 0.0
14 18 0.006
15 18 0.0
16 18 0.006
17 18 0.0
18 18 0.952
19 18 0.006
20 18 0.006
0 19 0.0
1 19 0.0
2 19 0.0
3 19 0.0
4 19 0.0
5 19 0.018
6 19 0.0
7 19 0.006
8 19 0.0
9 19 0.0
10 19 0.0
11 19 0.006
12 19 0.0
13 19 0.0
14 19 0.0
15 19 0.0
16 19 0.0
17 19 0.0
18 19 0.0
19 19 0.971
20 19 0.0
0 20 0.0
1 20 0.018
2 20 0.0
3 20 0.0
4 20 0.0
5 20 0.109
6 20 0.0
7 20 0.055
8 20 0.0
9 20 0.0
10 20 0.0
11 20 0.091
12 20 0.018
13 20 0.018
14 20 0.036
15 20 0.018
16 20 0.018
17 20 0.0
18 20 0.018
19 20 0.055
20 20 0.545
        };
    \end{axis}
\end{tikzpicture}
%
%\caption{I'm confused~5!}\label{tab:CM5}
%\end{figure*}
	}
\\
(a) CNN
& & 
(b) LSTM
\\ \\
	\resizebox{0.45\textwidth}{0.25\textheight}{%
	%\begin{tikzpicture}[scale=0.60,every node/.style={scale=0.9}]
%\begin{tikzpicture}[scale=0.55]
\begin{tikzpicture}[scale=0.60]
    \begin{axis}[%colorbar/width=2.5mm,
        width=20cm,
        height=20cm,
%        colormap={blackwhite}{gray(0cm)=(1); gray(1cm)=(0.5)},
%   colormap={bluewhite}{color=(white) color=(blue)},
%   colormap={bluewhite}{color=(white) rgb255=(0,191,255)},
    colormap={bluewhite}{color=(white) rgb255=(100,149,237)},
        xticklabels={Adload,Agent,Allaple,BHO,Bifrose,CeeInject,Cycbot,FakeRean,Hotbar,Injector,
        OnLineGames,Renos,Rimecud,Small,Toga,VB,VBinject,Vobfus,Vundo,Winwebsec,Zbot},
        xtick={0,...,20},
        xtick style={draw=none},
    xticklabel style={anchor=east,rotate=45,yshift=-5pt},
        yticklabels={Adload,Agent,Allaple,BHO,Bifrose,CeeInject,Cycbot,FakeRean,Hotbar,Injector,
        OnLineGames,Renos,Rimecud,Small,Toga,VB,VBinject,Vobfus,Vundo,Winwebsec,Zbot},
        ytick={0,...,20},
        ytick style={draw=none},
        enlargelimits=false,
        xticklabel style={font=\large},
        yticklabel style={font=\large},
        colorbar,
        colorbar style={
%           width=0.05*\pgfkeysvalueof{/pgfplots/parent axis width},%%% added this
%           height=0.5*\pgfkeysvalueof{/pgfplots/parent axis height},
%       plot graphics/node/.style={scale=1.33,anchor=south west,inner sep=0pt,}, %%% scale colorbar fill %%%
            ytick={0.0,0.2,0.4,0.6,0.8,1.0},
            yticklabels={0.0,0.2,0.4,0.6,0.8,1.0},
            yticklabel={\pgfmathprintnumber\tick},
            yticklabel style={font=\large,
                    /pgf/number format/fixed,
            /pgf/number format/precision=1}
        },
%        point meta min=0,
%        point meta max=100,
        point meta min=0.0,
        point meta max=1.0,
        nodes near coords={\pgfmathprintnumber\pgfplotspointmeta},
        % ---------------------------------------------------------------------
        % show `nodes near coords' but adapt the style so that values
        % above a threshold get another style
        % (adapted from <http://tex.stackexchange.com/a/141006/95441>)
        % #1: the THRESHOLD after which we switch to a special display.
        nodes near coords black white/.style={
            % define the style of the nodes with "small" values
            small value/.style={
                yshift=-7pt,
%                text=white,
                text=black,
                /pgf/number format/fixed,
                /pgf/number format/precision=3,
                /pgf/number format/zerofill
            },
            % define the style of the nodes with "large" values
            large value/.style={
                yshift=-7pt,
%                text=black,
                text=white,
                /pgf/number format/fixed,
                /pgf/number format/precision=3,
                /pgf/number format/zerofill
            },
            every node near coord/.style={
                check for zero/.code={
                    \pgfmathfloatifflags{\pgfplotspointmeta}{0}{
                        % If meta=0, make the node a coordinate
                        % (which doesn't have text)
                        \pgfkeys{/tikz/coordinate}
                    }{
                        \begingroup
                        % this group is merely to switch to FPU locally.
                        % Might be unnecessary, but who knows.
                        \pgfkeys{/pgf/fpu}
                        \pgfmathparse{\pgfplotspointmeta<#1}
                        \global\let\result=\pgfmathresult
                        \endgroup
                        %
                        % simplifies debugging:
                        %\show\result
                        %
                        \pgfmathfloatcreate{1}{1.0}{0}
                        \let\ONE=\pgfmathresult
                        \ifx\result\ONE
                            % AH: our condition 'y < #1' is met.
                            \pgfkeysalso{/pgfplots/small value}
                        \else
                            % ok, proceed as usual.
                            \pgfkeysalso{/pgfplots/large value}
                        \fi
                    }
                },
                check for zero,
            },
        },
        % asign a value to the new style which is the threshold at which
        % the two style `small value' or `large value' are used
%        nodes near coords black white=50,
        nodes near coords black white=0.5,
        % -----------------------------------------------------------------
    ]
        \addplot[
            matrix plot,
            mesh/cols=21,
            point meta=explicit,draw=gray
        ] table [meta=C] {
            x y C
0 0 0.976
1 0 0.0
2 0 0.0
3 0 0.024
4 0 0.0
5 0 0.0
6 0 0.0
7 0 0.0
8 0 0.0
9 0 0.0
10 0 0.0
11 0 0.0
12 0 0.0
13 0 0.0
14 0 0.0
15 0 0.0
16 0 0.0
17 0 0.0
18 0 0.0
19 0 0.0
20 0 0.0
0 1 0.0
1 1 0.771
2 1 0.0
3 1 0.0
4 1 0.0
5 1 0.0
6 1 0.0
7 1 0.029
8 1 0.0
9 1 0.0
10 1 0.0
11 1 0.029
12 1 0.0
13 1 0.029
14 1 0.0
15 1 0.0
16 1 0.057
17 1 0.029
18 1 0.057
19 1 0.0
20 1 0.0
0 2 0.0
1 2 0.0
2 2 1.0
3 2 0.0
4 2 0.0
5 2 0.0
6 2 0.0
7 2 0.0
8 2 0.0
9 2 0.0
10 2 0.0
11 2 0.0
12 2 0.0
13 2 0.0
14 2 0.0
15 2 0.0
16 2 0.0
17 2 0.0
18 2 0.0
19 2 0.0
20 2 0.0
0 3 0.042
1 3 0.0
2 3 0.0
3 3 0.915
4 3 0.0
5 3 0.0
6 3 0.0
7 3 0.0
8 3 0.0
9 3 0.0
10 3 0.0
11 3 0.0
12 3 0.0
13 3 0.0
14 3 0.0
15 3 0.0
16 3 0.014
17 3 0.0
18 3 0.028
19 3 0.0
20 3 0.0
0 4 0.0
1 4 0.026
2 4 0.0
3 4 0.0
4 4 0.342
5 4 0.158
6 4 0.0
7 4 0.0
8 4 0.0
9 4 0.026
10 4 0.0
11 4 0.0
12 4 0.0
13 4 0.0
14 4 0.0
15 4 0.0
16 4 0.395
17 4 0.0
18 4 0.0
19 4 0.053
20 4 0.0
0 5 0.0
1 5 0.0
2 5 0.0
3 5 0.0
4 5 0.0
5 5 0.956
6 5 0.0
7 5 0.0
8 5 0.0
9 5 0.0
10 5 0.0
11 5 0.019
12 5 0.0
13 5 0.0
14 5 0.0
15 5 0.0
16 5 0.006
17 5 0.0
18 5 0.006
19 5 0.006
20 5 0.006
0 6 0.0
1 6 0.0
2 6 0.0
3 6 0.0
4 6 0.0
5 6 0.0
6 6 0.984
7 6 0.008
8 6 0.0
9 6 0.0
10 6 0.0
11 6 0.0
12 6 0.0
13 6 0.0
14 6 0.0
15 6 0.0
16 6 0.0
17 6 0.0
18 6 0.0
19 6 0.008
20 6 0.0
0 7 0.0
1 7 0.0
2 7 0.0
3 7 0.0
4 7 0.0
5 7 0.0
6 7 0.0
7 7 0.989
8 7 0.0
9 7 0.0
10 7 0.0
11 7 0.0
12 7 0.0
13 7 0.0
14 7 0.011
15 7 0.0
16 7 0.0
17 7 0.0
18 7 0.0
19 7 0.0
20 7 0.0
0 8 0.0
1 8 0.0
2 8 0.0
3 8 0.0
4 8 0.0
5 8 0.0
6 8 0.0
7 8 0.0
8 8 1.0
9 8 0.0
10 8 0.0
11 8 0.0
12 8 0.0
13 8 0.0
14 8 0.0
15 8 0.0
16 8 0.0
17 8 0.0
18 8 0.0
19 8 0.0
20 8 0.0
0 9 0.0
1 9 0.0
2 9 0.0
3 9 0.0
4 9 0.032
5 9 0.161
6 9 0.0
7 9 0.0
8 9 0.0
9 9 0.516
10 9 0.0
11 9 0.0
12 9 0.0
13 9 0.0
14 9 0.0
15 9 0.0
16 9 0.194
17 9 0.0
18 9 0.065
19 9 0.032
20 9 0.0
0 10 0.0
1 10 0.021
2 10 0.0
3 10 0.0
4 10 0.0
5 10 0.0
6 10 0.0
7 10 0.021
8 10 0.0
9 10 0.0
10 10 0.894
11 10 0.0
12 10 0.0
13 10 0.0
14 10 0.0
15 10 0.0
16 10 0.0
17 10 0.0
18 10 0.064
19 10 0.0
20 10 0.0
0 11 0.0
1 11 0.0
2 11 0.0
3 11 0.01
4 11 0.0
5 11 0.0
6 11 0.0
7 11 0.0
8 11 0.0
9 11 0.0
10 11 0.0
11 11 0.99
12 11 0.0
13 11 0.0
14 11 0.0
15 11 0.0
16 11 0.0
17 11 0.0
18 11 0.0
19 11 0.0
20 11 0.0
0 12 0.0
1 12 0.0
2 12 0.0
3 12 0.0
4 12 0.0
5 12 0.0
6 12 0.0
7 12 0.0
8 12 0.0
9 12 0.0
10 12 0.0
11 12 0.0
12 12 0.971
13 12 0.0
14 12 0.0
15 12 0.0
16 12 0.029
17 12 0.0
18 12 0.0
19 12 0.0
20 12 0.0
0 13 0.0
1 13 0.057
2 13 0.0
3 13 0.0
4 13 0.0
5 13 0.0
6 13 0.0
7 13 0.0
8 13 0.0
9 13 0.0
10 13 0.0
11 13 0.0
12 13 0.0
13 13 0.743
14 13 0.0
15 13 0.0
16 13 0.114
17 13 0.0
18 13 0.029
19 13 0.057
20 13 0.0
0 14 0.0
1 14 0.0
2 14 0.0
3 14 0.0
4 14 0.013
5 14 0.0
6 14 0.0
7 14 0.013
8 14 0.0
9 14 0.0
10 14 0.0
11 14 0.0
12 14 0.0
13 14 0.0
14 14 0.658
15 14 0.0
16 14 0.316
17 14 0.0
18 14 0.0
19 14 0.0
20 14 0.0
0 15 0.0
1 15 0.0
2 15 0.0
3 15 0.0
4 15 0.0
5 15 0.0
6 15 0.0
7 15 0.052
8 15 0.0
9 15 0.017
10 15 0.0
11 15 0.0
12 15 0.0
13 15 0.0
14 15 0.0
15 15 0.672
16 15 0.224
17 15 0.034
18 15 0.0
19 15 0.0
20 15 0.0
0 16 0.0
1 16 0.0
2 16 0.0
3 16 0.0
4 16 0.0
5 16 0.0
6 16 0.0
7 16 0.006
8 16 0.0
9 16 0.0
10 16 0.006
11 16 0.0
12 16 0.0
13 16 0.0
14 16 0.0
15 16 0.0
16 16 0.966
17 16 0.0
18 16 0.006
19 16 0.006
20 16 0.011
0 17 0.0
1 17 0.0
2 17 0.0
3 17 0.0
4 17 0.0
5 17 0.0
6 17 0.0
7 17 0.0
8 17 0.0
9 17 0.0
10 17 0.0
11 17 0.0
12 17 0.0
13 17 0.0
14 17 0.0
15 17 0.0
16 17 0.005
17 17 0.995
18 17 0.0
19 17 0.0
20 17 0.0
0 18 0.0
1 18 0.0
2 18 0.0
3 18 0.0
4 18 0.0
5 18 0.0
6 18 0.0
7 18 0.012
8 18 0.0
9 18 0.0
10 18 0.0
11 18 0.0
12 18 0.0
13 18 0.0
14 18 0.0
15 18 0.0
16 18 0.0
17 18 0.0
18 18 0.964
19 18 0.018
20 18 0.006
0 19 0.0
1 19 0.0
2 19 0.0
3 19 0.0
4 19 0.0
5 19 0.006
6 19 0.0
7 19 0.006
8 19 0.0
9 19 0.0
10 19 0.0
11 19 0.0
12 19 0.0
13 19 0.0
14 19 0.006
15 19 0.0
16 19 0.0
17 19 0.0
18 19 0.0
19 19 0.982
20 19 0.0
0 20 0.0
1 20 0.0
2 20 0.0
3 20 0.0
4 20 0.0
5 20 0.018
6 20 0.0
7 20 0.036
8 20 0.0
9 20 0.0
10 20 0.0
11 20 0.018
12 20 0.0
13 20 0.0
14 20 0.055
15 20 0.0
16 20 0.055
17 20 0.0
18 20 0.0
19 20 0.018
20 20 0.8
        };
    \end{axis}
\end{tikzpicture}
%
%\caption{I'm confused~5!}\label{tab:CM5}
%\end{figure*}
	}
& &
	\resizebox{0.45\textwidth}{0.25\textheight}{%
	%\begin{tikzpicture}[scale=0.60,every node/.style={scale=0.9}]
%\begin{tikzpicture}[scale=0.55]
\begin{tikzpicture}[scale=0.60]
    \begin{axis}[%colorbar/width=2.5mm,
        width=20cm,
        height=20cm,
%        colormap={blackwhite}{gray(0cm)=(1); gray(1cm)=(0.5)},
%   colormap={bluewhite}{color=(white) color=(blue)},
%   colormap={bluewhite}{color=(white) rgb255=(0,191,255)},
    colormap={bluewhite}{color=(white) rgb255=(100,149,237)},
        xticklabels={Adload,Agent,Allaple,BHO,Bifrose,CeeInject,Cycbot,FakeRean,Hotbar,Injector,
        OnLineGames,Renos,Rimecud,Small,Toga,VB,VBinject,Vobfus,Vundo,Winwebsec,Zbot},
        xtick={0,...,20},
        xtick style={draw=none},
    xticklabel style={anchor=east,rotate=45,yshift=-5pt},
        yticklabels={Adload,Agent,Allaple,BHO,Bifrose,CeeInject,Cycbot,FakeRean,Hotbar,Injector,
        OnLineGames,Renos,Rimecud,Small,Toga,VB,VBinject,Vobfus,Vundo,Winwebsec,Zbot},
        ytick={0,...,20},
        ytick style={draw=none},
        enlargelimits=false,
        xticklabel style={font=\large},
        yticklabel style={font=\large},
        colorbar,
        colorbar style={
%           width=0.05*\pgfkeysvalueof{/pgfplots/parent axis width},%%% added this
%           height=0.5*\pgfkeysvalueof{/pgfplots/parent axis height},
%       plot graphics/node/.style={scale=1.33,anchor=south west,inner sep=0pt,}, %%% scale colorbar fill %%%
            ytick={0.0,0.2,0.4,0.6,0.8,1.0},
            yticklabels={0.0,0.2,0.4,0.6,0.8,1.0},
            yticklabel={\pgfmathprintnumber\tick},
            yticklabel style={font=\large,
                    /pgf/number format/fixed,
            /pgf/number format/precision=1}
        },
%        point meta min=0,
%        point meta max=100,
        point meta min=0.0,
        point meta max=1.0,
        nodes near coords={\pgfmathprintnumber\pgfplotspointmeta},
        % ---------------------------------------------------------------------
        % show `nodes near coords' but adapt the style so that values
        % above a threshold get another style
        % (adapted from <http://tex.stackexchange.com/a/141006/95441>)
        % #1: the THRESHOLD after which we switch to a special display.
        nodes near coords black white/.style={
            % define the style of the nodes with "small" values
            small value/.style={
                yshift=-7pt,
%                text=white,
                text=black,
                /pgf/number format/fixed,
                /pgf/number format/precision=3,
                /pgf/number format/zerofill
            },
            % define the style of the nodes with "large" values
            large value/.style={
                yshift=-7pt,
%                text=black,
                text=white,
                /pgf/number format/fixed,
                /pgf/number format/precision=3,
                /pgf/number format/zerofill
            },
            every node near coord/.style={
                check for zero/.code={
                    \pgfmathfloatifflags{\pgfplotspointmeta}{0}{
                        % If meta=0, make the node a coordinate
                        % (which doesn't have text)
                        \pgfkeys{/tikz/coordinate}
                    }{
                        \begingroup
                        % this group is merely to switch to FPU locally.
                        % Might be unnecessary, but who knows.
                        \pgfkeys{/pgf/fpu}
                        \pgfmathparse{\pgfplotspointmeta<#1}
                        \global\let\result=\pgfmathresult
                        \endgroup
                        %
                        % simplifies debugging:
                        %\show\result
                        %
                        \pgfmathfloatcreate{1}{1.0}{0}
                        \let\ONE=\pgfmathresult
                        \ifx\result\ONE
                            % AH: our condition 'y < #1' is met.
                            \pgfkeysalso{/pgfplots/small value}
                        \else
                            % ok, proceed as usual.
                            \pgfkeysalso{/pgfplots/large value}
                        \fi
                    }
                },
                check for zero,
            },
        },
        % asign a value to the new style which is the threshold at which
        % the two style `small value' or `large value' are used
%        nodes near coords black white=50,
        nodes near coords black white=0.5,
        % -----------------------------------------------------------------
    ]
        \addplot[
            matrix plot,
            mesh/cols=21,
            point meta=explicit,draw=gray
        ] table [meta=C] {
            x y C
0 0 0.952
1 0 0.0
2 0 0.0
3 0 0.024
4 0 0.0
5 0 0.0
6 0 0.0
7 0 0.0
8 0 0.0
9 0 0.0
10 0 0.0
11 0 0.024
12 0 0.0
13 0 0.0
14 0 0.0
15 0 0.0
16 0 0.0
17 0 0.0
18 0 0.0
19 0 0.0
20 0 0.0
0 1 0.0
1 1 0.514
2 1 0.0
3 1 0.029
4 1 0.0
5 1 0.0
6 1 0.0
7 1 0.086
8 1 0.0
9 1 0.057
10 1 0.0
11 1 0.057
12 1 0.0
13 1 0.0
14 1 0.0
15 1 0.057
16 1 0.086
17 1 0.029
18 1 0.029
19 1 0.057
20 1 0.0
0 2 0.0
1 2 0.0
2 2 1.0
3 2 0.0
4 2 0.0
5 2 0.0
6 2 0.0
7 2 0.0
8 2 0.0
9 2 0.0
10 2 0.0
11 2 0.0
12 2 0.0
13 2 0.0
14 2 0.0
15 2 0.0
16 2 0.0
17 2 0.0
18 2 0.0
19 2 0.0
20 2 0.0
0 3 0.042
1 3 0.0
2 3 0.0
3 3 0.915
4 3 0.0
5 3 0.014
6 3 0.0
7 3 0.0
8 3 0.0
9 3 0.0
10 3 0.0
11 3 0.0
12 3 0.0
13 3 0.0
14 3 0.0
15 3 0.014
16 3 0.0
17 3 0.0
18 3 0.014
19 3 0.0
20 3 0.0
0 4 0.0
1 4 0.026
2 4 0.053
3 4 0.0
4 4 0.5
5 4 0.053
6 4 0.0
7 4 0.0
8 4 0.0
9 4 0.053
10 4 0.0
11 4 0.0
12 4 0.026
13 4 0.0
14 4 0.026
15 4 0.026
16 4 0.211
17 4 0.0
18 4 0.0
19 4 0.026
20 4 0.0
0 5 0.0
1 5 0.019
2 5 0.013
3 5 0.0
4 5 0.025
5 5 0.892
6 5 0.0
7 5 0.006
8 5 0.0
9 5 0.013
10 5 0.0
11 5 0.006
12 5 0.0
13 5 0.006
14 5 0.0
15 5 0.0
16 5 0.006
17 5 0.0
18 5 0.0
19 5 0.006
20 5 0.006
0 6 0.0
1 6 0.0
2 6 0.0
3 6 0.0
4 6 0.0
5 6 0.0
6 6 0.984
7 6 0.0
8 6 0.0
9 6 0.0
10 6 0.0
11 6 0.0
12 6 0.0
13 6 0.0
14 6 0.0
15 6 0.0
16 6 0.0
17 6 0.0
18 6 0.0
19 6 0.016
20 6 0.0
0 7 0.0
1 7 0.011
2 7 0.011
3 7 0.0
4 7 0.0
5 7 0.0
6 7 0.032
7 7 0.926
8 7 0.0
9 7 0.0
10 7 0.0
11 7 0.011
12 7 0.0
13 7 0.0
14 7 0.0
15 7 0.0
16 7 0.011
17 7 0.0
18 7 0.0
19 7 0.0
20 7 0.0
0 8 0.0
1 8 0.0
2 8 0.0
3 8 0.0
4 8 0.0
5 8 0.0
6 8 0.0
7 8 0.0
8 8 1.0
9 8 0.0
10 8 0.0
11 8 0.0
12 8 0.0
13 8 0.0
14 8 0.0
15 8 0.0
16 8 0.0
17 8 0.0
18 8 0.0
19 8 0.0
20 8 0.0
0 9 0.0
1 9 0.032
2 9 0.032
3 9 0.0
4 9 0.0
5 9 0.097
6 9 0.0
7 9 0.065
8 9 0.0
9 9 0.484
10 9 0.032
11 9 0.032
12 9 0.0
13 9 0.0
14 9 0.0
15 9 0.032
16 9 0.161
17 9 0.0
18 9 0.032
19 9 0.0
20 9 0.0
0 10 0.0
1 10 0.0
2 10 0.043
3 10 0.0
4 10 0.0
5 10 0.0
6 10 0.0
7 10 0.0
8 10 0.0
9 10 0.021
10 10 0.809
11 10 0.043
12 10 0.0
13 10 0.0
14 10 0.0
15 10 0.0
16 10 0.0
17 10 0.0
18 10 0.064
19 10 0.021
20 10 0.0
0 11 0.0
1 11 0.0
2 11 0.01
3 11 0.0
4 11 0.0
5 11 0.0
6 11 0.0
7 11 0.0
8 11 0.0
9 11 0.0
10 11 0.0
11 11 0.943
12 11 0.01
13 11 0.0
14 11 0.01
15 11 0.0
16 11 0.0
17 11 0.0
18 11 0.029
19 11 0.0
20 11 0.0
0 12 0.0
1 12 0.0
2 12 0.088
3 12 0.0
4 12 0.0
5 12 0.0
6 12 0.0
7 12 0.0
8 12 0.0
9 12 0.0
10 12 0.0
11 12 0.029
12 12 0.882
13 12 0.0
14 12 0.0
15 12 0.0
16 12 0.0
17 12 0.0
18 12 0.0
19 12 0.0
20 12 0.0
0 13 0.029
1 13 0.057
2 13 0.057
3 13 0.029
4 13 0.0
5 13 0.057
6 13 0.0
7 13 0.0
8 13 0.0
9 13 0.029
10 13 0.0
11 13 0.057
12 13 0.0
13 13 0.543
14 13 0.029
15 13 0.0
16 13 0.057
17 13 0.0
18 13 0.0
19 13 0.057
20 13 0.0
0 14 0.0
1 14 0.013
2 14 0.0
3 14 0.0
4 14 0.025
5 14 0.0
6 14 0.013
7 14 0.0
8 14 0.0
9 14 0.013
10 14 0.025
11 14 0.013
12 14 0.0
13 14 0.0
14 14 0.696
15 14 0.0
16 14 0.177
17 14 0.013
18 14 0.0
19 14 0.013
20 14 0.0
0 15 0.0
1 15 0.034
2 15 0.017
3 15 0.0
4 15 0.0
5 15 0.0
6 15 0.017
7 15 0.0
8 15 0.0
9 15 0.0
10 15 0.0
11 15 0.0
12 15 0.0
13 15 0.0
14 15 0.0
15 15 0.569
16 15 0.172
17 15 0.155
18 15 0.0
19 15 0.034
20 15 0.0
0 16 0.0
1 16 0.0
2 16 0.006
3 16 0.0
4 16 0.011
5 16 0.0
6 16 0.0
7 16 0.006
8 16 0.0
9 16 0.0
10 16 0.006
11 16 0.006
12 16 0.006
13 16 0.0
14 16 0.011
15 16 0.034
16 16 0.82
17 16 0.067
18 16 0.017
19 16 0.011
20 16 0.0
0 17 0.0
1 17 0.0
2 17 0.0
3 17 0.0
4 17 0.0
5 17 0.005
6 17 0.0
7 17 0.0
8 17 0.0
9 17 0.0
10 17 0.0
11 17 0.0
12 17 0.0
13 17 0.0
14 17 0.005
15 17 0.005
16 17 0.026
17 17 0.958
18 17 0.0
19 17 0.0
20 17 0.0
0 18 0.0
1 18 0.006
2 18 0.0
3 18 0.0
4 18 0.0
5 18 0.0
6 18 0.0
7 18 0.018
8 18 0.0
9 18 0.0
10 18 0.0
11 18 0.0
12 18 0.0
13 18 0.0
14 18 0.0
15 18 0.0
16 18 0.0
17 18 0.0
18 18 0.958
19 18 0.012
20 18 0.006
0 19 0.0
1 19 0.0
2 19 0.0
3 19 0.0
4 19 0.0
5 19 0.0
6 19 0.012
7 19 0.006
8 19 0.0
9 19 0.0
10 19 0.0
11 19 0.0
12 19 0.0
13 19 0.0
14 19 0.0
15 19 0.0
16 19 0.0
17 19 0.0
18 19 0.0
19 19 0.982
20 19 0.0
0 20 0.0
1 20 0.0
2 20 0.018
3 20 0.0
4 20 0.0
5 20 0.0
6 20 0.0
7 20 0.018
8 20 0.0
9 20 0.018
10 20 0.0
11 20 0.036
12 20 0.018
13 20 0.0
14 20 0.145
15 20 0.0
16 20 0.018
17 20 0.036
18 20 0.0
19 20 0.055
20 20 0.636
        };
    \end{axis}
\end{tikzpicture}
%
%\caption{I'm confused~5!}\label{tab:CM5}
%\end{figure*}
	}
\\
(c) Bagged neural networks
& &
(d) Classic techniques
\\ \\
	\resizebox{0.45\textwidth}{0.25\textheight}{%
	%\begin{tikzpicture}[scale=0.60,every node/.style={scale=0.9}]
%\begin{tikzpicture}[scale=0.55]
\begin{tikzpicture}[scale=0.60]
    \begin{axis}[%colorbar/width=2.5mm,
        width=20cm,
        height=20cm,
%        colormap={blackwhite}{gray(0cm)=(1); gray(1cm)=(0.5)},
%   colormap={bluewhite}{color=(white) color=(blue)},
%   colormap={bluewhite}{color=(white) rgb255=(0,191,255)},
    colormap={bluewhite}{color=(white) rgb255=(100,149,237)},
        xticklabels={Adload,Agent,Allaple,BHO,Bifrose,CeeInject,Cycbot,FakeRean,Hotbar,Injector,
        OnLineGames,Renos,Rimecud,Small,Toga,VB,VBinject,Vobfus,Vundo,Winwebsec,Zbot},
        xtick={0,...,20},
        xtick style={draw=none},
    xticklabel style={anchor=east,rotate=45,yshift=-5pt},
        yticklabels={Adload,Agent,Allaple,BHO,Bifrose,CeeInject,Cycbot,FakeRean,Hotbar,Injector,
        OnLineGames,Renos,Rimecud,Small,Toga,VB,VBinject,Vobfus,Vundo,Winwebsec,Zbot},
        ytick={0,...,20},
        ytick style={draw=none},
        enlargelimits=false,
        xticklabel style={font=\large},
        yticklabel style={font=\large},
        colorbar,
        colorbar style={
%           width=0.05*\pgfkeysvalueof{/pgfplots/parent axis width},%%% added this
%           height=0.5*\pgfkeysvalueof{/pgfplots/parent axis height},
%       plot graphics/node/.style={scale=1.33,anchor=south west,inner sep=0pt,}, %%% scale colorbar fill %%%
            ytick={0.0,0.2,0.4,0.6,0.8,1.0},
            yticklabels={0.0,0.2,0.4,0.6,0.8,1.0},
            yticklabel={\pgfmathprintnumber\tick},
            yticklabel style={font=\large,
                    /pgf/number format/fixed,
            /pgf/number format/precision=1}
        },
%        point meta min=0,
%        point meta max=100,
        point meta min=0.0,
        point meta max=1.0,
        nodes near coords={\pgfmathprintnumber\pgfplotspointmeta},
        % ---------------------------------------------------------------------
        % show `nodes near coords' but adapt the style so that values
        % above a threshold get another style
        % (adapted from <http://tex.stackexchange.com/a/141006/95441>)
        % #1: the THRESHOLD after which we switch to a special display.
        nodes near coords black white/.style={
            % define the style of the nodes with "small" values
            small value/.style={
                yshift=-7pt,
%                text=white,
                text=black,
                /pgf/number format/fixed,
                /pgf/number format/precision=3,
                /pgf/number format/zerofill
            },
            % define the style of the nodes with "large" values
            large value/.style={
                yshift=-7pt,
%                text=black,
                text=white,
                /pgf/number format/fixed,
                /pgf/number format/precision=3,
                /pgf/number format/zerofill
            },
            every node near coord/.style={
                check for zero/.code={
                    \pgfmathfloatifflags{\pgfplotspointmeta}{0}{
                        % If meta=0, make the node a coordinate
                        % (which doesn't have text)
                        \pgfkeys{/tikz/coordinate}
                    }{
                        \begingroup
                        % this group is merely to switch to FPU locally.
                        % Might be unnecessary, but who knows.
                        \pgfkeys{/pgf/fpu}
                        \pgfmathparse{\pgfplotspointmeta<#1}
                        \global\let\result=\pgfmathresult
                        \endgroup
                        %
                        % simplifies debugging:
                        %\show\result
                        %
                        \pgfmathfloatcreate{1}{1.0}{0}
                        \let\ONE=\pgfmathresult
                        \ifx\result\ONE
                            % AH: our condition 'y < #1' is met.
                            \pgfkeysalso{/pgfplots/small value}
                        \else
                            % ok, proceed as usual.
                            \pgfkeysalso{/pgfplots/large value}
                        \fi
                    }
                },
                check for zero,
            },
        },
        % asign a value to the new style which is the threshold at which
        % the two style `small value' or `large value' are used
%        nodes near coords black white=50,
        nodes near coords black white=0.5,
        % -----------------------------------------------------------------
    ]
        \addplot[
            matrix plot,
            mesh/cols=21,
            point meta=explicit,draw=gray
        ] table [meta=C] {
            x y C
0 0 0.976
1 0 0.0
2 0 0.0
3 0 0.024
4 0 0.0
5 0 0.0
6 0 0.0
7 0 0.0
8 0 0.0
9 0 0.0
10 0 0.0
11 0 0.0
12 0 0.0
13 0 0.0
14 0 0.0
15 0 0.0
16 0 0.0
17 0 0.0
18 0 0.0
19 0 0.0
20 0 0.0
0 1 0.0
1 1 0.771
2 1 0.0
3 1 0.0
4 1 0.0
5 1 0.0
6 1 0.0
7 1 0.029
8 1 0.0
9 1 0.0
10 1 0.0
11 1 0.029
12 1 0.0
13 1 0.029
14 1 0.0
15 1 0.0
16 1 0.057
17 1 0.029
18 1 0.057
19 1 0.0
20 1 0.0
0 2 0.0
1 2 0.0
2 2 1.0
3 2 0.0
4 2 0.0
5 2 0.0
6 2 0.0
7 2 0.0
8 2 0.0
9 2 0.0
10 2 0.0
11 2 0.0
12 2 0.0
13 2 0.0
14 2 0.0
15 2 0.0
16 2 0.0
17 2 0.0
18 2 0.0
19 2 0.0
20 2 0.0
0 3 0.042
1 3 0.0
2 3 0.0
3 3 0.915
4 3 0.0
5 3 0.0
6 3 0.0
7 3 0.0
8 3 0.0
9 3 0.0
10 3 0.0
11 3 0.0
12 3 0.0
13 3 0.0
14 3 0.0
15 3 0.0
16 3 0.014
17 3 0.0
18 3 0.028
19 3 0.0
20 3 0.0
0 4 0.0
1 4 0.026
2 4 0.0
3 4 0.0
4 4 0.342
5 4 0.158
6 4 0.0
7 4 0.0
8 4 0.0
9 4 0.026
10 4 0.0
11 4 0.0
12 4 0.0
13 4 0.0
14 4 0.0
15 4 0.0
16 4 0.395
17 4 0.0
18 4 0.0
19 4 0.053
20 4 0.0
0 5 0.0
1 5 0.0
2 5 0.0
3 5 0.0
4 5 0.0
5 5 0.956
6 5 0.0
7 5 0.0
8 5 0.0
9 5 0.0
10 5 0.0
11 5 0.019
12 5 0.0
13 5 0.0
14 5 0.0
15 5 0.0
16 5 0.006
17 5 0.0
18 5 0.006
19 5 0.006
20 5 0.006
0 6 0.0
1 6 0.0
2 6 0.0
3 6 0.0
4 6 0.0
5 6 0.0
6 6 0.984
7 6 0.008
8 6 0.0
9 6 0.0
10 6 0.0
11 6 0.0
12 6 0.0
13 6 0.0
14 6 0.0
15 6 0.0
16 6 0.0
17 6 0.0
18 6 0.0
19 6 0.008
20 6 0.0
0 7 0.0
1 7 0.0
2 7 0.0
3 7 0.0
4 7 0.0
5 7 0.0
6 7 0.0
7 7 0.989
8 7 0.0
9 7 0.0
10 7 0.0
11 7 0.0
12 7 0.0
13 7 0.0
14 7 0.011
15 7 0.0
16 7 0.0
17 7 0.0
18 7 0.0
19 7 0.0
20 7 0.0
0 8 0.0
1 8 0.0
2 8 0.0
3 8 0.0
4 8 0.0
5 8 0.0
6 8 0.0
7 8 0.0
8 8 1.0
9 8 0.0
10 8 0.0
11 8 0.0
12 8 0.0
13 8 0.0
14 8 0.0
15 8 0.0
16 8 0.0
17 8 0.0
18 8 0.0
19 8 0.0
20 8 0.0
0 9 0.0
1 9 0.0
2 9 0.0
3 9 0.0
4 9 0.032
5 9 0.161
6 9 0.0
7 9 0.0
8 9 0.0
9 9 0.516
10 9 0.0
11 9 0.0
12 9 0.0
13 9 0.0
14 9 0.0
15 9 0.0
16 9 0.194
17 9 0.0
18 9 0.065
19 9 0.032
20 9 0.0
0 10 0.0
1 10 0.021
2 10 0.0
3 10 0.0
4 10 0.0
5 10 0.0
6 10 0.0
7 10 0.021
8 10 0.0
9 10 0.0
10 10 0.894
11 10 0.0
12 10 0.0
13 10 0.0
14 10 0.0
15 10 0.0
16 10 0.0
17 10 0.0
18 10 0.064
19 10 0.0
20 10 0.0
0 11 0.0
1 11 0.0
2 11 0.0
3 11 0.01
4 11 0.0
5 11 0.0
6 11 0.0
7 11 0.0
8 11 0.0
9 11 0.0
10 11 0.0
11 11 0.99
12 11 0.0
13 11 0.0
14 11 0.0
15 11 0.0
16 11 0.0
17 11 0.0
18 11 0.0
19 11 0.0
20 11 0.0
0 12 0.0
1 12 0.0
2 12 0.0
3 12 0.0
4 12 0.0
5 12 0.0
6 12 0.0
7 12 0.0
8 12 0.0
9 12 0.0
10 12 0.0
11 12 0.0
12 12 0.971
13 12 0.0
14 12 0.0
15 12 0.0
16 12 0.029
17 12 0.0
18 12 0.0
19 12 0.0
20 12 0.0
0 13 0.0
1 13 0.057
2 13 0.0
3 13 0.0
4 13 0.0
5 13 0.0
6 13 0.0
7 13 0.0
8 13 0.0
9 13 0.0
10 13 0.0
11 13 0.0
12 13 0.0
13 13 0.743
14 13 0.0
15 13 0.0
16 13 0.114
17 13 0.0
18 13 0.029
19 13 0.057
20 13 0.0
0 14 0.0
1 14 0.0
2 14 0.0
3 14 0.0
4 14 0.013
5 14 0.0
6 14 0.0
7 14 0.013
8 14 0.0
9 14 0.0
10 14 0.0
11 14 0.0
12 14 0.0
13 14 0.0
14 14 0.658
15 14 0.0
16 14 0.316
17 14 0.0
18 14 0.0
19 14 0.0
20 14 0.0
0 15 0.0
1 15 0.0
2 15 0.0
3 15 0.0
4 15 0.0
5 15 0.0
6 15 0.0
7 15 0.052
8 15 0.0
9 15 0.017
10 15 0.0
11 15 0.0
12 15 0.0
13 15 0.0
14 15 0.0
15 15 0.672
16 15 0.224
17 15 0.034
18 15 0.0
19 15 0.0
20 15 0.0
0 16 0.0
1 16 0.0
2 16 0.0
3 16 0.0
4 16 0.0
5 16 0.0
6 16 0.0
7 16 0.006
8 16 0.0
9 16 0.0
10 16 0.006
11 16 0.0
12 16 0.0
13 16 0.0
14 16 0.0
15 16 0.0
16 16 0.966
17 16 0.0
18 16 0.006
19 16 0.006
20 16 0.011
0 17 0.0
1 17 0.0
2 17 0.0
3 17 0.0
4 17 0.0
5 17 0.0
6 17 0.0
7 17 0.0
8 17 0.0
9 17 0.0
10 17 0.0
11 17 0.0
12 17 0.0
13 17 0.0
14 17 0.0
15 17 0.0
16 17 0.005
17 17 0.995
18 17 0.0
19 17 0.0
20 17 0.0
0 18 0.0
1 18 0.0
2 18 0.0
3 18 0.0
4 18 0.0
5 18 0.0
6 18 0.0
7 18 0.012
8 18 0.0
9 18 0.0
10 18 0.0
11 18 0.0
12 18 0.0
13 18 0.0
14 18 0.0
15 18 0.0
16 18 0.0
17 18 0.0
18 18 0.964
19 18 0.018
20 18 0.006
0 19 0.0
1 19 0.0
2 19 0.0
3 19 0.0
4 19 0.0
5 19 0.006
6 19 0.0
7 19 0.006
8 19 0.0
9 19 0.0
10 19 0.0
11 19 0.0
12 19 0.0
13 19 0.0
14 19 0.006
15 19 0.0
16 19 0.0
17 19 0.0
18 19 0.0
19 19 0.982
20 19 0.0
0 20 0.0
1 20 0.0
2 20 0.0
3 20 0.0
4 20 0.0
5 20 0.018
6 20 0.0
7 20 0.036
8 20 0.0
9 20 0.0
10 20 0.0
11 20 0.018
12 20 0.0
13 20 0.0
14 20 0.055
15 20 0.0
16 20 0.055
17 20 0.0
18 20 0.0
19 20 0.018
20 20 0.8
        };
    \end{axis}
\end{tikzpicture}
%
%\caption{I'm confused~5!}\label{tab:CM5}
%\end{figure*}
	}
& & 
	\resizebox{0.45\textwidth}{0.25\textheight}{%
	%\begin{tikzpicture}[scale=0.60,every node/.style={scale=0.9}]
%\begin{tikzpicture}[scale=0.55]
\begin{tikzpicture}[scale=0.60]
    \begin{axis}[%colorbar/width=2.5mm,
        width=20cm,
        height=20cm,
%        colormap={blackwhite}{gray(0cm)=(1); gray(1cm)=(0.5)},
%   colormap={bluewhite}{color=(white) color=(blue)},
%   colormap={bluewhite}{color=(white) rgb255=(0,191,255)},
    colormap={bluewhite}{color=(white) rgb255=(100,149,237)},
        xticklabels={Adload,Agent,Allaple,BHO,Bifrose,CeeInject,Cycbot,FakeRean,Hotbar,Injector,
        OnLineGames,Renos,Rimecud,Small,Toga,VB,VBinject,Vobfus,Vundo,Winwebsec,Zbot},
        xtick={0,...,20},
        xtick style={draw=none},
    xticklabel style={anchor=east,rotate=45,yshift=-5pt},
        yticklabels={Adload,Agent,Allaple,BHO,Bifrose,CeeInject,Cycbot,FakeRean,Hotbar,Injector,
        OnLineGames,Renos,Rimecud,Small,Toga,VB,VBinject,Vobfus,Vundo,Winwebsec,Zbot},
        ytick={0,...,20},
        ytick style={draw=none},
        enlargelimits=false,
        xticklabel style={font=\large},
        yticklabel style={font=\large},
        colorbar,
        colorbar style={
%           width=0.05*\pgfkeysvalueof{/pgfplots/parent axis width},%%% added this
%           height=0.5*\pgfkeysvalueof{/pgfplots/parent axis height},
%       plot graphics/node/.style={scale=1.33,anchor=south west,inner sep=0pt,}, %%% scale colorbar fill %%%
            ytick={0.0,0.2,0.4,0.6,0.8,1.0},
            yticklabels={0.0,0.2,0.4,0.6,0.8,1.0},
            yticklabel={\pgfmathprintnumber\tick},
            yticklabel style={font=\large,
                    /pgf/number format/fixed,
            /pgf/number format/precision=1}
        },
%        point meta min=0,
%        point meta max=100,
        point meta min=0.0,
        point meta max=1.0,
        nodes near coords={\pgfmathprintnumber\pgfplotspointmeta},
        % ---------------------------------------------------------------------
        % show `nodes near coords' but adapt the style so that values
        % above a threshold get another style
        % (adapted from <http://tex.stackexchange.com/a/141006/95441>)
        % #1: the THRESHOLD after which we switch to a special display.
        nodes near coords black white/.style={
            % define the style of the nodes with "small" values
            small value/.style={
                yshift=-7pt,
%                text=white,
                text=black,
                /pgf/number format/fixed,
                /pgf/number format/precision=3,
                /pgf/number format/zerofill
            },
            % define the style of the nodes with "large" values
            large value/.style={
                yshift=-7pt,
%                text=black,
                text=white,
                /pgf/number format/fixed,
                /pgf/number format/precision=3,
                /pgf/number format/zerofill
            },
            every node near coord/.style={
                check for zero/.code={
                    \pgfmathfloatifflags{\pgfplotspointmeta}{0}{
                        % If meta=0, make the node a coordinate
                        % (which doesn't have text)
                        \pgfkeys{/tikz/coordinate}
                    }{
                        \begingroup
                        % this group is merely to switch to FPU locally.
                        % Might be unnecessary, but who knows.
                        \pgfkeys{/pgf/fpu}
                        \pgfmathparse{\pgfplotspointmeta<#1}
                        \global\let\result=\pgfmathresult
                        \endgroup
                        %
                        % simplifies debugging:
                        %\show\result
                        %
                        \pgfmathfloatcreate{1}{1.0}{0}
                        \let\ONE=\pgfmathresult
                        \ifx\result\ONE
                            % AH: our condition 'y < #1' is met.
                            \pgfkeysalso{/pgfplots/small value}
                        \else
                            % ok, proceed as usual.
                            \pgfkeysalso{/pgfplots/large value}
                        \fi
                    }
                },
                check for zero,
            },
        },
        % asign a value to the new style which is the threshold at which
        % the two style `small value' or `large value' are used
%        nodes near coords black white=50,
        nodes near coords black white=0.5,
        % -----------------------------------------------------------------
    ]
        \addplot[
            matrix plot,
            mesh/cols=21,
            point meta=explicit,draw=gray
        ] table [meta=C] {
            x y C
0 0 0.976
1 0 0.0
2 0 0.0
3 0 0.024
4 0 0.0
5 0 0.0
6 0 0.0
7 0 0.0
8 0 0.0
9 0 0.0
10 0 0.0
11 0 0.0
12 0 0.0
13 0 0.0
14 0 0.0
15 0 0.0
16 0 0.0
17 0 0.0
18 0 0.0
19 0 0.0
20 0 0.0
0 1 0.0
1 1 0.6
2 1 0.0
3 1 0.0
4 1 0.0
5 1 0.0
6 1 0.0
7 1 0.057
8 1 0.0
9 1 0.0
10 1 0.029
11 1 0.057
12 1 0.0
13 1 0.029
14 1 0.0
15 1 0.029
16 1 0.057
17 1 0.029
18 1 0.057
19 1 0.057
20 1 0.0
0 2 0.0
1 2 0.0
2 2 1.0
3 2 0.0
4 2 0.0
5 2 0.0
6 2 0.0
7 2 0.0
8 2 0.0
9 2 0.0
10 2 0.0
11 2 0.0
12 2 0.0
13 2 0.0
14 2 0.0
15 2 0.0
16 2 0.0
17 2 0.0
18 2 0.0
19 2 0.0
20 2 0.0
0 3 0.042
1 3 0.0
2 3 0.0
3 3 0.901
4 3 0.0
5 3 0.0
6 3 0.0
7 3 0.0
8 3 0.0
9 3 0.0
10 3 0.0
11 3 0.0
12 3 0.0
13 3 0.0
14 3 0.0
15 3 0.0
16 3 0.014
17 3 0.0
18 3 0.042
19 3 0.0
20 3 0.0
0 4 0.0
1 4 0.026
2 4 0.0
3 4 0.0
4 4 0.447
5 4 0.079
6 4 0.0
7 4 0.0
8 4 0.0
9 4 0.053
10 4 0.0
11 4 0.0
12 4 0.0
13 4 0.0
14 4 0.0
15 4 0.0
16 4 0.368
17 4 0.0
18 4 0.0
19 4 0.026
20 4 0.0
0 5 0.0
1 5 0.0
2 5 0.0
3 5 0.0
4 5 0.0
5 5 0.968
6 5 0.0
7 5 0.0
8 5 0.0
9 5 0.006
10 5 0.0
11 5 0.013
12 5 0.0
13 5 0.0
14 5 0.0
15 5 0.0
16 5 0.006
17 5 0.0
18 5 0.0
19 5 0.006
20 5 0.0
0 6 0.0
1 6 0.0
2 6 0.0
3 6 0.0
4 6 0.0
5 6 0.0
6 6 0.984
7 6 0.008
8 6 0.0
9 6 0.0
10 6 0.0
11 6 0.0
12 6 0.0
13 6 0.0
14 6 0.0
15 6 0.0
16 6 0.0
17 6 0.0
18 6 0.0
19 6 0.008
20 6 0.0
0 7 0.0
1 7 0.0
2 7 0.0
3 7 0.0
4 7 0.0
5 7 0.0
6 7 0.0
7 7 1.0
8 7 0.0
9 7 0.0
10 7 0.0
11 7 0.0
12 7 0.0
13 7 0.0
14 7 0.0
15 7 0.0
16 7 0.0
17 7 0.0
18 7 0.0
19 7 0.0
20 7 0.0
0 8 0.0
1 8 0.0
2 8 0.0
3 8 0.0
4 8 0.0
5 8 0.0
6 8 0.0
7 8 0.0
8 8 1.0
9 8 0.0
10 8 0.0
11 8 0.0
12 8 0.0
13 8 0.0
14 8 0.0
15 8 0.0
16 8 0.0
17 8 0.0
18 8 0.0
19 8 0.0
20 8 0.0
0 9 0.0
1 9 0.0
2 9 0.0
3 9 0.0
4 9 0.0
5 9 0.129
6 9 0.0
7 9 0.0
8 9 0.0
9 9 0.548
10 9 0.032
11 9 0.0
12 9 0.0
13 9 0.0
14 9 0.0
15 9 0.0
16 9 0.194
17 9 0.0
18 9 0.065
19 9 0.032
20 9 0.0
0 10 0.0
1 10 0.0
2 10 0.0
3 10 0.0
4 10 0.0
5 10 0.0
6 10 0.0
7 10 0.021
8 10 0.0
9 10 0.0
10 10 0.872
11 10 0.021
12 10 0.0
13 10 0.0
14 10 0.0
15 10 0.0
16 10 0.0
17 10 0.0
18 10 0.064
19 10 0.021
20 10 0.0
0 11 0.0
1 11 0.0
2 11 0.0
3 11 0.01
4 11 0.0
5 11 0.0
6 11 0.0
7 11 0.0
8 11 0.0
9 11 0.0
10 11 0.0
11 11 0.99
12 11 0.0
13 11 0.0
14 11 0.0
15 11 0.0
16 11 0.0
17 11 0.0
18 11 0.0
19 11 0.0
20 11 0.0
0 12 0.0
1 12 0.0
2 12 0.0
3 12 0.0
4 12 0.0
5 12 0.0
6 12 0.0
7 12 0.0
8 12 0.0
9 12 0.0
10 12 0.0
11 12 0.0
12 12 0.971
13 12 0.0
14 12 0.0
15 12 0.0
16 12 0.029
17 12 0.0
18 12 0.0
19 12 0.0
20 12 0.0
0 13 0.0
1 13 0.029
2 13 0.0
3 13 0.0
4 13 0.0
5 13 0.0
6 13 0.0
7 13 0.029
8 13 0.0
9 13 0.0
10 13 0.0
11 13 0.029
12 13 0.0
13 13 0.771
14 13 0.0
15 13 0.0
16 13 0.086
17 13 0.0
18 13 0.029
19 13 0.029
20 13 0.0
0 14 0.0
1 14 0.0
2 14 0.0
3 14 0.0
4 14 0.013
5 14 0.0
6 14 0.0
7 14 0.013
8 14 0.0
9 14 0.0
10 14 0.0
11 14 0.013
12 14 0.0
13 14 0.0
14 14 0.658
15 14 0.0
16 14 0.304
17 14 0.0
18 14 0.0
19 14 0.0
20 14 0.0
0 15 0.0
1 15 0.017
2 15 0.0
3 15 0.0
4 15 0.0
5 15 0.0
6 15 0.0
7 15 0.034
8 15 0.0
9 15 0.0
10 15 0.0
11 15 0.0
12 15 0.0
13 15 0.0
14 15 0.0
15 15 0.707
16 15 0.172
17 15 0.052
18 15 0.0
19 15 0.017
20 15 0.0
0 16 0.0
1 16 0.0
2 16 0.0
3 16 0.0
4 16 0.011
5 16 0.0
6 16 0.0
7 16 0.0
8 16 0.0
9 16 0.0
10 16 0.006
11 16 0.0
12 16 0.0
13 16 0.0
14 16 0.006
15 16 0.0
16 16 0.961
17 16 0.006
18 16 0.006
19 16 0.006
20 16 0.0
0 17 0.0
1 17 0.0
2 17 0.0
3 17 0.0
4 17 0.0
5 17 0.0
6 17 0.0
7 17 0.0
8 17 0.0
9 17 0.0
10 17 0.0
11 17 0.0
12 17 0.0
13 17 0.0
14 17 0.0
15 17 0.0
16 17 0.016
17 17 0.984
18 17 0.0
19 17 0.0
20 17 0.0
0 18 0.0
1 18 0.0
2 18 0.0
3 18 0.0
4 18 0.0
5 18 0.0
6 18 0.0
7 18 0.018
8 18 0.0
9 18 0.0
10 18 0.0
11 18 0.0
12 18 0.0
13 18 0.0
14 18 0.0
15 18 0.0
16 18 0.0
17 18 0.0
18 18 0.958
19 18 0.018
20 18 0.006
0 19 0.0
1 19 0.0
2 19 0.0
3 19 0.0
4 19 0.0
5 19 0.006
6 19 0.0
7 19 0.0
8 19 0.0
9 19 0.0
10 19 0.0
11 19 0.0
12 19 0.0
13 19 0.0
14 19 0.0
15 19 0.0
16 19 0.0
17 19 0.0
18 19 0.0
19 19 0.994
20 19 0.0
0 20 0.0
1 20 0.0
2 20 0.0
3 20 0.0
4 20 0.0
5 20 0.018
6 20 0.0
7 20 0.036
8 20 0.0
9 20 0.018
10 20 0.0
11 20 0.018
12 20 0.0
13 20 0.0
14 20 0.091
15 20 0.0
16 20 0.055
17 20 0.0
18 20 0.0
19 20 0.055
20 20 0.709
        };
    \end{axis}
\end{tikzpicture}
%
%\caption{I'm confused~5!}\label{tab:CM5}
%\end{figure*}
	}
\\
(e) All neural networks
& & 
(f) All models
	\end{tabular}
	\caption{Confusion matrices for voting ensembles}
	\label{fig:conf_3}
\end{figure}

%
% Voting all
%

%\begin{figure}[!htb]
%	\centering
%	\input figures/conf_vote_all.tex
%	\caption{Confusion Matrix for Voting Based on All Models}
%	\label{fig:conf_vote_all}
%\end{figure}

%\input conf_large.tex

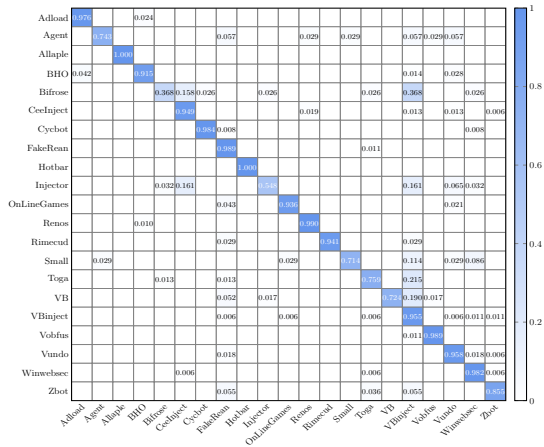
\end{document}